\documentclass[10pt,a4paper]{article}

\usepackage[english]{babel}
\usepackage{graphicx}
\usepackage{hyperref} 
\usepackage[hmargin=3.0cm,vmargin=3.0cm]{geometry}
\usepackage{amsmath}
\usepackage{amssymb}
\usepackage{amsfonts}
\usepackage{multicol}
\usepackage{slashed}
\usepackage{float}
\usepackage{caption}
\usepackage{subcaption}
\usepackage{amsthm}
\usepackage{cite}
\usepackage[compat=1.1.0]{tikz-feynman}

\newtheorem*{mydef}{Definition}

\title{\bf Bifurcations in the RG-Flow of QCD}
\author{Folkert~Kuipers$^1$\thanks{E-mail: F.Kuipers@sussex.ac.uk},
	$\ $
	Umut~G\"{u}rsoy$^1$\thanks{E-mail: U.Gursoy@uu.nl}
	$\ $
	and
	Yuri~Kuznetsov$^2$\thanks{E-mail: I.A.Kouzenetsov@uu.nl}
	\\
	$^1${\em Institute for Theoretical Physics, Utrecht University, 3584 CC Utrecht, The Netherlands}
	\\
	$^2${\em Mathematical Institute, Utrecht University, 3584 CC Utrecht, The Netherlands}
}

\begin{document}

\maketitle

\begin{abstract}
	Bifurcation analysis is used to study an effective model of QCD$_4$ with four-fermi interactions. Our analysis supports the scenario of a fixed point merger at the lower edge of the conformal window. This indicates square root scaling of the anomalous scaling dimensions of the fermion fields just above the lower edge and exponential scaling just below. We also predict existence of new fixed points in this model whose (dis)appearance may indicate transitions of the flow within the conformal window. Furthermore, we make new predictions for the critical value $(N_{f}/N_{c})_{\textrm{crit}}$ at the lower edge. We also obtain exotic spiraling flows that are generated by complex scaling dimensions of the effective four-fermi interactions. Finally, we extend the model by adding a  scalar field that couples with a Yukawa interaction term and study the modifications it causes to RG-flows. 
\end{abstract}

\section{Introduction}
Renormalization group describes the flow of a theory in the space spanned by the coupling constants called the ``theory space".  In quantum field theory renormalization group flows are governed by a set of beta functions that result in ordinary differential equations. The flow induced by these beta functions provide information about the change of coupling constants and anomalous field dimensions with the energy scale. An interesting aspect of the RG flows are the fixed points where the beta functions vanish. In particular one is interested in determining the positions of these fixed points and their stability properties. In addition one would like to obtain possible renormalized trajectories generated by the beta functions. Since the RG flows are generated by ordinary differential equations, it is natural to study them with the tools developed for the theory of dynamical systems. In this article, we use analytical and numerical\footnote{Throughout the article we make extensive use of the MATLAB package MatCont \cite{Matcont1,Matcont}} methods for bifurcation analysis to study {\em bifurcations} in renormalization group flows, as recently proposed by Gukov \cite{Gukov}.\\

Bifurcation theory, which we summarize in  Appendix \ref{Ap:Bifurcations}, categorizes topological changes in a dynamical system and is therefore an ideal tool to study the appearance and disappearance of fixed points. Using this categorization, we can not only determine easily when fixed points appear and disappear on the RG flow, but also obtain information about the nature of these transitions where the fixed points (dis)appear. As discussed in \cite{Kaplan}, there are three typical ways in which fixed points of an RG-flow (dis)appear:
\begin{itemize}
	\item Merger of a fixed point with non-trivial coupling constant and a fixed point with a trivial coupling constant. At this merger the stability of the non-trivial fixed point changes.
	\item Merger of two fixed point with non-trivial coupling constants. At this merger  both fixed points disappear.
	\item A divergent fixed-point as one or multiple coupling constants run off to infinity.
\end{itemize}
The first and the second scenarios are examples of standard bifurcations. The second is naturally described by the well known saddle-node bifurcation, while the first is called a transcritical bifurcation. It should be noted that in this scenario the fixed points usually do not merge as was stated in \cite{Kaplan} but collide and exchange stability. However, at this transition the coupling constant at the non-trivial fixed point generally becomes negative, hence moves into a physical uninteresting region. Furthermore, it is an easy exercise to show that the first scenario always induces a linear scaling of the anomalous dimension at the fixed point \cite{Gukov}, while the second scenario implies square root \cite{Gukov} and exponential (also known as BKT or Miransky scaling) scaling of the anomalous dimensions \cite{Kaplan} at the two sides of the merger. Finally, the third scenario above with diverging fixed points is much harder to describe using the techniques from the bifurcation theory. However, this scenario is mostly encountered in a perturbative analysis of the RG flows which becomes invalid anyway at the divergence. It turns out that in some cases this last scenario is replaced by the second one in an effective field theory.\\

In this paper, we apply bifurcation analysis to an effective model of QCD as suggested in \cite{Gukov}. We consider QCD$_4$ with gauge group $SU(N_c)$ with an arbitrary number of colors $N_c$ and flavors, $N_f$, as the bifurcation parameters. The Lagrangian is given by 
\begin{equation}\label{eq:Intro_Action}
	\mathcal{L} = -\frac{1}{4} G_{\mu\nu}^{A} G^{\mu\nu}_{A} + \sum_{i=1}^{N_{f}} \bar{\psi}_{a i} \left( i \slashed{D}_{b}^{a} - m_{i} \delta_{b}^{a} \right) \psi^{b i}\, ,
\end{equation}
where $G_A$ are the gluon field strengths and $\psi_i$ are the fermions of flavor $i$. Furthermore $a,b,c$ are used to label the color in the fundamental representation, and $A,B,C$ is used for the color in the adjoint representation. We have
\begin{equation*}
	G_{\mu\nu}^{A} = \partial_{\mu} A_{\nu}^{A} - \partial_{\nu} A_{\mu}^{A} + g f^{ABC} A_{\mu}^{B} A_{\nu}^{C} \quad \textrm{and} \quad \slashed{D} = \gamma^{\mu} \left( \partial_{\mu} - i g t_{A} A_{\mu}^{A}  \right).
\end{equation*}
We will consider the massless case, $m_i=0 \; \forall i$, where the Lagrangian becomes invariant under a global $SU(N_f)_L\times SU(N_f)_R \times U(1)$ symmetry which can be broken to $SU(N_f)_V$ by formation of a chiral condensate.\\

\begin{figure}[H]
	\centering
	\includegraphics[width=.7\linewidth]{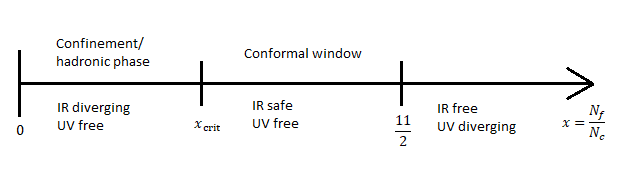}
	\caption{Phase diagram of QCD as a function of $x=N_f/N_c$ at zero temperature and zero chemical potential.}
	\label{fig:Intro_PhaseDiagQCD}
\end{figure}

The two-loop beta function for the gauge coupling is given by \cite{Caswell}
\begin{equation}\label{eq:Intro_2loopbeta}
	\beta_{g}^{[2]} \equiv \Lambda \frac{d \alpha_{g}}{d \Lambda} = -2 b_0 \alpha_{g}^{2} - 2 b_1 \alpha_{g}^{3}
\end{equation}
with $\alpha_g := \frac{g^2}{(4\pi)^2}$ and
\begin{align}
	b_0 &= \frac{11}{3} N_c - \frac{2}{3} N_f,\\
	b_1 &= \frac{34}{3} N_{c}^{2} - N_f \left( \frac{N_{c}^{2} -1}{N_c} + \frac{10}{3} N_c \right), 
\end{align}
are the scheme independent beta-function coefficients.
This beta function has a (degenerate) trivial root at $\alpha_g=0$, and a non-trivial root at $\alpha_g=-\frac{b_0}{b_1}$. The non-trivial fixed point becomes negative and therefore unphysical for
\begin{align}
	x\equiv \frac{N_f}{N_c} > \frac{11}{2}\, ,
\end{align}
and 
\begin{equation}\label{eq:Intro_pertloweredge}
	\frac{N_f}{N_c} < \frac{34 N_{c}^{2}}{13N_{c}^{2} - 3}\, ,
\end{equation}
where the latter bound runs between 3.4 and 2.6 for $1<N_c<\infty$, thus it is always below the first bound for any $N_c$. The region between these bounds, where a non-trivial fixed point exists, is known as the {\em conformal window} (cf. figure \ref{fig:Intro_PhaseDiagQCD}), and its existence was first discussed in \cite{Caswell,Banks}. Within the conformal window the non-trivial fixed point is attractive in the IR limit, while the trivial fixed point is attractive in the UV. Above the conformal window, the trivial fixed point becomes an IR attractor. Since $\alpha_{g}^{nt}$ becomes small close to $x=11/2$ higher order contributions to the beta function (\ref{eq:Intro_2loopbeta}) can be neglected hence the value of upper bound is robust. However, close to the lower edge $\alpha_g$ becomes very large, invalidates the above estimate and (\ref{eq:Intro_pertloweredge}) becomes unreliable. In particular, other fixed points resulting from higher order terms might influence this critical value.\\

The type of phase transition that occurs at the lower edge of the conformal window, that we call $x_{crit}$, and its value has been a central problem in the literature. Use of Schwinger-Dyson equations \cite{Silva, MiranskiYamawaki, AppelquistTerning2, Cohen} generically suggest that the theory is in the confining phase with chiral symmetry breaking for $x$ below $x_{crit}$, see figure \ref{fig:Intro_PhaseDiagQCD}. As for the value of $x_{crit}$, estimates for $N_c=3$ are typically in the range $6 \lessapprox x_{\textrm{crit}} \lessapprox 12$ \cite{Iwasaki,Velkovsky,GiesI,Ryttov,AppelquistFlemingI,AppelquistFlemingII,HasenfratzI,HasenfratzII,Kusafuka,Fodor,Silva,AokiAll,Cohen,AppelquistTerning1,AppelquistTerning2,MiranskiYamawaki,AppelquistAll}, and follow from the various techniques (cf. Table 6 of \cite{Gukov} for a recent overview of various studies and their prediction). A more recent study \cite{RychkovI,RychkovII} uses Potts models to discuss the behaviour at the lower edge of the conformal window. The transition at the lower edge could be first order \cite{Antipin,Lombardo} or infinite order \cite{Kaplan,MiranskiYamawaki}, and can be analyzed through several methods.  
This is an important open problem as characterization of this phase transition will lead to a better understanding of the strong interactions and might be relevant for model building beyond the standard model \cite{Lombardo}. Furthermore, functional RG and holographic approaches indicate that the conformal window is continuously connected to the quark gluon plasma phase at finite temperature \cite{BraunFisher,Alho,Jarvinen}. Therefore studies of the conformal window may provide crucial information on the quark gluon plasma and vice versa.\\

Several works in the literature \cite{GiesI, Lombardo, BraunFisher, RychkovI, RychkovII} point toward the possibility that the IR stable fixed point $\alpha_g = -b_0/b_1$ annihilates with another fixed point of the RG flow at $x_{crit}$ in the form of a saddle-node bifurcation (second option in the list above). This would in turn indicate that a UV irrelevant operator, such as a 4-fermi interaction, crosses marginality precisely at $x_{crit}$ \cite{Gukov} as already anticipated in the early literature \cite{AppelquistTerning2, AppelquistTerning1, AppelquistAll}. In this paper, we investigate this possibility by considering an effective model for QCD$_4$ that includes four-fermi interactions and determine all fixed points within this model in and around the conformal window. Furthermore, we discuss how they appear and disappear as the bifurcation parameters $N_c$ and $N_f$ are varied. Finally we study the effects of adding a scalar operator with a Yukawa type interaction to our results in this effective theory. \\

The paper is organized as follows. In the next section we introduce the effective model with the four-fermi interactions, derive the beta function equations both for finite values of $N_c$ and $N_f$ and in the Veneziano limit, and study the fixed point structure using bifurcation theory in detail. In section 4, we extend the model including the scalar operator. We conclude and discuss our results and list possible future work in the last section. Appendix A  provides a short introduction to bifurcation theory in dynamical systems and Appendix B details the calculation of the beta-functions in perturbation theory. 

\section{An effective model for QCD}

\subsection{Beta functions in an effective model for QCD}
We study the theory in the Wilsonian picture. For this, we write down the effective action, which is given by the Legendre transform of the Lagrangian (\ref{eq:Intro_Action}) with $m_i=0$. We impose the symmetries obeyed by the Lagrangian on the effective action and expand the effective action in the quark fields $\psi^{ai}$ up to the 4-quark interactions, as was done in \cite{Kusafuka}. This yields
\begin{equation}\label{eq:QCD_EffAction}
	\Gamma_{\Lambda} = \int d^{4}x \left( -\frac{1}{4g^{2}} G_{\mu\nu}^{A} G_{A}^{\mu\nu} + \bar{\psi}_{a i} i \slashed{D} \psi^{a i} + \mathcal{L}_{4f}\right),
\end{equation} 
where $\mathcal{L}_{4f}$ denote the four-fermi interactions and we redefined the gauge fields
\begin{equation}
	A\rightarrow A'= g A
\end{equation}	
such that 
\begin{equation}
	G_{\mu\nu}^{A} \rightarrow \frac{1}{g}G_{\mu\nu}^{A} \quad \textrm{and} \quad D_{\mu} = \partial_{\mu} - i t_{A} A^{\prime A}_{\mu}.
\end{equation}

One can write down four independent four-fermi operators \cite{GiesII,BraunGies,Miyashita,Terao} respecting the symmetries. We choose a basis as in \cite{Terao,Kusafuka}:
\begin{equation*}
	\mathcal{L}_{4f} = \frac{G_{S}}{\Lambda^{2(1+\eta)}}\mathcal{O}_{S} + \frac{G_{V}}{\Lambda^{2(1+\eta)}}\mathcal{O}_{V} + \frac{G_{V_1}}{\Lambda^{2(1+\eta)}}\mathcal{O}_{V_1} + \frac{G_{V_2}}{\Lambda^{2(1+\eta)}}\mathcal{O}_{V_2},
\end{equation*}
where $\eta=\gamma_{\psi}$ is the anomalous dimension of the fermion field and
\begin{align*}
	\mathcal {O}_{S} &= 2 \bar{L}_{i} R^{j} \bar{R}_{j} L^{i} = \frac{1}{2} [\bar{\psi}_{i} \psi^{j} \bar{\psi}_{j} \psi^{i} - \bar{\psi}_{i} \gamma_{5} \psi^{j} \bar{\psi}_{j} \gamma_{5} \psi^{i}],\\
	\mathcal{O}_{V} &= \bar{L}_{i} \gamma^{\mu} L^{j}  \bar{L}_{j} \gamma_{\mu} L^{i} + (L\leftrightarrow R) = \frac{1}{2} [\bar{\psi}_{i} \gamma^{\mu} \psi^{j} \bar{\psi}_{j} \gamma_{\mu} \psi^{i} + \bar{\psi}_{i} \gamma^{\mu} \gamma_{5} \psi^{j} \bar{\psi}_{j} \gamma_{\mu} \gamma_{5} \psi^{i}],\\
	\mathcal{O}_{V_1} &= 2\bar{L}_{i} \gamma^{\mu} L^{i}  \bar{R}_{j} \gamma_{\mu} R^{j} = \frac{1}{2} [(\bar{\psi}_{i} \gamma^{\mu} \psi^{i})^{2} - (\bar{\psi}_{i} \gamma^{\mu} \gamma_{5} \psi^{i})^{2}],\\
	\mathcal{O}_{V_2} &= (\bar{L}_{i} \gamma^{\mu} L^{i})^{2} + (L\leftrightarrow R) = \frac{1}{2} [(\bar{\psi}_{i} \gamma^{\mu} \psi^{i})^{2} + (\bar{\psi}_{i} \gamma^{\mu} \gamma_{5} \psi^{i})^{2}]
\end{align*}

One can calculate the beta functions of the four-fermi interactions using the Wetterich equation given in Appendix \ref{Ap:FourFermi}. The beta function for the gauge coupling $\alpha_g$ cannot be derived from the effective action (\ref{eq:QCD_EffAction}) easily as one needs an expansion in the gluon fields in the effective action \cite{Kusafuka} which complicates the calculation substantially. Instead, we can start from the perturbative equation (\ref{eq:Intro_2loopbeta}) and add the effective four-fermi interactions to it. We provide this calculation in appendix \ref{Ap:GaugeVertex} and obtain the same result as in \cite{Kusafuka}. The complete set of beta-functions is then given by\footnote{We find slightly different beta functions for the four-fermi interactions from the ones presented in \cite{Kusafuka}. They agree in the Veneziano limit however.}
\begin{equation}\label{eq:QCD_Model1}
	\begin{cases}
		\Lambda \frac{d \alpha_{g}}{d\Lambda} = & -\frac{2}{3} (11N_c - 2N_f) \alpha_{g}^{2} - \frac{2}{3} \left(34N_{c}^{2} - 13N_{c}N_{f} + 3\frac{N_f}{N_c} \right) \alpha_{g}^{3} + 2N_{c}N_{f}g_{V}\alpha_{g}^{2},\\
		\Lambda \frac{d g_{S}}{d\Lambda} =  &2g_{S} - 2N_{c} g_{S}^{2} + 2N_{f} g_{S}g_{V} + 6 g_{S}g_{V_1} + 2g_{S}g_{V_2}\\
		&-6 \left(N_c - \frac{1}{N_c}\right) g_{S}\alpha_{g} + 12g_{V_1} \alpha_{g} - \frac{3}{2} \left(3N_{c} -\frac{8}{N_{c}}\right) \alpha_{g}^{2},\\
		\Lambda \frac{d g_{V}}{d\Lambda} =  &2g_{V} + \frac{N_{f}}{4} g_{S}^{2} + (N_{c} + N_{f}) g_{V}^{2} - 6g_{V}g_{V_2}\\
		&-\frac{6}{N_{c}} g_{V} + 6 g_{V_2} \alpha_{g} - \frac{3}{4} \left( N_{c} - \frac{8}{N_{c}} \right) \alpha_{g}^{2},\\
		\Lambda \frac{d g_{V_1}}{d\Lambda} = &2g_{V_1} - \frac{1}{4} g_{S}^{2} - g_{S} g_{V} - 3 g_{V_1}^{2} - N_{f} g_{S} g_{V_2} + 2(N_{c}+N_{f}) g_{V}g_{V_1}\\
		&+ 2(N_{c}N_{f}+1) g_{V_1} g_{V_2} + \frac{6}{N_{c}} g_{V_1} \alpha_{g} + \frac{3}{4} \left(1 + \frac{4}{N_{c}^{2}}\right) \alpha_{g}^{2},\\
		\Lambda \frac{d g_{V_2}}{d\Lambda} = &2g_{V_2} - 3 g_{V}^{2} - N_{c} N_{f} g_{V_1}^{2} + (N_{c}N_{f}-2)g_{V_2}^{2} - N_{f} g_{S} g_{V_1}\\
		&+2(N_{c} + N_{f})g_{V} g_{V_2} + 6 g_{V} \alpha_g - \frac{6}{N_c} g_{V_2} \alpha_{g} - \frac{3}{4}  \left(3+ \frac{4}{N_{c}^{2}}\right)\alpha_{g}^{2},
	\end{cases}
\end{equation}
where we defined the rescaled variables
\begin{equation}
	\alpha_g := \frac{g^2}{(4\pi)^2}, \quad g_i :=\frac{G_i}{4\pi^2}.
\end{equation}
Let us further define 
\begin{equation*}
	x := \frac{N_f}{N_c}, \quad N := N_c,
\end{equation*} 
and perform the rescaling
\begin{align}
	N \alpha_{g} &\rightarrow \alpha_{g}\nonumber\\
	N g_{S} &\rightarrow g_{S}\nonumber\\
	N g_{V} &\rightarrow g_{V}\label{eq:QCD_rescaling}\\
	N^{2} g_{V_1} &\rightarrow g_{V_1}\nonumber\\
	N^{2} g_{V_2} &\rightarrow g_{V_2}\nonumber,
\end{align}
which makes the set (\ref{eq:QCD_Model1})  amenable to the Veneziano limit where $N\rightarrow\infty$ and $x$ remains finite.  After the rescaling we obtain
\begin{equation}\label{eq:QCD_Model2}
	\begin{cases}
		\Lambda \frac{d \alpha_{g}}{d\Lambda} = & -\frac{2}{3} (11 - 2x) \alpha_{g}^{2} - \frac{2}{3} (34 - 13 x) \alpha_{g}^{3} + 2 x g_V \alpha_{g}^{2}\\
		& + N^{-2} \left(-2x \alpha_{g}^{3}\right),\\
		\Lambda \frac{d g_{S}}{d\Lambda} =  &2g_{S} - 2 g_{S}^{2} + 2x g_{S}g_{V} - 6g_{S}\alpha_{g} - \frac{9}{2} \alpha_{g}^{2}\\
		& + N^{-2} \left(6g_{S}g_{V_1} + 2g_{S}g_{V_2} + 6g_{S}\alpha_{g} + 12g_{V_1}\alpha_{g} + 12\alpha_{g}^{2}\right),\\
		\Lambda \frac{d g_{V}}{d\Lambda} =  &2g_{V} + \frac{1}{4}xg_{S}^{2} + (1 + x) g_{V}^{2} - \frac{3}{4} \alpha_{g}^{2}\\
		& + N^{-2} \left(-6g_{V}g_{V_2} - 6g_{V}\alpha_{g} + 6g_{V_2}\alpha_{g} + 6\alpha_{g}^{2}\right),\\
		\Lambda \frac{d g_{V_1}}{d\Lambda} = &2g_{V_1} - \frac{1}{4} g_{s}^{2} - g_{S} g_{V} - x g_{S} g_{V_2} + 2(1+x)g_{V}g_{V_1} + 2xg_{V_1}g_{V_2} + \frac{3}{4} \alpha_{g}^{2}\\
		& + N^{-2} \left(-3g_{V_1}^{2} + 2g_{V_1}g_{V_2} + 6g_{V_1}\alpha_{g} + 3\alpha_{g}^{2} \right),\\
		\Lambda \frac{d g_{V_2}}{d\Lambda} = &2g_{V_2} - 3 g_{V}^{2} - x g_{V_1}^{2} + x g_{V_2}^{2} - x g_{S} g_{V_1} + 2 (1+x) g_{V}g_{V_2} + 6g_{V}\alpha_{g} - \frac{9}{4} \alpha_{g}^{2}\\
		& + N^{-2} \left(-2g_{V_2}^{2} - 6 g_{V_2}\alpha_{g} - 3 \alpha_{g}^{2} \right).
	\end{cases}
\end{equation}

\subsection{Veneziano Limit}
In this section, we study the Veneziano limit of the model (\ref{eq:QCD_Model2}) as was done in \cite{Gukov} and \cite{Kusafuka}. The Veneziano limit, $N\rightarrow\infty$ yields
\begin{equation}
	\begin{cases}
		\frac{d \alpha_{g}}{d t} &= \frac{2}{3} (11 - 2x) \alpha_{g}^{2} + \frac{2}{3} (34 - 13 x) \alpha_{g}^{3} - 2 x g_V \alpha_{g}^2,\\
		\frac{d g_{S}}{d t} &= -2g_{S} + 2 g_{S}^{2} - 2x g_{S}g_{V} + 6g_{S}\alpha_{g} + \frac{9}{2} \alpha_{g}^{2},\\
		\frac{d g_{V}}{d t} &= -2g_{V} - \frac{1}{4}xg_{S}^{2} - (1 + x) g_{V}^{2} + \frac{3}{4} \alpha_{g}^{2},\\
		\frac{d g_{V_1}}{d t} &= -2g_{V_1} + \frac{1}{4} g_{s}^{2} + g_{S} g_{V} + x g_{S} g_{V_2} - 2(1+x)g_{V}g_{V_1} - 2xg_{V_1}g_{V_2} - \frac{3}{4} \alpha_{g}^{2},\\
		\frac{d g_{V_2}}{d t} &= -2g_{V_2} + 3 g_{V}^{2} + x g_{V_1}^{2} - x g_{V_2}^{2} + x g_{S} g_{V_1} - 2(1+x) g_{V}g_{V_2} - 6g_{V}\alpha_{g} + \frac{9}{4} \alpha_{g}^{2},\\
	\end{cases}
\end{equation}
where we have defined $t = - \ln(\Lambda/\Lambda_0)$. It is important to note that all higher order contributions in the perturbative theory are either suppressed or give constant contributions in the Veneziano limit with the rescaled variables. We now observe that the first three equations decouple from the last two. Therefore, we can reduce analysis to the system\footnote{The orbits still have components in the $g_{V_1}$ and $g_{V_2}$ direction, which depend on the coordinates $(\alpha_g,g_S,g_V)$. However, the flow projected on the $(\alpha_g,g_S,g_V)$ is independent of $g_{V_1}$ and $g_{V_2}$.}

\begin{equation}\label{eq:QCD_Toymodel}
	\begin{cases}
		\dot{\alpha_{g}} &= \frac{2}{3} (11 - 2x) \alpha_{g}^{2} + \frac{2}{3} (34 - 13 x) \alpha_{g}^{3} - 2 x g_V \alpha_{g}^2,\\
		\dot{g_{S}} &= -2g_{S} + 2 g_{S}^{2} - 2x g_{S}g_{V} + 6g_{S}\alpha_{g} + \frac{9}{2} \alpha_{g}^{2},\\
		\dot{g_{V}} &= -2g_{V} - \frac{1}{4}xg_{S}^{2} - (1 + x) g_{V}^{2} + \frac{3}{4} \alpha_{g}^{2}.\\
	\end{cases}
\end{equation}
The domains of the variables and the parameters are given by $x\in\mathbb{R}^{+}$, $\alpha \in\mathbb{R}_{0}^{+}$ and $g_{S}, g_{V} \in\mathbb{R}$. Due to the fact that the model (\ref{eq:QCD_Model1}) is perturbative plus a few higher order corrections, we expect better results for small parameter values. As a rough bound we will use $\alpha_g<1$, $|g_i|<1$, and notice when this bound is exceeded.\\

The beta function for $\alpha_g$ has a double root for $\alpha_g^{t}=0$ and another root for $\alpha_g^{nt}=\frac{11-2x-3x g_V}{13x-34}$. Therefore, the fixed points of the RG-flow lie on these manifolds. The manifold defined by $\alpha_g=0$ is an invariant set of the system, since $\dot{\alpha}_{g}$ vanishes for $\alpha_g = 0$.\\

\begin{figure}
	\begin{minipage}{0.45\linewidth}
		\includegraphics[width=3.0in]{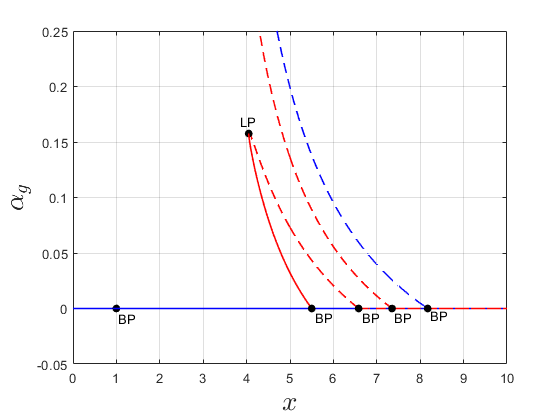}
	\end{minipage}
	\hspace{0.1\linewidth}
	\begin{minipage}{0.45\linewidth}
		\includegraphics[width=3.0in]{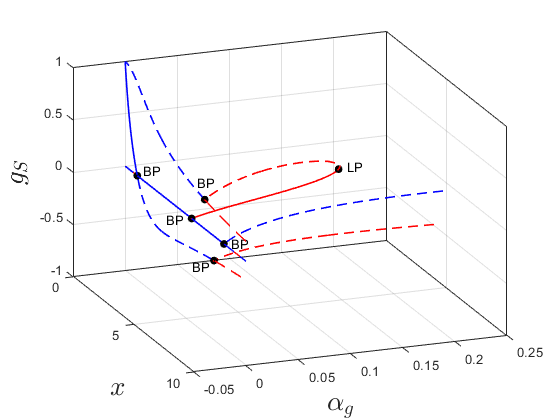}
	\end{minipage}
	\\
	\begin{minipage}{0.45\linewidth}
		\includegraphics[width=3.0in]{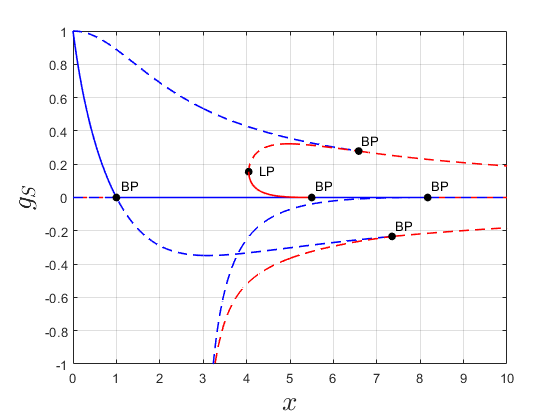}
	\end{minipage}
	\hspace{0.1\linewidth}
	\begin{minipage}{0.45\linewidth}
		\includegraphics[width=3.0in]{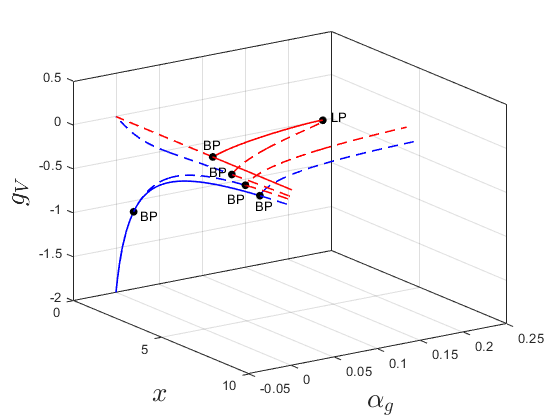}
	\end{minipage}
	\\
	\begin{minipage}{0.45\linewidth}
		\includegraphics[width=3.0in]{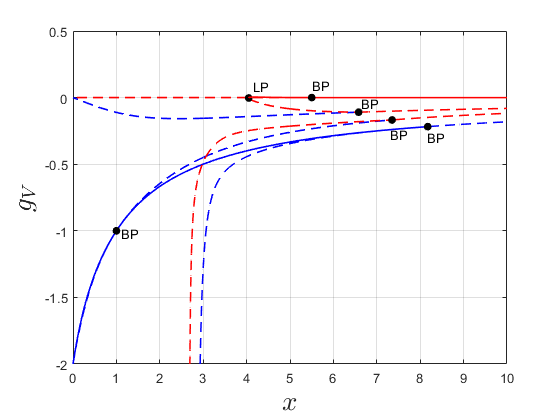}
	\end{minipage}
	\hspace{0.1\linewidth}
	\begin{minipage}{0.45\linewidth}
		\includegraphics[width=3.0in]{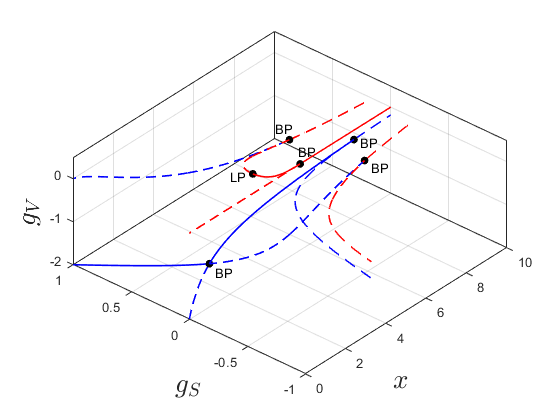}
	\end{minipage}
	\begin{minipage}{1\linewidth}
		\centering
		\includegraphics[width=3.0in]{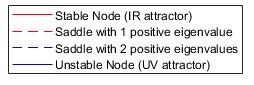}
	\end{minipage}
	\caption{The fixed points of the RG-flow of (\ref{eq:QCD_Toymodel}) projected on the $(x,\alpha_g)$-plane, $(x,g_S)$-plane, $(x,g_V)$-plane, $(x,\alpha_g,g_S)$-space, $(x,\alpha_g,g_V)$-space and the $(x,g_S,g_V)$-space. LP: Limit Point. BP: Branching Point.}
	\label{fig:QCD_Ven_FixedPoints}
\end{figure}

The behavior of the fixed points projected on the $(x,\alpha_g)$, $(x,g_S)$, $(x,g_V)$ planes and $(x,\alpha_g,g_S)$, $(x,\alpha_g,g_V)$ and  $(x,g_S,g_V)$ spaces are shown in Figure \ref{fig:QCD_Ven_FixedPoints}. We use the following color coding in this figure:
\begin{itemize}
	\item Solid red line: stable node, 3 negative eigenvalues (IR attractor).
	\item Dashed red line: saddle point, 2 negative and 1 positive eigenvalue.
	\item Dashed blue line: saddle point, 1 negative and 2 positive eigenvalues.
	\item Solid blue line: unstable node, 3 positive eigenvalues (UV attractor).
\end{itemize}
 
The equilibria on the $\alpha_{g}=0$ manifold all have a trivial eigenvalue, and therefore do not fit the color coding as described above. In order to use the color coding, we define the sign of a trivial eigenvalue on this manifold by approaching the equilibrium along a line of constant $g_S$ and $g_V$ from positive but small $\alpha_g$. The trivial eigenvalue will then approach $0$ from either positive or negative values. If the trivial eigenvalue is slightly positive for $0<\alpha_g\ll1$, we define the trivial eigenvalue to be positive and vice versa.\footnote{This definition makes physical sense, since negative values of $\alpha_g$ are unphysical.}

In Figure \ref{fig:QCD_Ven_FixedPoints}, 4 up to 8 equilibria are shown, depending on the value of $x$. Furthermore, a few bifurcations are found. The parameter and coordinate values of these bifurcations are reported in Table \ref{tab:QCD_Ven_Bif}. The Branching points all correspond to transcritical bifurcations. The transcritical bifurcation at $x=1$ is located in the subspace $\alpha_{g}=0$, while the other transcritical bifurcations are located on the intersection of the subspaces $\alpha_{g}=0$ and $\alpha_{g}=\alpha_{g}^{nt}$. In the latter case, both a trivial and a non-trivial equilibrium\footnote{We call a fixed point trivial, if it has a trivial $\alpha_g$ value and non-trivial otherwise.} are involved. At all the non-trivial equilibria, the equilibrium has a positive eigenvalue, when $\alpha_{g}<0$, that becomes negative when $\alpha_{g}>0$, while the trivial equilibrium has a negative eigenvalue that becomes positive. At the branching point at $x=5.5$ a stable node with $\alpha_g>0$ is created. This indicates the upper edge of the conformal window as discussed in the introduction of this section. Finally, at $x=4.05$ two non-trivial equilibria collide and disappear through a saddle-node bifurcation (limit point). This saddle-node bifurcation could indicate the lower edge of the conformal window through a fixed point merger, as was discussed in \cite{Kusafuka} and \cite{Gukov}. The two other non-trivial fixed points diverge for $x\lesssim 2.5$, close to the divergence of the large $N_c$ limit in equation (\ref{eq:Intro_pertloweredge}).

\begin{table}[H]
	\centering
	\caption{Locations of bifurcations found in model (\ref{eq:QCD_Toymodel}) for QCD$_4$ in the Veneziano limit.}
	\label{tab:QCD_Ven_Bif}
	\begin{tabular}{|l|r|r|r|r|r|r|}
		\hline
		Bifurcation   & $x$ & $\alpha_g$ & $g_S$ & $g_V$  \\
		\hline
		Saddle-node   & $4.049$ & $0.158$ & $0.155$  & $-0.003$ \\
		Transcritical & $1.000$ & $0$	  & $0$      & $-1.000$ \\
		Transcritical & $5.501$ & $0$     & $0$      & $0$      \\
		Transcritical & $6.582$ & $0$     & $0.279$  & $-0.110$ \\
		Transcritical & $7.351$ & $0$     & $-0.234$ & $-0.168$ \\
		Transcritical & $8.173$ & $0$     & $0$      & $-0.218$ \\
		\hline
	\end{tabular}
\end{table}

\begin{figure}[H]
	\begin{minipage}{0.45\linewidth}
		\includegraphics[width=3.0in]{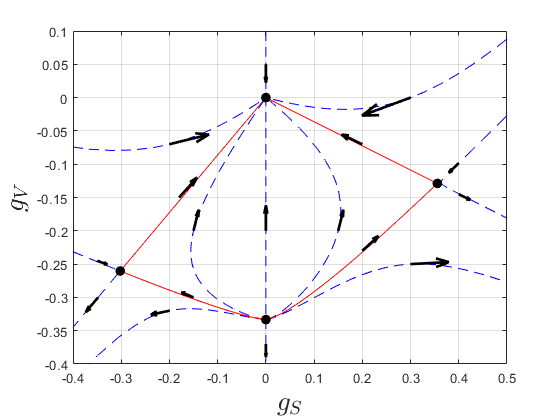}
		\subcaption{Phase portrait of the RG-flow in the $(g_S,g_V)$-plane for $\alpha_g=0$ in the Veneziano limit at $x=5$. Solid red lines indicate the boundaries of an invariant set.}
	\label{fig:QCD_Ven_Flowx5a0}
	\end{minipage}
	\\
	\begin{minipage}{0.45\linewidth}
		\includegraphics[width=3.0in]{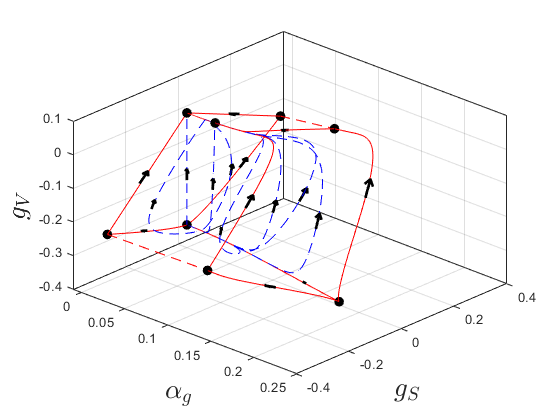}
		\subcaption{Phase portrait of the RG-flow in the $(\alpha_g,g_S,g_V)$-space in the Veneziano limit at $x=5$. Red lines indicates the skeleton of an invariant set.}
		\label{fig:QCD_Ven_Flowx53d}
	\end{minipage}
	\hspace{0.1\linewidth}
	\begin{minipage}{0.45\linewidth}
		\includegraphics[width=3.0in]{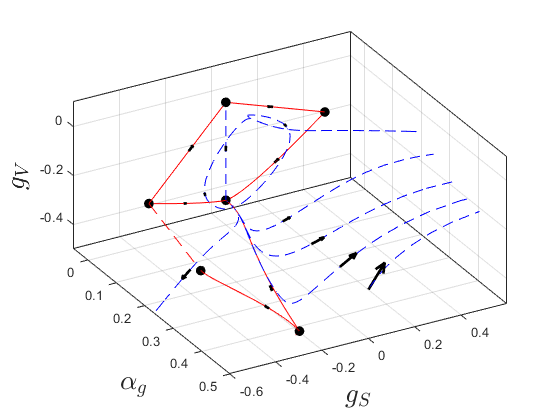}
		\subcaption{Phase portrait of the RG-flow in the $(\alpha_g,g_S,g_V)$-space in the Veneziano limit at $x=4$. Red lines indicate the boundaries of 2-dimensional invariant sets.}
		\label{fig:QCD_Ven_Flowx43d}
	\end{minipage}
\end{figure}

A few characteristic orbits of the RG-flow in the Veneziano limit are visualized in Figures \ref{fig:QCD_Ven_Flowx5a0}, \ref{fig:QCD_Ven_Flowx53d} and \ref{fig:QCD_Ven_Flowx43d}.\\

In Figure \ref{fig:QCD_Ven_Flowx5a0}, a phase portrait on the invariant plane $\alpha_g=0$ is shown at $x=5$. We find 4 fixed points, and 4 red critical heteroclinic orbits, connecting those fixed points, and marking the boundaries of an invariant set. All orbits inside this set have a finite UV and finite IR fixed point, while orbits outside the invariant set have a diverging IR limit or diverging UV limit. The structure of this flow does not change, when $x$ is varied in the region $x>1$. Only the location of the fixed points and connecting orbits will change continuously. At $x=1$, the two fixed points with smallest value of $g_V$ collide and exchange stability, via a transcritical bifurcation, yielding a similar phase portrait.\\

In Figure \ref{fig:QCD_Ven_Flowx53d}, a phase portrait of the flow in the 3-dimensional space is shown at $x=5$. Here, all 8 fixed points are shown, together with a few heteroclinic orbits connecting them. The red heteroclinic orbits form a skeleton of a 3-dimensional invariant set. This invariant set is bound by six invariant planes. The solid red lines lie on the intersections of these planes. The dashed red lines indicate the presence of other intersections, of which the exact locations haven't been found numerically. All orbits inside this set have a finite UV and IR fixed point, while orbits outside the invariant set have a diverging IR limit, a diverging UV limit or both. More precisely, the bulk of the invariant set consists of structurally stable orbits going from the equilibrium at $(\alpha_g,g_S,g_V) = (0,0,-0.333)$ to the equilibrium at $(\alpha_g,g_S,g_V) = (0.032,0.003,0.000)$, and the boundary planes consist of non-stable heteroclinic orbits between the other fixed points. Within the invariant set the theory is in a chirally symmetric phase, while outside the set the chiral symmetry is broken. The structure of this flow does not change when $x$ is varied in the region $4.05<x<5.5$. Only the location of the fixed points and connecting orbits will change continuously. At $x=4.05$ the two fixed points shown at $(\alpha_g,g_S,g_V) = (0.032,0.003,0.000)$ and $(\alpha_g,g_S,g_V) = (0.073,0.322,-0.084)$ disappear through a saddle-node bifurcation.\\

In Figure \ref{fig:QCD_Ven_Flowx43d}, a phase portrait in the whole space is shown at $x=4$. Here, only 6 fixed points are left. The six fixed points together span two unstable bounded 2-dimensional invariant sets. The solid red lines border these sets, and the dashed red line indicates the location, where the last border is expected, which hasn't been found numerically. The 3-dimensional invariant set, which was present at $x=5$, has been broken down at the saddle-node bifurcation to two 2-dimensional invariant planes. The structure of the flow does not change when $x$ is varied in the region $0<x<4.05$, but the two non-trivial fixed points diverge when $x$ is lowered towards $x=2.5$. The orbits outside of the invariant planes all diverge in the IR limit. The eigenvector with $0$ eigenvalue at the saddle-node bifurcation\footnote{This eigenvector is directly related to the operator that crosses marginality.} is mainly directed in the $g_{S}$ direction, as can be found in Table \ref{tab:QCD_Ven_CrossOperators}. Therefore, the divergence is the strongest in the $g_S$ direction, indicating that the chiral symmetry is broken due to the scalar part of the four-fermi interaction, as discussed in \cite{Aoki}. Furthermore, we notice that there still exist chirally symmetric trajectories with a zero UV and finite IR limit, but these are unstable, and therefore need a very precise fine-tuning within this model.\\

\begin{table}[H]
	\centering
	\caption{Eigenvalues and eigenvectors at the saddle-node bifurcation at $x=4.049$.}
	\label{tab:QCD_Ven_CrossOperators}
	\begin{tabular}{|c|c|}
		\hline
		Eigenvalue & Normalized eigenvector w.r.t the basis $\{e_{\alpha},e_{S},e_{V}\}$ \\
		\hline
		$0$     & $(0.09,\; 0.98, -0.15)$\\
		$-0.56$ & $(0.19, -0.95,\; 0.24)$\\
		$-2.13$ & $(0.10,\; 0.50,\; 0.86)$\\
		\hline
	\end{tabular}
\end{table}

\subsection{Small $N_c$ regime}
\subsubsection{Fixed points and Codim-1 bifurcations}
\begin{figure}[H]
	\begin{minipage}{0.45\linewidth}
		\centering
		\includegraphics[width=3.0in]{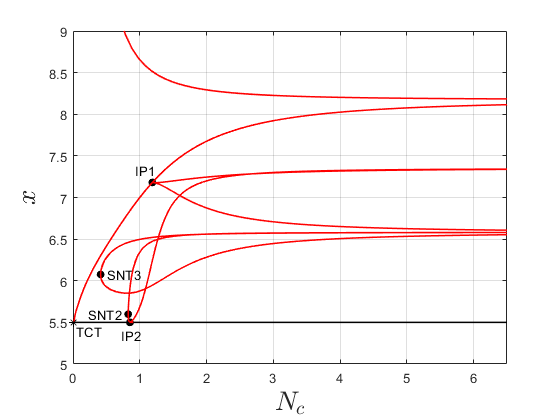}
		\subcaption{Transcritical bifurcations on the intersection of the $(\alpha_g = 0)$ and $(\alpha_g = \alpha_{g}^{nt})$ subspaces. Black line: transcritical bifurcation marking the upper edge of the conformal window bifurcation curves.}
		\label{fig:QCD_BifDiag_TC_Nc_x}
	\end{minipage}
	\hspace{0.1\linewidth}
	\begin{minipage}{0.45\linewidth}
		\centering
		\includegraphics[width=3.0in]{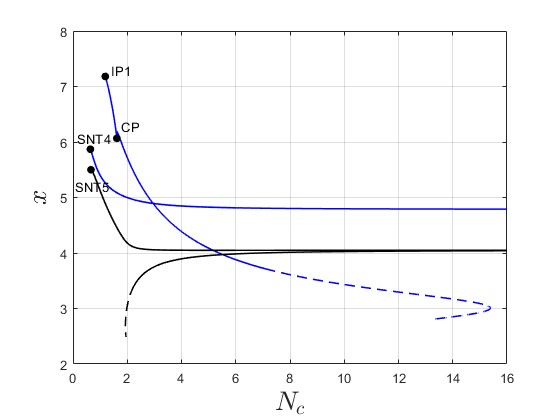}
		\subcaption{Saddle-node bifurcations in the $(\alpha_{g}=\alpha_{g}^{nt}\geq 0)$-subspace. Black lines: saddle-node bifurcation, that was also found in the Veneziano limit.}
		\label{fig:QCD_BifDiag_SN_Nc_x}
	\end{minipage}
	\caption{Bifurcations of the QCD$_4$ model (\ref{eq:QCD_Model2}), shown in the $(N_{c},x)$-plane. Red Lines: transcritical bifurcations.  Blue lines: saddle-node bifurcations. Dashed lines: bifurcation curves, with at least one of the coupling constants $|g_i|>1$. CP: Cusp bifurcation. SNT: Saddle-Node-Transcritical bifurcation. TCT: Transcritical-Transcritical bifurcation. IP: degenerate bifurcation point.}
	\label{fig:QCD_BifDiag_Nc_x}
\end{figure}

In the model out of the Veneziano limit there are 2 bifurcation parameters $N_c$ and $x = N_f/N_c$, which allows for more interesting types of bifurcations. Again, we are interested in the renormalized trajectories of the theory and the invariant sets enclosed by the renormalized trajectories, indicating the existence of chirally symmetric theories. We have numerically found that the model contains up to 20 different real fixed points with $\alpha_g\geq0$ for integer values of $(N_c,N_f)\in[1,10]\times[1,100]$. The beta function for $\alpha_g$ yields two roots
\begin{equation*}
\alpha_{g} = 0, \quad \alpha_g = \alpha_{g}^{nt}:= N_{c}^{2} \frac{-11 + 2 x + 3 x g_V}{34N_{c}^{2} -13 x N_{c}^{2} + 3 x},
\end{equation*}
where the trivial root has multiplicity two. The $(\alpha_g=0)$-hyperplane is an invariant set of the model, which allows to study the model without gauge interactions separately. Setting $\alpha_g=0$, we find up to 12 fixed points, which can be continued using MatCont. Two of those disappear through a saddle-node bifurcation at $x>7.5$, which is above the conformal window. The 10 remaining fixed points play an important role for the behavior for $\alpha_g\neq0$. For every of these ten fixed points, we find a related fixed point on the manifold $\alpha_{g} = \alpha_{g}^{nt}$. For large values of $x$ all the non-trivial fixed points have $\alpha_g<0$, but for smaller values of $x$ they have $\alpha_g>0$, and become physically relevant. On the intersection of the two hyperplanes, where $\alpha_{g}=\alpha_{g}^{nt}=0$, i.e.  when $x = \frac{11}{2+3g_V}$, all non-trivial fixed points collide with a trivial fixed point, through a transcritical bifurcation. The lines at which these transcritical bifurcations occur are shown in Figure \ref{fig:QCD_BifDiag_TC_Nc_x}. Using the transcritical bifurcations, we relate all the non-trivial fixed points to a trivial fixed point.\\

The 10 non-trivial fixed points with positive $\alpha_g$ exist for large values of $x$. However, for small values, they either disappear through a saddle-node bifurcation or diverge in one or multiple couplings. In Figure \ref{fig:QCD_BifDiag_SN_Nc_x} the saddle-node bifurcation lines are shown. Here, we see that two pairs (black lines) of fixed points will disappear through a saddle-node bifurcation which is connected to the saddle-node bifurcation that was found in the Veneziano limit\footnote{There were in fact two saddle-node bifurcations in the Veneziano limit with differing $(g_{V_1},g_{V_2})$. They were reported as one since the beta functions for $g_{V_1}$ and $g_{V_2}$ decoupled and weren't considered}. The two bifurcation curves are close together for large $N_c$, but split up at small $N_c$. One of the curves starts to increase steeply around $N_c=2$, while the other starts decreasing around $N_c=6$, and diverges in the coupling constants at $N_c=2$. Therefore the saddle-node bifurcation of this pair does not exist for $N_c=1$, and the fixed points diverge instead. Then there is a third pair of fixed points (upper blue line) that disappears through a saddle-node bifurcation for all $N_c$ around $x=4.8$. However this value starts to increase around $N_c=4$. A fourth pair disappears through a saddle-node bifurcation when $1<N_c<16$, diverges for $N_c\geq16$, while at $N_c=1$ the fixed points will not exist for positive $\alpha_g$. The last two fixed points do not disappear through a saddle-node bifurcation, but will diverge in multiple couplings at small $x$.\\

\begin{figure}[H]
	\centering
	\includegraphics[width=.9\linewidth]{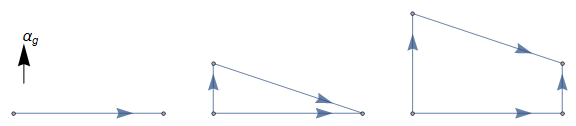}
	\caption{Topology of the RG-flow between 4 fixed points in a group. Left (Large N): the 2 non-trivial fixed points have $\alpha_g<0$. Middle (intemediate N): One of the non-trivial fixed points has $\alpha_g>0$. Right (small N): Both of the non-trivial fixed points have $\alpha_g>0$.}
	\label{fig:QCD_FixedPointStruc}
\end{figure}

Using these saddle-node bifurcation curves, we can partition the 10 non-trivial fixed points into 5 pairs. Since all 10 non-trivial fixed points are related to a trivial fixed point through a transcritical bifurcation curve, we can partition the 20 fixed points into groups of 4 consisting of 2 trivial and 2 non-trivial fixed points. It turns out that the heteroclinic orbits between the fixed points within each of these sets have a similar topology, which is as shown in Figure \ref{fig:QCD_FixedPointStruc}: at first, for large $x$, the two non-trivial fixed points lie below the $\alpha_g=0$ hyperplane, and the $\alpha_{g}=0$ hyperplane is locally attractive in the IR limit, then, at a specific value of $x$, one of the non-trivial and one of the trivial fixed points collide through a transcritical bifurcation and the non-trivial fixed point gets a positive $\alpha_{g}$ creating a locally attractive IR limit set bounded by orbits between the 3 fixed points with $\alpha_g\geq0$. Next, at smaller specific $x$, the other non-trivial fixed point will collide with the other trivial fixed point and also get positive $\alpha_g$ value. After this, at small $x$, the non-trivial fixed points will either disappear through a saddle-node bifurcation or diverge, making the $\alpha_{g}=0$ hyperplane locally repulsive in the IR limit.\\

\begin{figure}[H]
	\begin{minipage}{0.45\linewidth}
		\centering
		\includegraphics[width=3.0in]{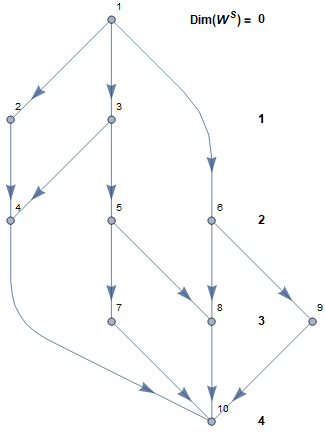}
	\end{minipage}
	\hspace{0.1\linewidth}
	\begin{minipage}{0.45\linewidth}
		\centering
		\includegraphics[width=2.3in]{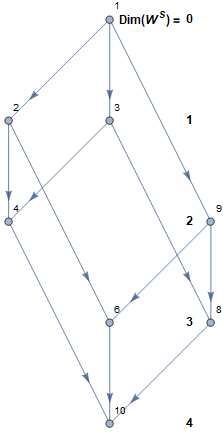}
	\end{minipage}
	\caption{Stability of the fixed points and topology of the RG-flow between the fixed points on the four dimensional invariant set $\alpha_g=0$ at $(N_c,N_f)=(3,15)$ (left), and at $(N_c,N_f)=(3,3)$ (right). Bold numbers correspond to the dimension of the stable manifold at the fixed point.}
	\label{fig:QCD_FixedPointStruc_fund315}
\end{figure}

\begin{figure}[H]
	\begin{minipage}{0.45\linewidth}
		\includegraphics[width=3.0in]{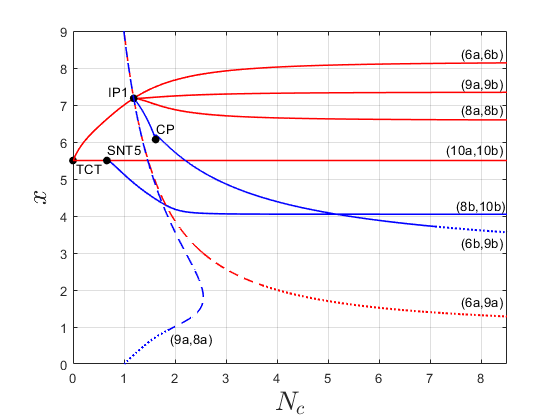}
		\subcaption{Fixed points 6,8,9,10; $N_c\in[0,8.5]$.}
		\label{fig:QCD_BifDiag_Nc_x_a}
	\end{minipage}
	\hspace{0.1\linewidth}
	\begin{minipage}{0.45\linewidth}
		\includegraphics[width=3.0in]{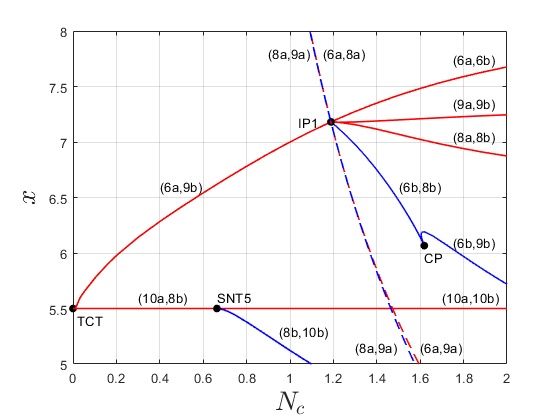}
		\subcaption{Fixed points 6,8,9,10; $N_c\in[0,2]$.}
		\label{fig:QCD_BifDiag_Nc_x_b}
	\end{minipage}
	\\
	\begin{minipage}{0.45\linewidth}
		\includegraphics[width=3.0in]{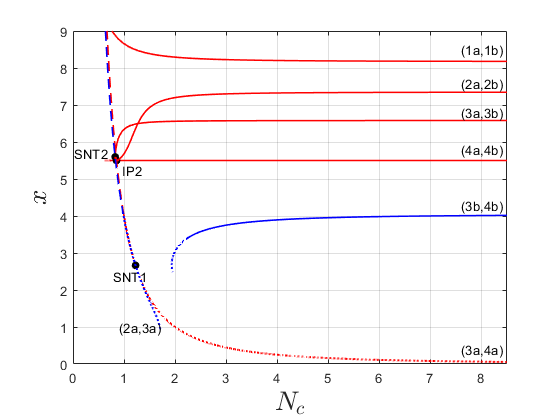}
		\subcaption{Fixed points 1,2,3,4; $N_c\in[0,8.5]$.}
		\label{fig:QCD_BifDiag_Nc_x_c}
	\end{minipage}
	\hspace{0.1\linewidth}
	\begin{minipage}{0.45\linewidth}
		\includegraphics[width=3.0in]{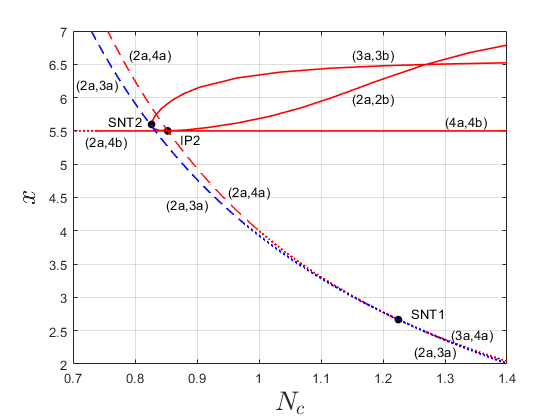}
		\subcaption{Fixed points 1,2,3,4; $N_c\in[0.7,1.4]$.}
		\label{fig:QCD_BifDiag_Nc_x_d}
	\end{minipage}
	\\
	\begin{minipage}{0.45\linewidth}
		\includegraphics[width=3.0in]{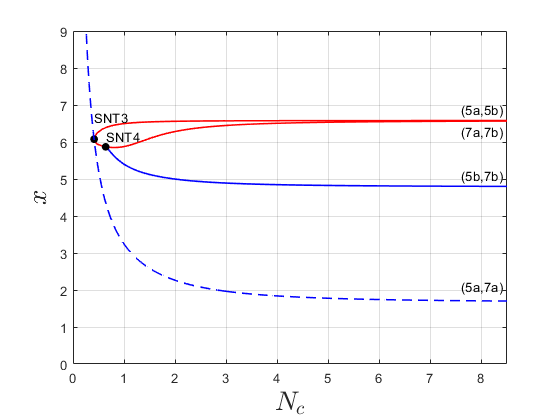}
		\subcaption{Fixed points 5,7; $N_c\in[0,8.5]$.}
		\label{fig:QCD_BifDiag_Nc_x_e}
	\end{minipage}
	\hspace{0.1\linewidth}
	\begin{minipage}{0.45\linewidth}
		\includegraphics[width=3.0in]{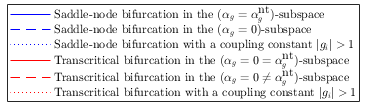}
	\end{minipage}
	\caption{Bifurcations of various sets of fixed points in the QCD$_4$ model (\ref{eq:QCD_Model2}) in the $(N_{c},x)$-plane. CP: Cusp bifurcation. SNT: Saddle-Node-Transcritical bifurcation. TCT: Transcritical-Transcritical bifurcation. IP: degenerate bifurcation point. Numbers indicate which equilibria are involved in the bifurcation, and correspond to the numbers in Figure \ref{fig:QCD_FixedPointStruc_fund315}.}
	\label{fig:QCD_BifDiag_Nc_x_ser}
\end{figure}

We can also visualize the structure of the flow on the invariant $(\alpha_g=0)$-space. A graphical representation of the fixed points and a few critical orbits is shown in Figure \ref{fig:QCD_FixedPointStruc_fund315} for $(N_c,N_f)=(3,15)$ and $(N_c,N_f)=(3,3)$. We notice that two points (labeled by 5 and 7) do exist for $(N_c,N_f)=(3,15)$, but do not exist for $(N_c,N_f)=(3,3)$. This is due to a saddle-node bifurcation in the $(\alpha_g=0)$-space at $(N_c,N_f)\approx(3,5.9)$. Furthermore two fixed points (labeled by 6 and 9) have exchanged stability through a transcritical bifurcation at  $(N_c,N_f)\approx(3,7.7)$. The curves corresponding to these two bifurcations are shown in Figure \ref{fig:QCD_BifDiag_Nc_x_e} and \ref{fig:QCD_BifDiag_Nc_x_a} respectively. In the remainder of this section, we'll use the numbers as in Figure \ref{fig:QCD_FixedPointStruc_fund315} to label the fixed points. Since we have related every trivial fixed point uniquely to a non-trivial fixed point, we give the non-trivial fixed point the same number as its trivial counterpart. However, we add the label `a' to the trivial fixed points and the label `b' to the non-trivial fixed points.\\

We have numerically found all local bifurcations\footnote{We have performed a quite extensive study, and therefore have great confidence that we found all local bifurcations in this region. Furthermore, we have not found any global bifurcations, and believe that there are none in the studied region. However, this is numerically harder to verify.} involving these 20 fixed points for $\alpha_g\geq0$, and these are shown in Figure \ref{fig:QCD_BifDiag_Nc_x_ser}. This figure also includes the saddle-node and transcritical bifurcation curves shown in Figure \ref{fig:QCD_BifDiag_Nc_x}.\\

\begin{figure}[H]
	\begin{minipage}{0.45\linewidth}
		\includegraphics[width=3.0in]{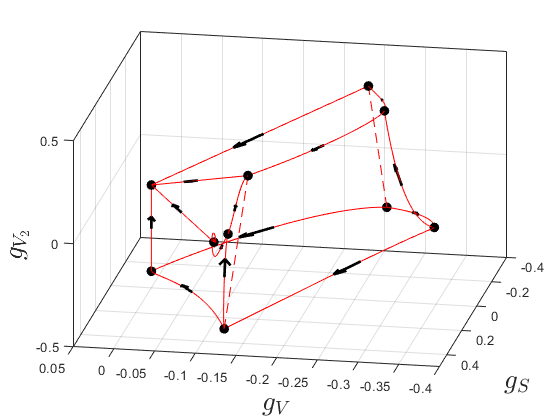}
	\end{minipage}
	\hspace{0.1\linewidth}
	\begin{minipage}{0.45\linewidth}
		\includegraphics[width=3.0in]{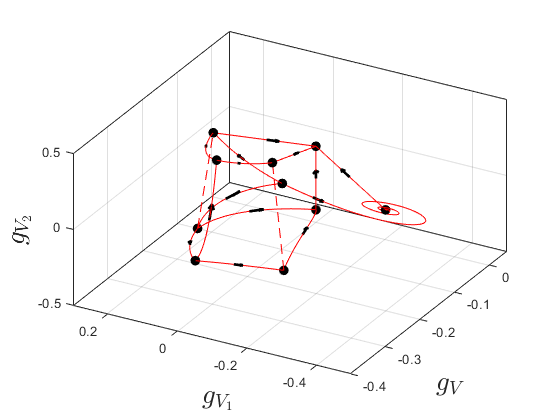}
	\end{minipage}
	\caption{Fixed points and critical heteroclinic orbits in the QCD$_4$ model (\ref{eq:QCD_Model2}) at $(N_c,N_f)=(3,15)$ and $\alpha_{g}=0$. Solid line: heteroclinic connections. Dashed lines: heteroclinic connections that are expected to exist, but haven't been found numerically. Left: projection on the $(g_{S},g_{V},g_{V_2})$-space. Right: projection on the $(g_{V},g_{V_1},g_{V_2})$-space.}
	\label{fig:QCD_ag0_Nc3_Nf15_3D}
\end{figure}

\begin{figure}[H]
	\begin{minipage}{0.45\linewidth}
		\includegraphics[width=3.0in]{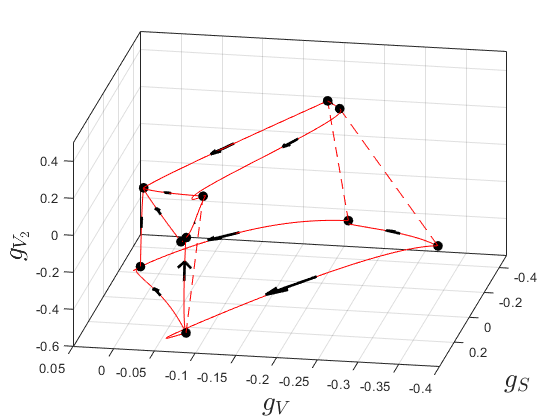}
	\end{minipage}
	\hspace{0.1\linewidth}
	\begin{minipage}{0.45\linewidth}
		\includegraphics[width=3.0in]{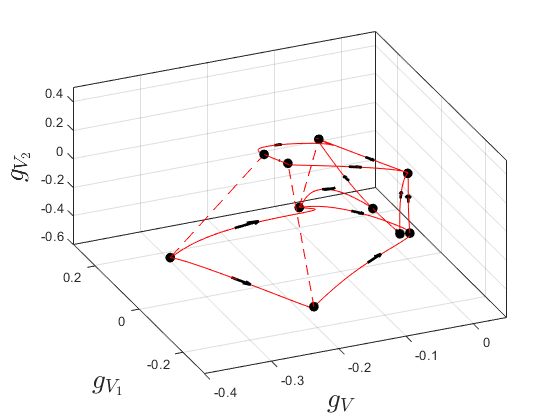}
	\end{minipage}
	\caption{Fixed points and critical heteroclinic orbits in the QCD$_4$ model (\ref{eq:QCD_Model2}) at $(N_c,N_f)=(3,15)$ and $\alpha_{g}=\alpha_{g}^{nt}$. Solid line: heteroclinic connections. Dashed lines: heteroclinic connections that are expected to exist, but haven't been found numerically. Left: projection on the $(g_{S},g_{V},g_{V_2})$-space. Right: projection on the $(g_{V},g_{V_1},g_{V_2})$-space.}
	\label{fig:QCD_Nc3_Nf15_3D}
\end{figure}

\begin{figure}[H]
	\begin{minipage}{0.45\linewidth}
		\includegraphics[width=3.0in]{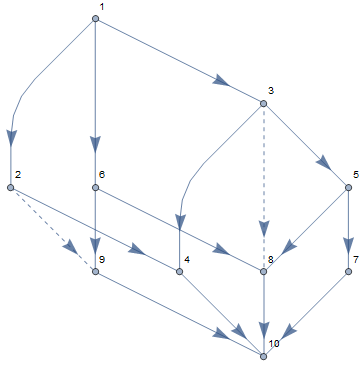}
		\subcaption{Topology of the RG-flow at $\alpha_g=0$ corresponding to figure \ref{fig:QCD_ag0_Nc3_Nf15_3D}.}
		\label{fig:QCD_FixedPointStruc_a0_3_15}
	\end{minipage}
	\hspace{0.1\linewidth}
	\begin{minipage}{0.45\linewidth}
		\includegraphics[width=3.0in]{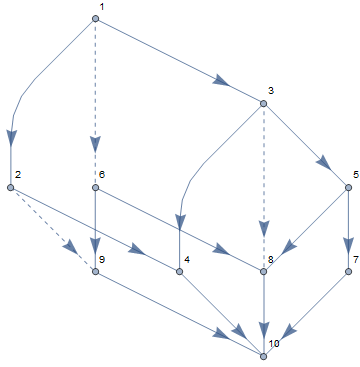}
		\subcaption{Topology of the RG-flow at $\alpha_g=\alpha_g^{nt}$ corresponding to figure \ref{fig:QCD_Nc3_Nf15_3D}.}
		\label{fig:QCD_FixedPointStruc_3_15}
	\end{minipage}
	\caption{Graphical representation of the fixed points and the skeleton of the invariant set at $(N_c,N_f)=(3,15)$ both on the $\alpha_g=0$ and the $\alpha_g=\alpha_{g}^{nt}$ subspaces.}
	\label{fig:QCD_FixedPointStruc_1}
\end{figure}

\begin{figure}[H]
	\begin{minipage}{0.45\linewidth}
		\includegraphics[width=3.0in]{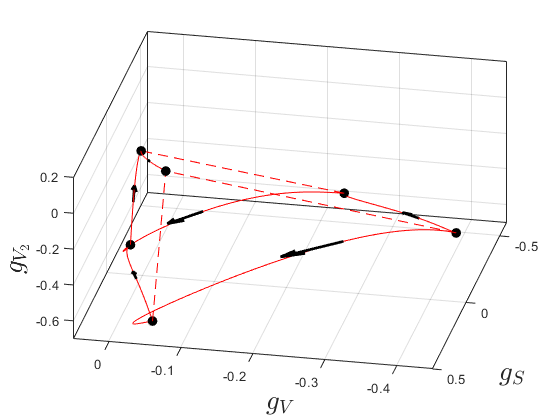}
	\end{minipage}
	\hspace{0.1\linewidth}
	\begin{minipage}{0.45\linewidth}
		\includegraphics[width=3.0in]{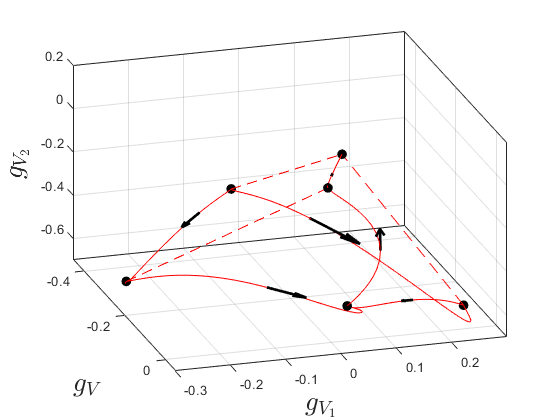}
	\end{minipage}
	\caption{Fixed points and critical heteroclinic orbits of the QCD$_4$ model (\ref{eq:QCD_Model2}) at $(N_c,N_f)=(3,13)$ and $\alpha_{g}=\alpha_{g}^{nt}$. Solid line: heteroclinic connections. Dashed lines: heteroclinic connections that are expected to exist, but haven't been found numerically. Left: projection on the $(g_{S},g_{V},g_{V_2})$-space. Right: projection on the $(g_{V},g_{V_1},g_{V_2})$-space.}
	\label{fig:QCD_Nc3_Nf13_3D}
\end{figure}

\begin{figure}[H]
	\begin{minipage}{0.45\linewidth}
		\includegraphics[width=3.0in]{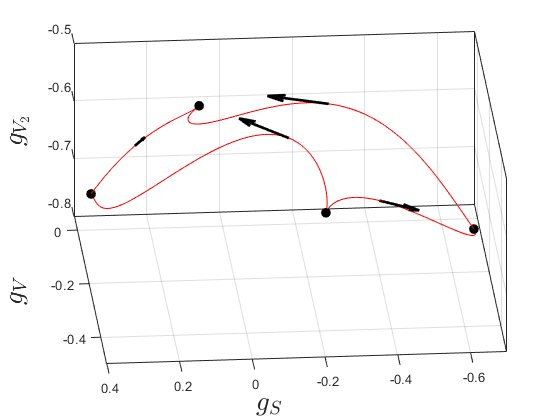}
	\end{minipage}
	\hspace{0.1\linewidth}
	\begin{minipage}{0.45\linewidth}
		\includegraphics[width=3.0in]{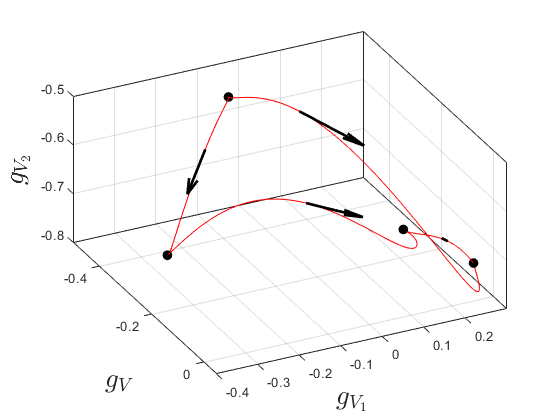}
	\end{minipage}
	\caption{Fixed points and critical heteroclinic orbits of the QCD$_4$ model (\ref{eq:QCD_Model2}) at $(N_c,N_f)=(3,12)$ and $\alpha_{g}=\alpha_{g}^{nt}$. Left: projection on the $(g_{S},g_{V},g_{V_2})$-space. Right: projection on the $(g_{V},g_{V_1},g_{V_2})$-space.}
	\label{fig:QCD_Nc3_Nf12_3D}
\end{figure}

\begin{figure}[H]
	\begin{minipage}{0.45\linewidth}
		\includegraphics[width=3.0in]{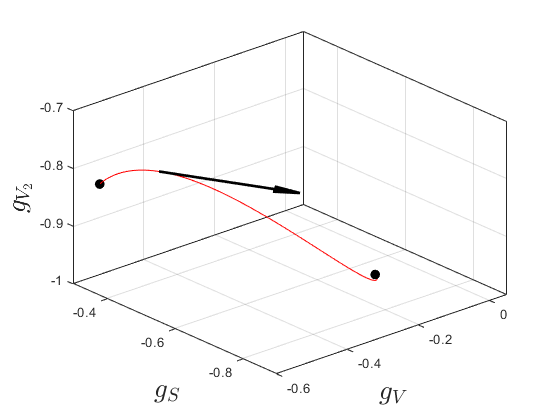}
	\end{minipage}
	\hspace{0.1\linewidth}
	\begin{minipage}{0.45\linewidth}
		\includegraphics[width=3.0in]{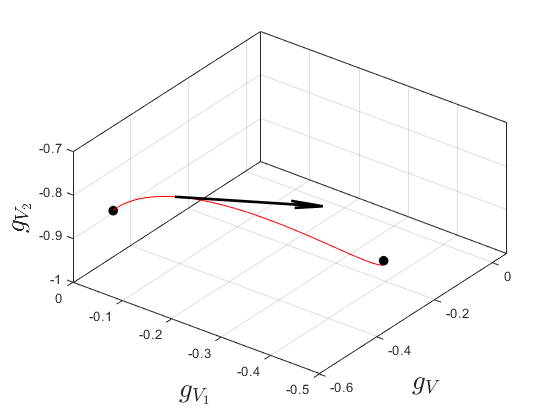}
	\end{minipage}
	\caption{Fixed points and a critical heteroclinic orbit of the QCD$_4$ model (\ref{eq:QCD_Model2}) at $(N_c,N_f)=(3,11)$ and $\alpha_{g}=\alpha_{g}^{nt}$. Left: projection on the $(g_{S},g_{V},g_{V_2})$-space. Right: projection on the $(g_{V},g_{V_1},g_{V_2})$-space.}
	\label{fig:QCD_Nc3_Nf11_3D}
\end{figure}

\begin{figure}[H]
	\begin{minipage}{0.25\linewidth}
		\includegraphics[width=1.5in]{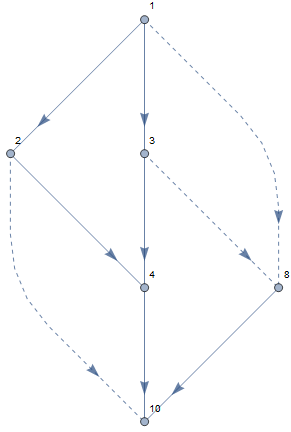}
		\subcaption{Topology of the RG-flow $N_f=13$ corresponding to figure \ref{fig:QCD_Nc3_Nf13_3D}.}
		\label{fig:QCD_FixedPointStruc_a0_3_15}
	\end{minipage}
	\hspace{0.05\linewidth}
	\begin{minipage}{0.25\linewidth}
		\includegraphics[width=1.2in]{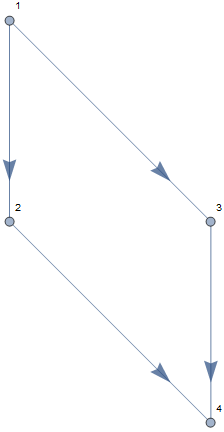}
		\subcaption{Topology of the RG-flow $N_f=12$ corresponding to figure \ref{fig:QCD_Nc3_Nf12_3D}.}
		\label{fig:QCD_FixedPointStruc_3_15}
	\end{minipage}
	\hspace{0.05\linewidth}
	\begin{minipage}{0.25\linewidth}
		\includegraphics[width=0.1in]{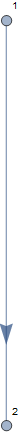}
		\subcaption{Topology of the RG-flow $N_f=11$ corresponding to figure \ref{fig:QCD_Nc3_Nf11_3D}.}
		\label{fig:QCD_FixedPointStruc_3_15}
	\end{minipage}
	\caption{Graphical representation of the fixed points and the skeleton of the invariant set on the $\alpha_g=\alpha_{g}^{nt}$ subspace at $N_c=3$ and various values of $N_f$.}
	\label{fig:QCD_FixedPointStruc_2}
\end{figure}

We use the bifurcation diagram \ref{fig:QCD_BifDiag_Nc_x_ser} to distinguish and visualize several phases within the RG-flow generated by our model (\ref{eq:QCD_Model1}). For this, we set $N_c=3$, and take $N_f\in\{11,12,13,15\}$. The topology on the invariant $(\alpha_g=0)$-space will be the same for these four values of $x$, since there are no bifurcations in this region. Only the location of the fixed points will change continuously. Projections of the fixed points and a few heteroclinic orbits in this 4-dimensional invariant subspace, at $\alpha_g=0$, are shown in Figure \ref{fig:QCD_ag0_Nc3_Nf15_3D}. The fixed points and a few heteroclinic orbits in the $\alpha_g=\alpha_{g}^{nt}$ subspace are shown in Figures \ref{fig:QCD_Nc3_Nf15_3D}, \ref{fig:QCD_Nc3_Nf13_3D}, \ref{fig:QCD_Nc3_Nf12_3D} and \ref{fig:QCD_Nc3_Nf11_3D}. In Figures \ref{fig:QCD_FixedPointStruc_1} and \ref{fig:QCD_FixedPointStruc_2} these figures are shown in a graphical representation.\\

At $N_f=15$, we see 10 fixed points and a few heteroclinic orbits connecting them, both in the $\alpha_g=0$ and the $\alpha_g=\alpha_{g}^{nt}$ hyperplane. The red lines shown in the figures form part of a skeleton of a closed set in the 4-dimensional subspaces $\alpha_g=0$ and $\alpha_g=\alpha_{g}^{nt}$. These two 4-dimensional sets on $\alpha_g=0$ and $\alpha_g=\alpha_{g}^{nt}$ form a 4-dimensional boundary of a larger 5-dimensional invariant subspace of the full 5-dimensional space. The other 4-dimensional subspaces that bound the full 5-dimensional invariant set are spanned by orbits going from $\alpha_g=0$ towards $\alpha_g=\alpha_{g}^{nt}$. Dashed red lines indicate parts of the skeleton that haven't been found numerically, but are expected to exist. The flow within the 5-dimensional invariant set is everywhere directed from the $\alpha_g=0$ towards the $\alpha_g=\alpha_{g}^{nt}>0$ subspace. This set consists of structurally stable renormalized trajectories that have finite UV (with trivial $\alpha_g$) and IR limits (with strictly positive $\alpha_g$) indicating the theory to be in a symmetric phase. Outside the set, orbits have diverging IR and/or UV limit, and theories will have a broken chiral symmetry. The boundaries consist of unstable orbits between other fixed points, and therefore indicate unstable symmetric theories.\\

When the flavor number is reduced just below $N_f=15$, at $N_f=14.7$ and $N_f=14.6$, two pairs of fixed points disappear through a saddle-node bifurcation. Because of these bifurcations, the invariant set of orbits indicating symmetric theories is reduced. However, the most attractive IR fixed point with $\dim(W^{s})=5$ still exists. This indicates that the invariant set is locally attractive in the IR limit. Reducing the flavor number even further to $N_f=12$ and $N_f=11$ more fixed points disappear through a saddle-node bifurcations. The invariant set of orbits at $\alpha_g=\alpha_{g}^{nt}$ is reduced at every step, and therefore the set of asymptotically stable orbits with finite UV and IR limit becomes smaller, making the set of possible symmetric theories smaller. For $N_f\leq12.2$, there is no fixed point left with $\dim(W^{s})=5$. Therefore, the remaining invariant set is not an IR attractor anymore. Furthermore, we expect that the dimension of the invariant set with finite trajectories reduces, when more fixed points have disappeared through a saddle-node bifurcation. When the invariant set splits up into multiple invariant sets with dimension smaller than $5$, symmetric theories only exist within this model under extreme fine-tuning. This is a strong indication that the lower edge of the conformal window has been crossed.\\

In Figure \ref{fig:QCD_ag0_Nc3_Nf15_3D}, there is another peculiarity visible. Here we see a heteroclinic orbit with spiraling behavior indicating the existence of renormalized trajectories with complex scaling behavior in and close to the $\alpha_{g}=0$ hyperplane. Equilibrium 5 and 7 are associated to this behavior.\\

\subsubsection{Operators crossing marginality and degenerate bifurcations}
We have found 4 saddle-node bifurcation curves (cf. figure \ref{fig:QCD_BifDiag_SN_Nc_x}, \ref{fig:QCD_BifDiag_Nc_x_ser}) with non-trivial value of $\alpha_g$. All four indicate the disappearance of two fixed points when the value of $x$ is decreased. Along these curves we can identify the eigenvector corresponding to a zero eigenvalue as the relevant operator that crosses marginality. These critical eigenvectors at $N_c=3$ are reported in Table \ref{tab:QCD_CrossOperators}. We have found that these operators are very dependent on $x$, but not on $N_c$. From the table we conclude that 3 of the saddle-node bifurcations have 2 complex eigenvalues at $N_c=3$, while one has only real eigenvalues. We notice that the normalized eigenvectors corresponding to the complex eigenvalues are all directed along the four-fermi couplings, and have very small contributions ($|\eta_{\alpha_g}|<0.007$) along the $e_{\alpha_g}$ direction, meaning that this complex scaling behavior involves operators, which are linear combinations of the effective four-fermi interactions. Furthermore, we see that the operator crossing marginality is a linear combination of all operators. However the scalar four-fermi interaction is most relevant in three of the saddle-node bifurcations, while in the saddle-node bifurcation of equilibria 5b and 7b the $V_1$ interaction is most relevant.\\

\begin{table}[H]
	\centering
	\caption{Eigenvalues and critical eigenvectors crossing marginality of the saddle-node bifurcation with non-trivial $\alpha_g$ at $N_c=3$.}
	\label{tab:QCD_CrossOperators}
	\begin{tabular}{|c|c|c|c|c|}
		\hline
		Equilibria  & $(N_c,x)$ & Eigenvalues & Critical eigenvector \\
		\hline
		8, 10 & $(3.00,4.06)$ & $\{-2.88, -2.10, -1.36, -0.64, 0\}$ & $(0.12,0.94,-0.17,0.12,\; 0.22)$ \\
		6, 9  & $(3.00,4.86)$ & $\{-3.62\pm0.27i, 1.79, -0.62, 0\}$ & $(0.11,0.86,-0.21,0.35,\; 0.28)$ \\
		3, 4  & $(3.00,3.75)$ & $\{3.38\pm1.05i, -2.57, -0.81, 0\}$ & $(0.07,0.77,-0.08,0.50,-0.38)$\\
		5, 7  & $(3.00,4.89)$ & $\{-1.66\pm0.09i, 1.59, -0.27, 0\}$ & $(0.04,0.06,-0.07,0.95,\; 0.28)$ \\
		\hline
	\end{tabular}
\end{table}

In the complete model we have encountered nine bifurcation points with higher codimension. An overview can be found in Table \ref{tab:QCD_BifComplete}. Six of these points lie in the physical uninteresting region $N_c\in(0,1)$, while three of them (CP, SNT1 and IP1) may have physical relevance, since $N_c\geq1$. Those points will be discussed below. The point labeled with `CP' corresponds to a cusp bifurcation and the points labeled with `SNT' correspond to saddle-node-transcritical bifurcations, as discussed in Appendix \ref{Ap:Bifurcations}. The point `TCT' labels an intersection of two independent transcritical bifurcation curves and corresponds to the normal form
\begin{equation}
	\begin{cases}
	\dot{x} = \alpha x + x^{2},\\
	\dot{y} = \beta y + y^{2}. 
	\end{cases}
\end{equation}
The points `IP' label more degenerate bifurcation points.\\

\begin{table}[H]
	\centering
	\caption{Locations of bifurcations found in the model of QCD$_4$ (\ref{eq:QCD_Toymodel}) with effective four-fermi interactions.}
	\label{tab:QCD_BifComplete}
	\begin{tabular}{|l|r|r|r|r|r|r|r|r|}
		\hline
		Bifurcation & $N_c$ & $N_f$ & $x$ & $\alpha_g$ & $g_S$ & $g_V$ & $g_{V_1}$ & $g_{V_2}$\\
		\hline
		CP   & $1.62$ & $9.82$ & $6.07$ & $0.05$ & $0.03$ & $-0.17$  & $0.01$  & $0.22$ \\
		SNT1 & $1.22$ & $3.27$ & $2.67$ & $0$    & $0$    & $0$      & $0$     & $-1.50$\\
		SNT2 & $0.83$ & $4.63$ & $5.60$ & $0$    & $0.28$ & $-0.01$  & $0.12$  & $-0.82$\\
		SNT3 & $0.41$ & $2.51$ & $6.08$ & $0$    & $0.28$ & $-0.06$  & $-0.02$ & $-0.01$\\
		SNT4 & $0.64$ & $3.76$ & $5.87$ & $0$    & $0.26$ & $-0.04$  & $-0.05$ & $-0.04$\\
		SNT5 & $0.66$ & $3.65$ & $5.50$ & $0$    & $0$    & $0$      & $0$     & $0$    \\
		TCT  & $0$    & $0$    & $0$    & $0$    & $0$    & $0$      & $0$     & $0$    \\
		IP1  & $1.19$ & $8.54$ & $7.18$ & $0$    & $0$    & $-0.16$  & $0$     & $0.17$ \\
		IP2  & $0.85$ & $4.69$ & $5.50$ & $0$    & $0$    & $0$      & $0$     & $-0.73$\\	
		\hline
	\end{tabular}
\end{table}

At the cusp bifurcation 2 saddle-node bifurcation curves with non-trivial value of $\alpha_g$ meet. The first curve is associated to the pair of equilibria (6b,8b) and the second to (6b,9b).\\

The point SNT1, describes a saddle-node-transcritical bifurcation, of a saddle-node curve that lies in the subspace $\alpha_{g}=0$ and involves the equilibria (2a,3a), and two transcritical curves that also lie in the $\alpha_{g}=0$ subspace and involve the equilibria (2a,4a) and (3a,4a).\\

IP1 is a highly degenerate bifurcation point. The model reduced on the invariant subspace $\alpha_g=0$ has a pitchfork bifurcation at this point. However, in the 5-dimensional model the branches of the equilibrium curves meeting at this pitchfork bifurcation are all transcritical bifurcation curves of the equilibria (6a,6b), (8a,8b), (9a,9b) and (6a,9b). In addition, to this pitchfork bifurcation we find a saddle-node-transcritical bifurcation at this point. The curves associated to this saddle-node-transcritical bifurcation all lie in the $\alpha_g=0$ subspace. The saddle-node curve involves the equilibria (8a,9a) and the branches of the transcritical bifurcation involve (6a,8a) and (6a,9a).\footnote{This is still a bit speculative. We expect the pitchfork bifurcation and the saddle-node-transcritical bifurcation to overlap, but this is hard to verify numerically.} On top of this, there is a saddle-node bifurcation curve with non-trivial $\alpha_g$ intersecting at this point. This curve involves the equilibria (6b,8b). Together with the branches of transcritical bifurcation curves of (6a,6b) and (8a,8b), that were contained in the pitchfork bifurcation, we could also see this point as saddle-node-transcritical bifurcation of these points. In short this point is highly degenerate and can be unfolded into two saddle-node-transcritical bifurcation and a pitchfork bifurcation of transcritical bifurcation curves.\\

\subsection{Discussion of the Results}
In the complete model we can distinguish 10 pairs of fixed points, that may be relevant for the physics in and below the conformal window. These points can be divided into three different sets: $\{1,2,3,4\}$, $\{5,7\}$ and $\{6,8,9,10\}$. Here, we used the numbering as in Figure \ref{fig:QCD_FixedPointStruc_fund315}. The first set contains the most repulsive fixed points, while the last set contains the most attractive fixed points.\\

The set of pairs of fixed points $\{6,8,9,10\}$ consists of points that were also found in the Veneziano limit. The non-trivial points of the pairs 8 and 10 disappear through a saddle-node bifurcation at $x\approx4$ for large values of $N_c$ like in the Veneziano limit. Furthermore, the non-trivial points of the pairs 6 and 9 diverge for $N_c\geq16$, and the trivial points of the pairs 6 and 9 have a transcritical bifurcation at $x=1$ for large values of $N_c$. This behavior was also found in the Veneziano limit. Equilibrium 10 is the most attractive and the point with non-trivial $\alpha_g$ in this pair describes the IR limit of the structurally stable renormalized trajectories that exist within the invariant set. We see that like in the Veneziano limit the non-trivial IR point disappears through a saddle-node bifurcation with the non-trivial equilibrium of pair 8 at $(N_c,x)=(3,4.06)$ or $(N_c,N_f)=(3,12.2)$. This saddle-node bifurcation is a good candidate for the lower edge of the conformal window in this model. This prediction is higher than but comparable to the predictions summarized in \cite{Gukov}. Furthermore, it is close to a recent prediction on a similar model including effective interactions \cite{Terao}. In addition, we see that the transcritical bifurcation of the equilibria of pair 10, where the most attractive equilibrium gets a positive $\alpha_g$ is constant at the ratio $x=5.5$. This line indicates the upper edge of the conformal window, and the same value was found in the Veneziano limit. This is consistent with earlier predictions, as was expected, since $\alpha_g$ is small, so the two-loop perturbative beta function of $\alpha_g$ without four-fermi interactions should give a good approximation. However, we have also found a few other non-trivial fixed points which already exist above the ratio $x=5.5$ in our model. This might indicate that the non-perturbative beta function for $\alpha_g$ contains fixed points for ratios above $x=5.5$, which are not present in the 2-loop beta function.\\

If we take the saddle-node bifurcation of the non-trivial equilibria of the pairs $8$ and $10$ as the lower edge of the conformal window, we see in Figure \ref{fig:QCD_BifDiag_Nc_x_a} that the the ratio $x$, indicating the lower edge of the conformal window, increases for $N_c\leq2$. Furthermore, we see that the two fixed points that were diverging in the Veneziano limit for small values of $x$ will disappear through a saddle-node bifurcation for values $1<N_c<16$. This saddle-node bifurcation takes place for ratios below the conformal window for $N_c>5$, and for $1<N_c\leq5$ it takes place in or above the conformal window, while at $N_c=1$ these fixed points do not exist for strictly positive $\alpha_g$. This saddle-node bifurcation curve affects the closed invariant set of structurally stable orbits, and could therefore be relevant for the physical behavior of QCD in the conformal window: for $N_c\geq16$, we expect similar behavior as in the conformal window, where these two fixed points diverge around $x\approx2.5$. Here, the fixed points form extremal points of a 5-dimensional closed invariant set, which is discontinuously broken down to a smaller invariant set, when the saddle-node bifurcation of the pairs 8 and 10 takes place. On the other hand, for $5<N_c<16$, the points no longer diverge, but also disappear through a saddle-node bifurcation. However, this happens for a value of $x$ which is below the lower edge of the conformal window. Therefore, the behavior within the conformal window won't be changed much by this saddle-node bifurcation. In the region $3\leq N_c \leq 5$, this saddle-node bifurcation takes place within the conformal window, resulting in a discontinuous change in size of the closed invariant set of structurally stable orbits (or renormalized trajectories with finite IR and UV limit) at the bifurcation. I.e. it does not affect the range of the conformal window, but it induces an discontinuous change in the set of symmetric theories, which could indicate a transition within the conformal window in the sense that a non-empty set of finite renormalized trajectories, which are present near the upper edge of the conformal window, are not present close to the lower edge and have diverging IR limits instead. At $N_c=2$, the saddle-node bifurcation takes place above the conformal window, and therefore this discontinuous transition won't take place within the conformal window. However, the set of symmetric theories in the conformal window will be smaller than at larger values of $N_c$. At $N_c=1$, this saddle-node bifurcation only takes place for negative $\alpha_g$, but not for positive $\alpha_g$. Furthermore, the lower edge of the conformal window has increased at $N_c=1$ making the conformal window smaller than at higher values.\\

The set of pairs of fixed points $\{1,2,3,4\}$ contains the four pairs of fixed points that are the most repulsive of the 10, and are therefore most relevant for the UV physics. Pair 1 is the most repulsive pair and therefore describes the UV limit of the renormalized trajectories within the invariant set of structurally stable orbits. At a curve $x>8$ the equilibria within this pair undergo a transcritical bifurcation, as can be seen in Figures \ref{fig:QCD_BifDiag_Nc_x_c}. Below this curve the most repulsive of the pair is the trivial equilibrium making the theory asymptotically free. Above this curve the $\alpha_{g}=0$ hyperplane is repulsive, and therefore the theory always diverges in the UV limit. The non-trivial fixed points of the pairs 1 and 2 both diverge when $x$ is decreased. Furthermore, we see that two of the non-trivial equilibria, the ones from the pairs 3 and 4, disappear through a saddle-node bifurcation for $N_c\geq2$ this bifurcation happens for smaller values of $x$ than the saddle-node bifurcation which we took as the lower edge of the conformal window. We note that this bifurcation curve also converges to $x\approx4$ in the Veneziano limit, where the saddle-node bifurcation was found in the Veneziano limit. We find that both this curve and the saddle-node bifurcation curve of the non-trivial fixed points 8 and 10 correspond to the saddle-node bifurcation that was found in the Veneziano limit, since the beta functions for $g_{V_1}$ and $g_{V_2}$ have two fixed points at this point, which shows that there are actually 2 saddle-node bifurcations in the Veneziano limit, that differ in the values $g_{V_1}$ and $g_{V_2}$. This saddle-node bifurcation curve of the non-trivial fixed points of the pairs 3 and 4 is another candidate for the lower edge of the conformal window. If we take the saddle-node bifurcation curve of points 3 and 4 as the lower edge of the conformal window, the saddle-node bifurcation curve would indicate a discontinuous change in the invariant set of structurally stable orbits indicating the symmetric theories. Furthermore, this would imply that the lower edge of the conformal window is just below $N_f=12$ at $N_c=3$, as was found in lattice studies such as\cite{AppelquistFlemingI}, and from Schwinger-Dyson equations with ladder resummations in \cite{Cohen,AppelquistTerning1,AppelquistTerning2,MiranskiYamawaki,AppelquistAll}, where the lower edge was predicted at
\begin{equation*}
	x_{\textrm{crit}} = \frac{100 N_c^2 - 66}{25 N_c^2 - 15}.
\end{equation*}

Finally, we notice that there is a last set of two pairs of fixed points, $\{5,7\}$, which wasn't found in the Veneziano limit. These fixed points have complex eigenvalues with relatively large complex part compared to the real part. This induces a spiraling behavior in the system. These points disappear through a saddle-node bifurcation both on the $\alpha_{g}=\alpha_{g}^{nt}$ and the $\alpha_{g}=0$ subspace. These fixed points might have a physical interpretation or might be a consequence of the approximate nature of our model. Furthermore, the other fixed points (except for the pairs 8 and 10) can have complex eigenvalues, but for those the complex part is smaller than the real part, making the spiraling behavior less visible. Also, the eigenvectors corresponding to the complex eigenvalues are largely directed into the direction of the effective four-fermi couplings. The complex eigenvalues indicate that when the orbits move away (towards) the fixed point there is both a scaling, indicated by the real part of the eigenvalue and a rotation, indicated by the complex part of the eigenvalue. The scaling determines the velocity of flowing away from (towards) the fixed point, while the rotation indicates that the relevant operator, which is a linear combination of the basis operators rotates in theory space. In our case this rotation only happens in the subspace spanned by effective operators, since the operators with complex eigenvectors are mostly directed along the effective operators.\\

In conclusion, we find two saddle-node bifurcations that could indicate the lower edge of the conformal window, in the sense that both involve the merger of a non-trivial fixed that is created through a transcritical bifurcation at $x=5.5$, and thus corresponds to the pertubative Banks-Zaks fixed point. These two fixed points merge with other non-trivial fixed points, which are created at values $x>5.5$ through transcritical bifurcations. These other fixed points could become apparent in higher order perturbation theory, when the beta function becomes polynomials of higher order in $\alpha_g$. The two possible lower edges of the conformal window are then given by $(N_c,N_f)=(3.00,12.2)$ and $(N_c,N_f)=(3.00,11.3)$. Apart from these 2 pairs non-trivial fixed points we find 3 more pairs of non-trivial fixed points that are all created through transcritical bifurcations at $x>5.5$ and either disappear through a saddle node bifurcation or diverge. Furthermore, we find that most bifurcations occur at almost constant $x(N)$, for values of $N\geq3$.

\section{QCD$_4$ with a Scalar Field}

In this section we extend our previous model with a meson-like scalar field $\Phi$ that couples to the fermion fields with the coupling constant $y$. The field $\Phi$ is invariant under $SU(N_c)$ and transforms in the adjoint representation of $SU(N_f)_L\times SU(N_f)_R$, i.e. as $\Phi\rightarrow g_L \Phi g_{R}^{\dagger}$, where $g_i\in SU(N_f)_i$. This extension is based on an extension mentioned in \cite{Terao}. The Lagrangian is extended with a massless scalar field and a Yukawa coupling with the fermion fields yielding extra terms in the Lagrangian:
\begin{equation}\label{eq:QCDYukA_Lag}
\delta \mathcal{L} = \frac{1}{2} \partial_{\mu} \Phi \partial^{\mu} \Phi - y \sum_{i,j=1}^{N_f}\left( \bar{L}_{i} \Phi_{j}^{i} R^{j} + \bar{R}_{i} (\Phi^{\dagger})_{j}^{i} L^{j}\right).
\end{equation}

The set of beta functions is then extended to a set of 6 differential equations in 6 variables and 2 parameters, which is calculated in appendix \ref{Ap:QCD6}:

\begin{equation}\label{eq:QCDYukA_Model1}
\begin{cases}
\Lambda \frac{d \alpha_{g}}{d\Lambda} = & -\frac{2}{3} (11N_c - 2N_f) \alpha_{g}^{2} - \frac{2}{3} \left(34N_{c}^{2} - 13N_{c}N_{f} + 3\frac{N_f}{N_c} \right) \alpha_{g}^{3} - 2 N_{f}^{2} \alpha_{y} \alpha_g^{2} + 2N_{c}N_{f}g_{V}\alpha_{g}^{2},\\
\Lambda \frac{d \alpha_{y}}{d\Lambda} = & 2 \alpha_{y} \left( 2 N_c \alpha_y + N_f \alpha_y - 3 \left( N_c - \frac{1}{N_c} \right) \alpha_g - 2 N_c g_S + 8 g_{V_1} \right),\\
\Lambda \frac{d g_{S}}{d\Lambda} =  &2 (1 + \alpha_y) g_{S} - 2N_{c} g_{S}^{2} + 2N_{f} g_{S}g_{V} + 6 g_{S}g_{V_1} + 2g_{S}g_{V_2} + 4 g_V \alpha_y\\
&-6 \left(N_c - \frac{1}{N_c}\right) g_{S}\alpha_{g} + 12g_{V_1} \alpha_{g} - \frac{3}{2} \left(3N_{c} -\frac{8}{N_{c}}\right) \alpha_{g}^{2},\\
\Lambda \frac{d g_{V}}{d\Lambda} =  &2 (1 + \alpha_y) g_{V} + \frac{N_{f}}{4} g_{S}^{2} + (N_{c} + N_{f}) g_{V}^{2} - 6g_{V}g_{V_2} -\frac{6}{N_{c}} g_{V} + g_{S} \alpha_y\\
&+ 6 g_{V_2} \alpha_{g} - \frac{3}{4} \left( N_{c} - \frac{8}{N_{c}} \right) \alpha_{g}^{2},\\
\Lambda \frac{d g_{V_1}}{d\Lambda} = &2 (1 + \alpha_y) g_{V_1} - \frac{1}{4} g_{S}^{2} - g_{S} g_{V} - 3 g_{V_1}^{2} - N_{f} g_{S} g_{V_2} + 2(N_{c}+N_{f}) g_{V}g_{V_1} \\
&+ 2(N_{c}N_{f}+1) g_{V_1} g_{V_2} - 2 g_{V_2} \alpha_y + 2 \alpha_{y}^{2} + \frac{6}{N_{c}} g_{V_1} \alpha_{g} + \frac{3}{4} \left(1 + \frac{4}{N_{c}^{2}}\right) \alpha_{g}^{2},\\
\Lambda \frac{d g_{V_2}}{d\Lambda} = &2 (1 + \alpha_y) g_{V_2} - 3 g_{V}^{2} - N_{c} N_{f} g_{V_1}^{2} + (N_{c}N_{f}-2)g_{V_2}^{2} - N_{f} g_{S} g_{V_1}\\
&+2(N_{c} + N_{f})g_{V} g_{V_2} - 2 g_{V_1} \alpha_y + 2 \alpha_{y}^{2} + 6 g_{V} \alpha_g - \frac{6}{N_c} g_{V_2} \alpha_{g} - \frac{3}{4}  \left(3+ \frac{4}{N_{c}^{2}}\right)\alpha_{g}^{2},
\end{cases}
\end{equation}
where we have defined $\alpha_y = \frac{y^2}{(4\pi)^2}$. We rescale the couplings as in (\ref{eq:QCD_rescaling}), together with the rescaling $N_c \alpha_y \rightarrow \alpha_y$, and we define the ratio $x := \frac{N_f}{N_c}$, and $N:=N_c$. We find
\begin{equation}\label{eq:QCDYukA_Model2}
\begin{cases}
\Lambda \frac{d \alpha_{g}}{d\Lambda} = & -\frac{2}{3} (11 - 2x) \alpha_{g}^{2} - \frac{2}{3} (34 - 13 x) \alpha_{g}^{3} - 2 x^2 \alpha_y \alpha_{g}^{2} + 2 x g_V \alpha_{g}^{2}\\
& + N^{-2} \left(-2x \alpha_{g}^{3}\right),\\
\Lambda \frac{d \alpha_{y}}{d\Lambda} = & 2 (2+x) \alpha_{y}^{2} - 6 \alpha_y \alpha_g - 4 g_S \alpha_y\\
&  + N^{-2} \left( 6 \alpha_{y} \alpha_g + 16 g_{V_1} \alpha_y \right),\\
\Lambda \frac{d g_{S}}{d\Lambda} =  &2g_{S} - 2 g_{S}^{2} + 2x g_{S}g_{V} - 6g_{S}\alpha_{g} - \frac{9}{2} \alpha_{g}^{2}\\
& + N^{-1} \left( 2 g_S \alpha_{y} + 4 g_{V} \alpha_y \right) + N^{-2} \left(6g_{S}g_{V_1} + 2g_{S}g_{V_2} + 6g_{S}\alpha_{g} + 12g_{V_1}\alpha_{g} + 12\alpha_{g}^{2}\right),\\
\Lambda \frac{d g_{V}}{d\Lambda} =  &2g_{V} + \frac{1}{4}xg_{S}^{2} + (1 + x) g_{V}^{2} - \frac{3}{4} \alpha_{g}^{2}\\
& + N^{-1} \left( 2 g_V \alpha_{y} + g_{S} \alpha_y \right) + N^{-2} \left(-6g_{V}g_{V_2} - 6g_{V}\alpha_{g} + 6g_{V_2}\alpha_{g} + 6\alpha_{g}^{2}\right),\\
\Lambda \frac{d g_{V_1}}{d\Lambda} = &2g_{V_1} - \frac{1}{4} g_{s}^{2} - g_{S} g_{V} - x g_{S} g_{V_2} + 2(1+x)g_{V}g_{V_1} + 2xg_{V_1}g_{V_2} + 2\alpha_{y}^{2} + \frac{3}{4} \alpha_{g}^{2}\\
& + N^{-1} \left( 2 g_{V_1} \alpha_{y} - 2 g_{V_2} \alpha_y \right) + N^{-2} \left(-3g_{V_1}^{2} + 2g_{V_1}g_{V_2} + 6g_{V_1}\alpha_{g} + 3\alpha_{g}^{2} \right),\\
\Lambda \frac{d g_{V_2}}{d\Lambda} = &2g_{V_2} - 3 g_{V}^{2} - x g_{V_1}^{2} + x g_{V_2}^{2} - x g_{S} g_{V_1} + 2 (1 + x) g_{V}g_{V_2} + 6g_{V}\alpha_{g} + 2\alpha_{y}^{2} - \frac{9}{4} \alpha_{g}^{2}\\
& + N^{-1} \left( 2 g_{V_2} \alpha_{y} - 2 g_{V_1} \alpha_y \right) + N^{-2} \left(-2g_{V_2}^{2} - 6 g_{V_2}\alpha_{g} - 3 \alpha_{g}^{2} \right).
\end{cases}
\end{equation}

\subsection{The Veneziano Limit}
We take the Veneziano limit ($N \rightarrow \infty$), which yields
\begin{equation}\label{eq:QCDYukA_Ven_Model1}
\begin{cases}
\Lambda \frac{d \alpha_{g}}{d\Lambda} = & -\frac{2}{3} (11 - 2x) \alpha_{g}^{2} - \frac{2}{3} (34 - 13 x) \alpha_{g}^{3} - 2 x^2 \alpha_y \alpha_{g}^{2} + 2 x g_V \alpha_{g}^{2},\\
\Lambda \frac{d \alpha_{y}}{d\Lambda} = & 2 (2+x) \alpha_{y}^{2} - 6 \alpha_y \alpha_g - 4 g_S \alpha_y,\\
\Lambda \frac{d g_{S}}{d\Lambda} =  &2g_{S} - 2 g_{S}^{2} + 2x g_{S}g_{V} - 6g_{S}\alpha_{g} - \frac{9}{2} \alpha_{g}^{2},\\
\Lambda \frac{d g_{V}}{d\Lambda} =  &2g_{V} + \frac{1}{4}xg_{S}^{2} + (1 + x) g_{V}^{2} - \frac{3}{4} \alpha_{g}^{2},\\
\Lambda \frac{d g_{V_1}}{d\Lambda} = &2g_{V_1} - \frac{1}{4} g_{s}^{2} - g_{S} g_{V} - x g_{S} g_{V_2} + 2(1+x)g_{V}g_{V_1} + 2xg_{V_1}g_{V_2} + 2\alpha{y}^{2} + \frac{3}{4} \alpha_{g}^{2},\\
\Lambda \frac{d g_{V_2}}{d\Lambda} = &2g_{V_2} - 3 g_{V}^{2} - x g_{V_1}^{2} + x g_{V_2}^{2} - x g_{S} g_{V_1} + 2 (1 + x) g_{V}g_{V_2} + 6g_{V}\alpha_{g} + 2\alpha{y}^{2} - \frac{9}{4} \alpha_{g}^{2}.\\
\end{cases}
\end{equation}
The first 4 equations decouple from the last two and we can reduce analysis to 
\begin{equation}\label{eq:QCDYukA_Ven_Model2}
\begin{cases}
\Lambda \frac{d \alpha_{g}}{d\Lambda} = & -\frac{2}{3} (11 - 2x) \alpha_{g}^{2} - \frac{2}{3} (34 - 13 x) \alpha_{g}^{3} - 2 x^2 \alpha_y \alpha_{g}^{2} + 2 x g_V \alpha_{g}^{2},\\
\Lambda \frac{d \alpha_{y}}{d\Lambda} = & 2 (2+x) \alpha_{y}^{2} - 6 \alpha_y \alpha_g - 4 g_S \alpha_y,\\
\Lambda \frac{d g_{S}}{d\Lambda} =  &2g_{S} - 2 g_{S}^{2} + 2x g_{S}g_{V} - 6g_{S}\alpha_{g} - \frac{9}{2} \alpha_{g}^{2},\\
\Lambda \frac{d g_{V}}{d\Lambda} =  &2g_{V} + \frac{1}{4}xg_{S}^{2} + (1 + x) g_{V}^{2} - \frac{3}{4} \alpha_{g}^{2}.\\
\end{cases}
\end{equation}

The domain of the variables and parameters is given by  $x \in\mathbb{R}^{+}$, $\alpha_g, \alpha_y \in\mathbb{R}_{0}^{+}$ and $g_{S}, g_{V} \in\mathbb{R}$. Due to the fact that the model (\ref{eq:QCDYukA_Model2}) is perturbative plus a few higher order corrections, we expect better results for small parameter values. As a rough bound we will use $\alpha_i<1$, $|g_i|<1$, and notice when this bound is exceeded.\\

The beta function for $\alpha_g$ has a double root for $\alpha_g^{t}=0$ and another root for $\alpha_g^{nt}=\frac{11 - 2x + 3x^2 \alpha_y - 3x g_V}{13x-34}$. Therefore, the fixed points of the RG-flow lie on these manifolds. The manifold defined by $\alpha_g=0$ is an invariant set of the system, since $\dot{\alpha}_{g}$ vanishes for $\alpha_g = 0$.\\

In addition, the beta function for $\alpha_y$ has a root for $\alpha_y^{t}=0$ and another root for $\alpha_y^{nt}=\frac{3\alpha_g + 2 g_S}{2 + x}$. Therefore, the fixed points of the RG-flow lie on these manifolds. The manifold defined by $\alpha_y=0$ is an invariant set of the system, since $\dot{\alpha}_{y}$ vanishes for $\alpha_y = 0$.\\

\begin{figure}
	\begin{minipage}{0.45\linewidth}
		\includegraphics[width=3.0in]{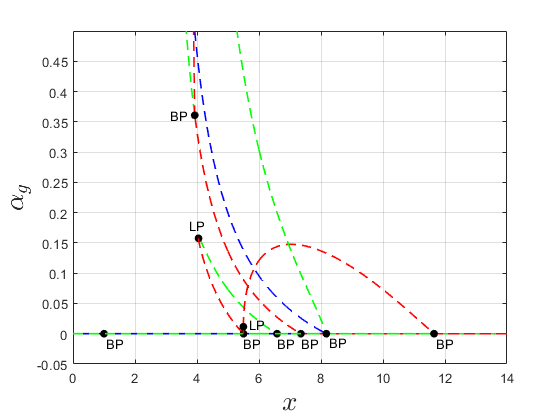}
	\end{minipage}
	\hspace{0.1\linewidth}
	\begin{minipage}{0.45\linewidth}
		\includegraphics[width=3.0in]{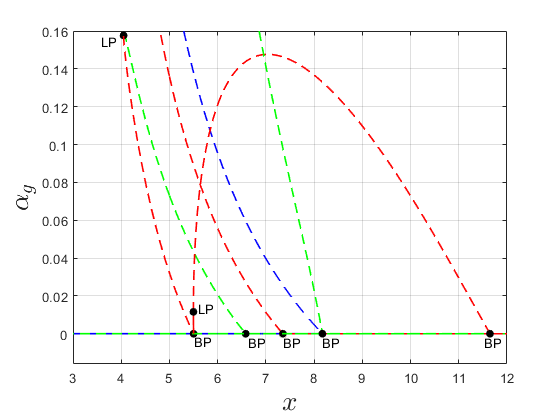}
	\end{minipage}
	\\
	\begin{minipage}{0.45\linewidth}
		\includegraphics[width=3.0in]{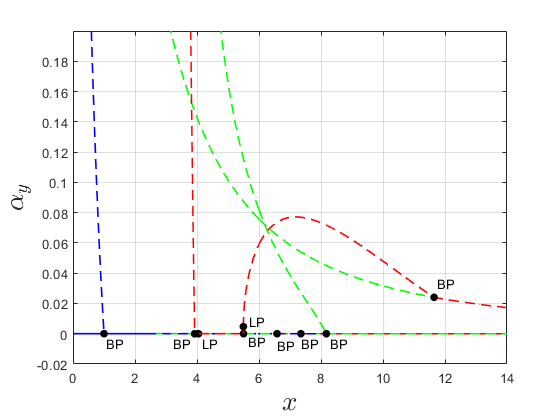}
	\end{minipage}
	\hspace{0.1\linewidth}
	\begin{minipage}{0.45\linewidth}
		\includegraphics[width=3.0in]{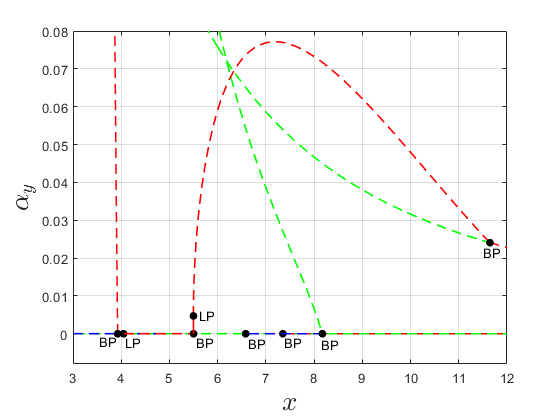}
	\end{minipage}
	\\
	\begin{minipage}{0.45\linewidth}
		\includegraphics[width=3.0in]{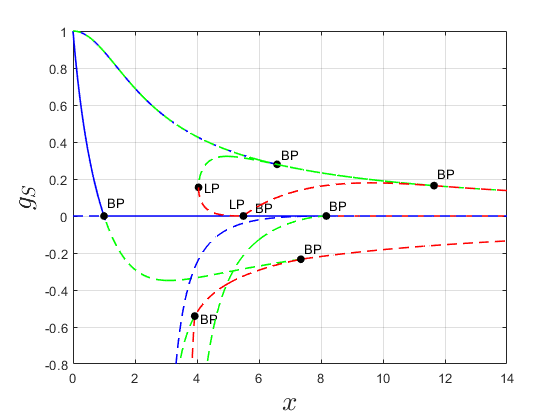}
	\end{minipage}
	\hspace{0.1\linewidth}
	\begin{minipage}{0.45\linewidth}
		\includegraphics[width=3.0in]{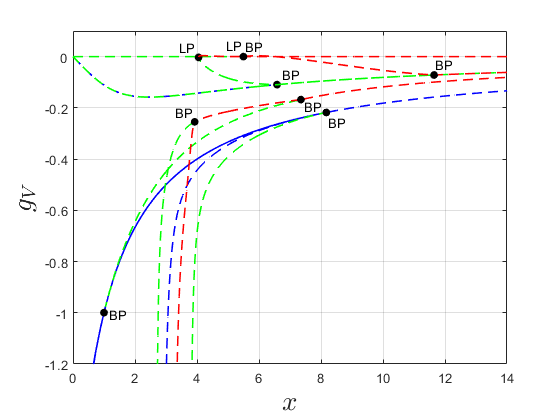}
	\end{minipage}
	\\
	\begin{minipage}{1\linewidth}
		\centering
		\includegraphics[width=3.0in]{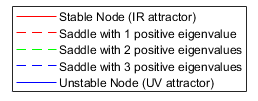}
	\end{minipage}
	\caption{The equilibria of the RG-flow of (\ref{eq:QCDYukA_Ven_Model2}) projected on the $(x,\alpha_g)$-plane, $(x,\alpha_y)$-plane, $(x,g_S)$-plane and the $(x,g_V)$-plane. LP: Limit Point. BP: Branching Point.}
	\label{fig:QCDYukA_Ven_2D}
\end{figure}

\begin{figure}
	\begin{minipage}{0.45\linewidth}
		\includegraphics[width=3.0in]{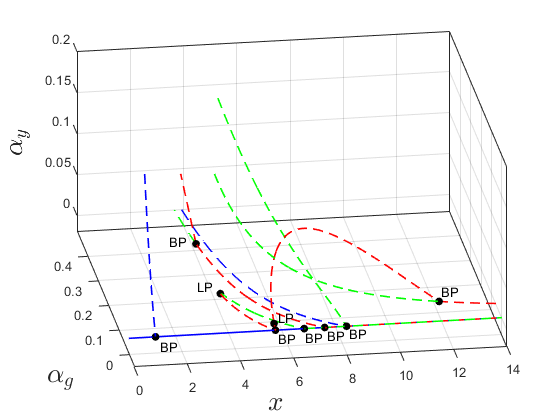}
	\end{minipage}
	\hspace{0.1\linewidth}
	\begin{minipage}{0.45\linewidth}
		\includegraphics[width=3.0in]{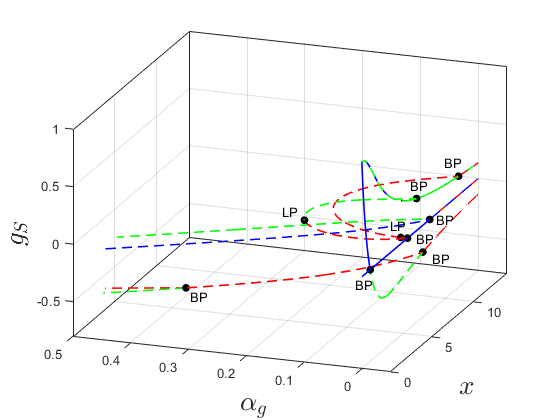}
	\end{minipage}
	\\
	\begin{minipage}{0.45\linewidth}
		\includegraphics[width=3.0in]{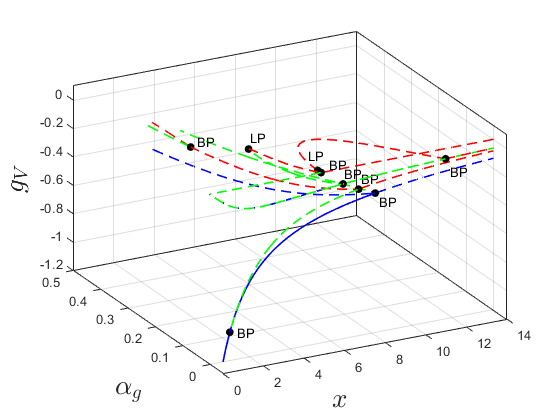}
	\end{minipage}
	\hspace{0.1\linewidth}
	\begin{minipage}{0.45\linewidth}
		\includegraphics[width=3.0in]{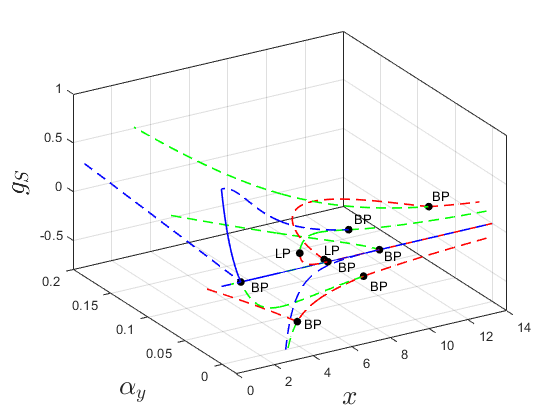}
	\end{minipage}
	\\
	\begin{minipage}{0.45\linewidth}
		\includegraphics[width=3.0in]{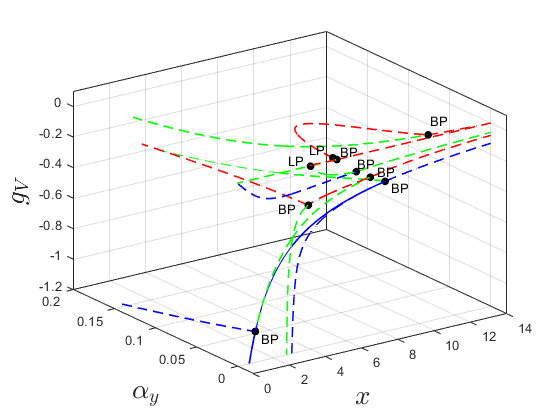}
	\end{minipage}
	\hspace{0.1\linewidth}
	\begin{minipage}{0.45\linewidth}
		\includegraphics[width=3.0in]{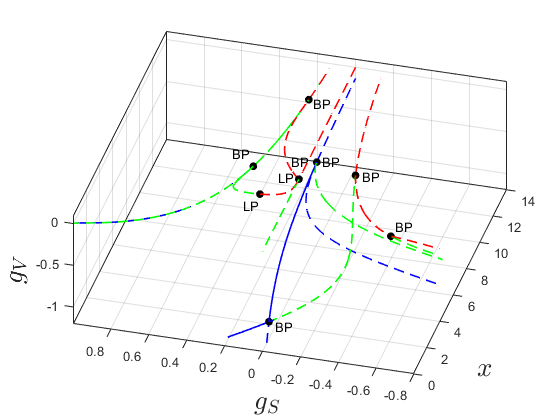}
	\end{minipage}
	\\
	\begin{minipage}{1\linewidth}
		\centering
		\includegraphics[width=3.0in]{QCD6/legend_ven_6d.png}
	\end{minipage}
	\caption{The equilibria of the RG-flow of (\ref{eq:QCDYukA_Ven_Model2}) projected on the $(x,\alpha_g,\alpha_y)$-space, $(x,\alpha_g,g_S)$-space, $(x,\alpha_g,g_V)$-space, $(x,\alpha_y,g_S)$-space, $(x,\alpha_y,g_V)$-space and the $(x,g_S,g_V)$-space. LP: Limit Point. BP: Branching Point.}
	\label{fig:QCDYukA_Ven_3D}
\end{figure}

The location of the fixed points projected on the $(x,\alpha_g)$-plane, $(x,\alpha_y)$-plane, $(x,g_S)$-plane and the $(x,g_V)$-plane is shown in Figure \ref{fig:QCDYukA_Ven_2D}, and the projections on the $(x,\alpha_g,\alpha_y)$-space, $(x,\alpha_g,g_S)$-space, $(x,\alpha_g,g_V)$-space, $(x,\alpha_y,g_S)$-space, $(x,\alpha_g,g_V)$-space and the $(x,g_S,g_V)$-space in Figure \ref{fig:QCDYukA_Ven_3D}. Here, we use the following color coding:

\begin{itemize}
	\item Solid red line: stable node, 4 negative eigenvalues (IR attractor).
	\item Dashed red line: saddle point, 3 negative and 1 positive eigenvalue.
	\item Dashed green line: saddle point, 2 negative and 2 positive eigenvalues.
	\item Dashed blue line: saddle point, 1 negative and 3 positive eigenvalues.
	\item Solid blue line: unstable node, 4 positive eigenvalues (UV attractor).
\end{itemize}

The equilibria on the $\alpha_{g}=0$ manifold all have a trivial eigenvalue, and therefore do not fit the color coding as described above. In order to use the color coding, we define the sign of a trivial eigenvalue on this manifold by approaching the equilibrium along a line of constant $\alpha_y$, $g_S$ and $g_V$ from positive but small $\alpha_g$. The trivial eigenvalue will then approach $0$ from either positive or negative values. If the trivial eigenvalue is slightly positive for $0<\alpha_g\ll1$, we define the trivial eigenvalue to be positive and vice versa.\footnote{This definition makes physical sense, since negative values of $\alpha_g$ are unphysical.}\\

Since $\beta_{\alpha_{y}}=0$ if $\alpha_y=0$, we find the same fixed points and bifurcations as in the previous model without scalar field (cf. Figure \ref{fig:QCD_Ven_FixedPoints}). In addition, we now find five fixed points with non-trivial value of $\alpha_y$. We find two additional branching points and one additional limit point, as can be seen in Table \ref{tab:QCDYukA_Ven_Bif}. Furthermore, at 3 transcritical bifurcations, that were found in the $\alpha_y=0$ model, we now find three equilibria that intersect and exchange stability, corresponding to the normal form
\begin{equation}
\begin{cases}
	\dot{x} = \alpha x + x^{2},\\
	\dot{y} = \alpha y + y^{2}. 
\end{cases}
\end{equation}

One of the additional fixed points branches from the transcritical bifurcation at $x=1.000$ that was already found in the model without a scalar, and diverges rapidly in $\alpha_y$. In addition, one new fixed point branches from one of the equilibria with non-trivial value of $\alpha_g$ at $x=3.926$, which also diverges rapidly in $\alpha_y$. Furthermore, at the branching point at $x=8.173$, an extra fixed point branches of with a non-trivial value of $\alpha_g$ and $\alpha_y$, which diverges when $x$ is lowered towards $x \approx 4$. Another new fixed point with non-trivial $\alpha_y$ exists in the scalar model. For this point $\alpha_y \rightarrow 0$ if $x \rightarrow \infty$  and $\alpha_y\rightarrow \infty$ when $x$ is lowered towards $x \approx 2$. From this point another fixed point branches at $x=11.647$. This point disappears trough a saddle-node bifurcation at $x=5.497$. The other point that disappears at this saddle-node bifurcation branches at $x=5.5$ from the trivial fixed point $(\alpha_g,\alpha_y,g_S,g_V)=(0,0,0,0)$. This point is the stable node in the system. This stable node disappears through a saddle-node bifurcation at $5.497$. We therefore find a conformal window with non-trivial for $\alpha_y$ at $x\in(5.497,5.501)$, which is smaller than a prediction on a model without effective four-fermi interactions, where $x\in(5.24,5.5)$ \cite{Terao}.\\

\begin{table}[H]
	\centering
	\caption{Bifurcations found in the Veneziano limit of QCD$_4$ with a scalar field (\ref{eq:QCDYukA_Ven_Model2}).}
	\label{tab:QCDYukA_Ven_Bif}
	\begin{tabular}{|l|r|r|r|r|r|r|r|}
		\hline
		Bifurcation   & $x$ & $\alpha_g$ & $\alpha_y$ & $g_S$    & $g_V$    \\
		\hline
		Saddle-node   & $4.049$  & $0$      & $0.158$ & $0.155$  & $-0.003$ \\
		Saddle-node   & $5.497$  & $0.012$  & $0.005$ & $0$      & $0$      \\ 
		Transcritical & $1.000$  & $0$      & $0$	  & $0$      & $-1.000$ \\
		Transcritical & $3.926$  & $0.361$  & $0$     & $-0.541$ & $-0.255$ \\ 
		Transcritical & $5.501$  & $0$      & $0$     & $0$      & $0$      \\
		Transcritical & $6.582$  & $0$      & $0$     & $0.279$  & $-0.110$ \\
		Transcritical & $7.351$  & $0$      & $0$     & $-0.234$ & $-0.168$ \\
		Transcritical & $8.173$  & $0$      & $0$     & $0$      & $-0.218$ \\
		Transcritical & $11.647$ & $0$      & $0.024$ & $0.164$  & $-0.072$ \\
		\hline
	\end{tabular}
\end{table}

\subsection{Small $N_c$ regime}

In this section, we analyze the complete model (\ref{eq:QCDYukA_Model2}) outside the Veneziano limit. The beta function for $\alpha_g$ has two roots:
\begin{equation*}
	\alpha_{g} = 0, \quad \alpha_g = \alpha_{g}^{nt}:= N_{c}^2 \frac{-11 + 2 x - 3 x^2 \alpha_y + 3 x g_V}{34N_{c}^{2} - 13 x N_c + 3 x},
\end{equation*}
where the trivial root has multiplicity 2. The trivial root makes the $\alpha_g=0$ hyperplane an invariant set of the model (\ref{eq:QCDYukA_Model2}). The beta function for $\alpha_y$ also has two roots:
\begin{equation*}
	\alpha_{y} = 0, \quad \alpha_y = \alpha_{y}^{nt}:= \frac{1}{N_c} \frac{N_{c}^{2} (3\alpha_g + 2g_S) - (3\alpha_g + 8 g_{V_1})}{2 + x N_{c}}.
\end{equation*}

\begin{figure}[H]
	\centering
	\includegraphics[width=.4\linewidth]{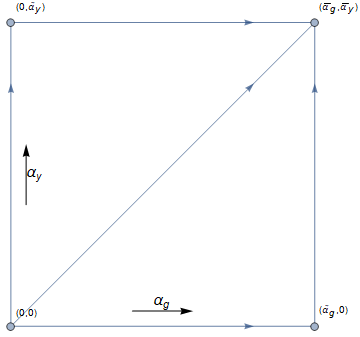}
	\caption{Topology of the RG-flow between 4 fixed points.}
	\label{fig:QCDYukA_FixedPointStruc}
\end{figure}

\begin{figure}[H]
	\begin{minipage}{0.45\linewidth}
		\includegraphics[width=3.0in]{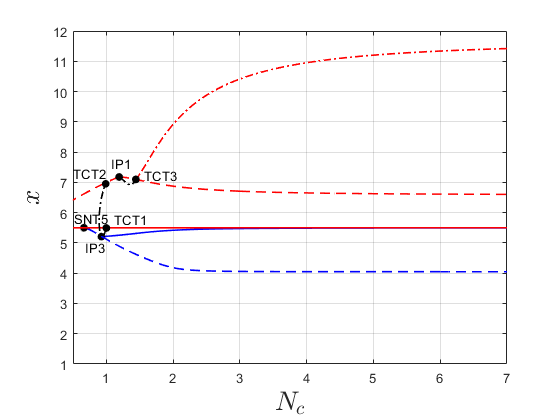}
		\subcaption{}
		\label{fig:QCDYukA_SN_TC_3}
	\end{minipage}
	\hspace{0.1\linewidth}
	\begin{minipage}{0.45\linewidth}
		\includegraphics[width=3.0in]{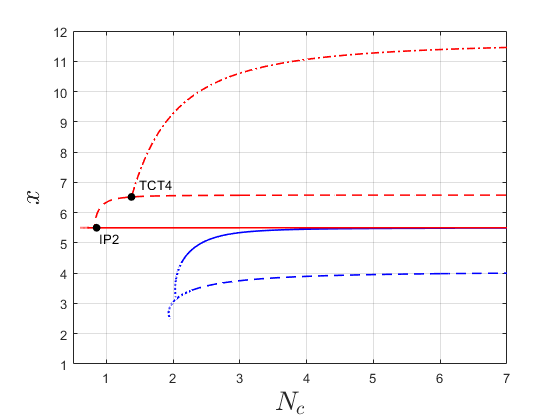}
		\subcaption{}
		\label{fig:QCDYukA_SN_TC_2}
	\end{minipage}
	\\
	\begin{minipage}{0.45\linewidth}
		\includegraphics[width=3.0in]{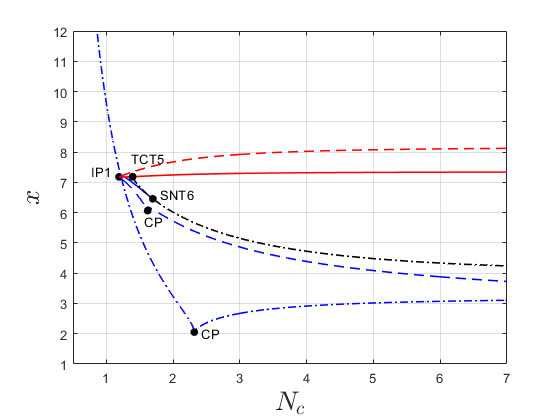}
		\subcaption{}
		\label{fig:QCDYukA_SN_TC_1}
	\end{minipage}
	\hspace{0.1\linewidth}
	\begin{minipage}{0.45\linewidth}
		\includegraphics[width=3.0in]{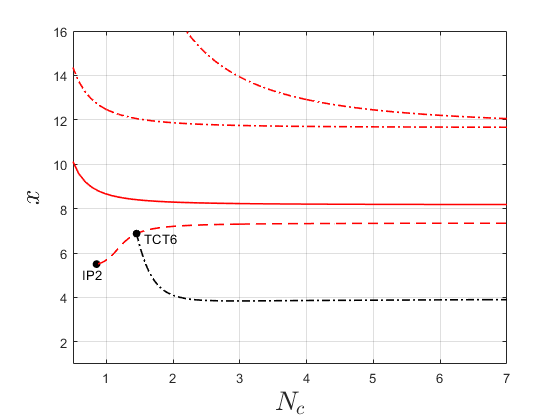}
		\subcaption{}
		\label{fig:QCDYukA_SN_TC_4}
	\end{minipage}
	\\
	\begin{minipage}{1\linewidth}
		\centering
		\includegraphics[width=3.0in]{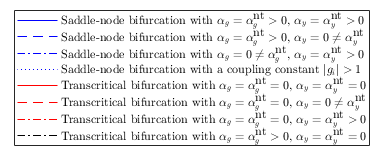}
	\end{minipage}
	\caption{Bifurcations in the RG-flow of (\ref{eq:QCDYukA_Model2}) in the $(N_c,x)$-plane. CP: Cusp bifurcation. SNT: Saddle-Node-Transcritical bifurcation. TCT: Transcritical-Transcritical bifurcation. IP: degenerate bifurcation point.}
	\label{fig:QCDYukA_SN_TC}
\end{figure}

The trivial root makes the $\alpha_y=0$ hyperplane an invariant set of the model (\ref{eq:QCDYukA_Model2}). Since $\alpha_y=0$ is an invariant hyperplane, we find the same bifurcations as in previous section. However, on top of those, there exist fixed points and bifurcations in the $\alpha_y\neq0$ space. We find up to six fixed points with $\alpha_g,\alpha_y \neq 0$ and up to six fixed points with $\alpha_g=0,\alpha_y \neq 0$, depending on the values of $(N_c,x)$. As in the previous section, we find that fixed points with non-trivial $\alpha_g$ always collide on the intersection $\alpha_{g}^{nt}=0$ with a fixed point with trivial value $\alpha_g=0$ through a transcritical bifurcation. This happens at
\begin{equation*}
	x = \frac{ 2+3g_V \pm \sqrt{(2+3g_V)^2 - 132 \alpha_{y}}}{6\alpha_y} \quad \textrm{or} \quad \alpha_{y}=0, \, x =  \frac{11}{2+3g_V},
\end{equation*}
where the second condition is a reduction to the model without scalar field. Similarly, fixed points with non-trivial $\alpha_y$ always collide on the intersection $\alpha_{y}^{nt}=0$ with a fixed point with trivial value $\alpha_y=0$ through a transcritical bifurcation. This happens at
\begin{equation*}
	N_c = \pm \sqrt{\frac{3\alpha_g + 8 g_{V_1}}{3\alpha_g + 2 g_S}} \quad \textrm{or} \quad g_S = - \frac{3}{2} \alpha_g, \, g_{V_1} = - \frac{3}{8} \alpha_g,
\end{equation*}
where only the positive root is a physically relevant solution. In this way, one can relate most fixed points with $(\alpha_g>0,\alpha_y>0)$ to a fixed point with $(\alpha_g=0,\alpha_y>0)$, a fixed point with $(\alpha_g>0,\alpha_y=0)$ and a fixed point with $(\alpha_g=0,\alpha_y=0)$. We find that the orbits between these points are always as indicated in Figure \ref{fig:QCDYukA_FixedPointStruc}. Furthermore, a non-trivial fixed point\footnote{We call a fixed point non-trivial if $(\alpha_g\neq0$, $\alpha_y\neq0)$} with positive $\alpha_g$ and $\alpha_y$ can appear through a transcritical bifurcation at $(\alpha_g=\alpha_{g}^{nt}=0,\alpha_y>0)$, at $(\alpha_g>0,\alpha_y=\alpha_{y}^{nt}=0)$ or at $(\alpha_g=\alpha_{g}^{nt}=0,\alpha_y=\alpha_{y}^{nt}=0)$. These transcritical bifurcation curves are shown in Figure \ref{fig:QCDYukA_SN_TC} by dash-dotted red lines, dash-dotted black lines and solid red lines respectively. When $x$ decreases non-trivial fixed points will either diverge in one or multiple couplings or disappear through a saddle-node bifurcation with another fixed point. We find three saddle-node bifurcation curves with $(\alpha_g \neq 0,\alpha_y \neq 0)$ as shown in Figure \ref{fig:QCDYukA_SN_TC_3}, \ref{fig:QCDYukA_SN_TC_2} and \ref{fig:QCDYukA_SN_TC_1}. They're shown together with the related\footnote{We say that saddle-node bifurcation curves with ($\alpha_g>0, \alpha_y>0)$ and $(\alpha_g>0, \alpha_y=0)$ or $(\alpha_g=0, \alpha_y>0)$ are related, if the equilibria that are involved can be related through a transcritical bifurcation in the $\alpha_g=\alpha_{g}^{nt}=0$ or $\alpha_y=\alpha_{y}^{nt}=0$ subspace.} saddle-node bifurcation curves that were found for $\alpha_y=0$ in the previous section. In addition, a saddle-node bifurcation curve with $(\alpha_g=0, \alpha_y=0)$ related to the saddle-node bifurcation in Figure \ref{fig:QCDYukA_SN_TC_1} is shown.\\

Figure \ref{fig:QCDYukA_SN_TC_3} shows the bifurcations of the sets of fixed points which were indicated as pairs 8 and 10 in the previous section. Here, we see the transcritical and saddle-node bifurcations that were found in the previous section. In addition, there is an extra transcritical bifurcation at the solid red line, where a non-trivial fixed point branches of. The solid red line converges to the branching point at $x=5.5$ that was found in the Veneziano limit. Furthermore, we see a transcritical bifurcation at non-trivial value of $\alpha_y$ (dash-dotted red line), where a non-trivial fixed point splits of. This transcritical bifurcation converges to the branching point found at $x=11.6$ in the Veneziano limit. The two non-trivial fixed points disappear through a saddle-node bifurcation at the solid blue line, which is related to the blue dashed line representing the saddle-node bifurcation that was found in the $\alpha_y=0$ space. The solid blue line converges to limit point that was found at $x=5.5$ in the Veneziano limit.\\

Figure \ref{fig:QCDYukA_SN_TC_2} shows the bifurcations of the sets of fixed points which were indicated by the pairs 3 and 4 in the previous section. Here, we see the transcritical and saddle-node bifurcations that were found in the previous section. In addition, there is an extra transcritical bifurcation at the solid red line, where a non-trivial fixed point branches of. The solid red line converges to the branching point at $x=5.5$ that was found in the Veneziano limit. Furthermore, we see a trancritical bifurcation at non-trivial value of $\alpha_y$ (dash-dotted red line), where a non-trivial fixed point splits of. This transcritical bifurcation converges to the branching point found at $x=11.6$ in the Veneziano limit. The two non-trivial fixed points disappear through a saddle-node bifurcation at the solid blue line, which is related to the blue dashed line representing the saddle-node bifurcation that was found in the $\alpha_y=0$ space. The solid blue line converges to limit point that was found at $x=5.5$ in the Veneziano limit.\\

Figure \ref{fig:QCDYukA_SN_TC_1} shows the bifurcations of the sets of fixed points which were indicated by the pairs 6 and 9 in the previous section. Here, we see the transcritical and saddle-node bifurcations that were found in the previous section. In addition there is an extra transcritical bifurcation at the solid red line, where a non-trivial fixed point branches of. The solid red line converges to the branching point at $x=7.4$ that was found in the Veneziano limit. Furthermore, we see a transcritical bifurcation at non-trivial value of $\alpha_g$ (dash-dotted black line), where a non-trivial fixed point splits of. This transcritical bifurcation diverges in the Veneziano limit. The two non-trivial fixed points disappear through a saddle-node bifurcation at the solid blue line, which is related to the blue dashed line representing the saddle-node bifurcation that was found in the $\alpha_y=0$ space. The solid blue line intersects the black dash-dotted line at SNT6. Therefore only in a small interval $N_c\in(1.2,1.5)$ the non-trivial fixed points disappear through a saddle-node bifurcation, while at larger $N_c$ the non-trivial fixed points that branch of from the solid red line and the black dash-dotted line diverge for small values of $x$. In addition, for these fixed points we find a saddle-node bifurcation curve not only in the $\alpha_y=0$ space, as was the case in the previous figures, but also in the $\alpha_g=0$ space (dash-dotted blue line).\\

Figure \ref{fig:QCDYukA_SN_TC_4} shows a few more transcritical bifurcation curves lying in the $\alpha_{g}^{nt}=0$ and/or $\alpha_{y}^{nt}=0$ subspace. Non-trivial fixed points that are related to these transcritical bifurcation curves, all diverge in one or multiple couplings for small $x$ instead of disappearing through a saddle-node bifurcation. We notice the the black, solid red and dashed red line involve the sets of equilibria 1 and 2, as labeled in the previous section.\\

The eigenvectors that cross zero at the saddle-node bifurcations shown in Figure \ref{fig:QCDYukA_SN_TC_3} and \ref{fig:QCDYukA_SN_TC_2} are shown in Table \ref{tab:QCDYukA_CrossOperators} for $N_c=3$ along with the eigenvalues of the other eigenvectors. We see that in this model the bifurcations are strongly triggered by the $\alpha_g$ interaction. Furthermore, as in the previous, model we find two complex eigenvalues for the saddle-node bifurcation of point the non-trivial fixed points 3 and 4. Eigenvectors corresponding to the complex eigenvalues are directed along the four-fermion couplings, and have very small contributions ($|\eta_{\alpha_{g}}|<0.004,|\eta_{\alpha_{y}}|<0.016$) along the $e_{\alpha_g}$ and $e_{\alpha_y}$ direction, meaning that this complex scaling behavior involves operators, which are linear combinations of the effective four-fermi interactions.\\

\begin{table}[H]
	\centering
	\caption{Eigenvalues and critical eigenvectors crossing marginality of the saddle-node bifurcation with non-trivial $\alpha_g, \alpha_y$ at $N_c=3$.}
	\label{tab:QCDYukA_CrossOperators}
	\begin{tabular}{|c|c|c|c|c|}
		\hline
		Equilibria  & $(N_c,x)$ & Eigenvalues & Critical eigenvector \\
		\hline
		8, 10 & $(3.00,5.47)$ & $\{-2.06, -2.04, -1.93, -1.83, -0.20, 0\}$ & $(0.92, 0.36, 0.11, 0.00, -0.04,\; 0.07)$ \\
		3, 4  & $(3.00,5.34)$ & $\{2.38\pm0.28i, -2.34, -1.36, -1.09, 0\}$ & $(0.69, 0.29, 0.32, 0.09,\; 0.41, -0.39)$\\
		\hline
	\end{tabular}
\end{table}

Furthermore, we find multiple bifurcation points in Figure \ref{fig:QCDYukA_SN_TC}, which are reported in Table \ref{tab:QCDYukA_Complete_Bif}. The first cusp bifurcation and the points SNT5, IP1 and IP2 were already found in the previous model. In addition, we find an extra cusp point in the saddle-node bifurcation curve in the $\alpha_g=0$ space. This curve has another cusp bifurcation at IP1, making IP1 even more degenerate. The points TCT1 up to TCT6 all describe the intersection of two transcritical bifurcation curves, and can be described by the normal form
\begin{equation}
	\begin{cases}
		\dot{x} = \alpha x + x^{2},\\
		\dot{y} = \beta y + y^{2}. 
	\end{cases}
\end{equation}

Finally, IP3 describes a point where 2 saddle-node bifurcation curves and 2 branches of a transcritical bifurcation intersect, and can therefore be described as a saddle-node-transcritical bifurcation, where an additional saddle-node
bifurcation intersects. Four fixed points are involved in this bifurcation.

\begin{table}[H]
	\centering
	\caption{Bifurcations found in the model of QCD$_4$ (\ref{eq:QCDYukA_Model2}) shown in Figure \ref{fig:QCDYukA_SN_TC}.}
	\label{tab:QCDYukA_Complete_Bif}
	\begin{tabular}{|l|r|r|r|r|r|r|r|r|r|}
		\hline
		Bifurcation & $N_c$ & $N_f$ & $x$ & $\alpha_g$ & $\alpha_y$ & $g_S$ & $g_V$ & $g_{V_1}$ & $g_{V_2}$\\
		\hline
		CP   & $1.62$ & $9.82$  & $6.07$ & $0.05$ & $0$    & $0.03$  & $-0.17$ & $0.01$  & $0.22$ \\
		CP   & $2.32$ & $4.75$  & $2.05$ & $0$ 	  & $0.20$ & $0.55 $ & $-0.32$ & $0.20$  & $0.51$ \\
		SNT5 & $0.66$ & $3.65$  & $5.50$ & $0$    & $0$    & $0$     & $0$     & $0$     & $0$    \\
		SNT6 & $1.70$ & $10.95$ & $6.46$ & $0.03$ & $0$    & $-0.10$ & $-0.17$ & $-0.05$ & $0.23$ \\
		TCT1 & $1.00$ & $5.50$  & $5.50$ & $0$    & $0$    & $0$     & $0$     & $0$     & $0$    \\
		TCT2 & $1.00$ & $6.95$  & $6.95$ & $0$    & $0$    & $0$     & $-0.14$ & $0$     & $0.14$ \\
		TCT3 & $1.44$ & $10.23$ & $7.10$ & $0$    & $0$    & $0.13$  & $-0.15$ & $0.07$  & $0.20$ \\
		TCT4 & $1.38$ & $8.97$  & $6.52$ & $0$    & $0$    & $0.40$  & $-0.10$ & $0.19$  & $-0.42$\\
		TCT5 & $1.39$ & $10.02$ & $7.19$ & $0$    & $0$    & $-0.10$ & $-0.16$ & $-0.05$ & $0.20$ \\
		TCT6 & $1.45$ & $9.98$  & $6.87$ & $0$    & $0$    & $-0.40$ & $-0.13$ & $-0.21$ & $0.39$ \\
		IP1  & $1.19$ & $8.54$  & $7.18$ & $0$    & $0$    & $0$     & $-0.16$ & $0$     & $0.17$ \\
		IP2  & $0.85$ & $4.69$  & $5.50$ & $0$    & $0$    & $0$     & $0$     & $0$     & $-0.73$\\
		IP3  & $0.93$ & $4.82$  & $5.21$ & $0.05$ & $0$    & $-0.01$ & $-0.01$ & $0.00$  & $0.01$ \\	
		\hline
	\end{tabular}
\end{table}

\subsection{Discussion of the Results}
In this section, we have seen how the addition of a scalar field alters the bifurcations found in the previous section. In particular, we would like to know whether we find similar behavior in the $(\alpha_g>0, \alpha_y>0)$ space as we found in the $(\alpha_g>0$, $\alpha_y=0)$ space. We find the transcritical bifurcation at $x=5.5$, which marks the upper edge of the conformal window. This transcritical bifurcation now indicates the intersection of three fixed points: one on the $(\alpha_g=0, \alpha_y=0)$ subspace, one on the $(\alpha_g>0, \alpha_y=0)$ subspace, and one on the $(\alpha_g>0, \alpha_y>0)$. Furthermore, at this line there are two different transcritical bifurcation curves for different values of $(g_S,g_V,g_{V_1},g_{V_2})$ as was also found in the previous model. One of these bifurcations is shown in Figure \ref{fig:QCDYukA_SN_TC_3} and the other in Figure \ref{fig:QCDYukA_SN_TC_2}. We see that both saddle-node bifurcations that could indicate the lower edge of the conformal window in the model with $\alpha_y=0$ have a related saddle-node bifurcation in the $(\alpha_g>0, \alpha_y>0)$ space, indicating that the phase transition at the lower edge of the conformal window is similar to the previous section and involves the merger of fixed points, followed by a walking behavior of the coupling constant, that changes continuously to a running behavior. However, the non-trivial saddle-node bifurcations are found at a larger value of $x$ than the ones in the $\alpha_y=0$ subspace, and are triggered by the $\alpha_g$ interaction instead of the $g_S$ interaction. This curve is very close but slightly below $x=5.5$ at large values of $N_c$ and slightly decreases towards smaller $N_c$. Therefore, the conformal window is much smaller in the theory that contains a scalar field in the adjoint representation, although fixed points and renormalized trajectories between those points remain to exist for smaller values of $x$, but these are very unstable and require a large degree of fine-tuning within this model.\\

Furthermore, we find that the fixed points, which were labeled by 5 and 7 have no related fixed points with non-trivial value of $\alpha_y$. Therefore, we do not have the spiraling behavior related to those fixed points as we had in the previous section. For the saddle-node bifurcation of the pair labeled by 6 and 9 there is only a related bifurcation with non-trivial and positive $\alpha_g$ and $\alpha_y$ in a small interval $N_c\in(1.2,1.5)$, which makes it irrelevant for the physical model, since we require integer values of $N_c$. Therefore, we find that the non-trivial fixed points of the sets 1, 2, 6 and 9 all diverge for small $x$.

\section{Conclusions and Outlook}
We have evaluated an effective model for QCD$_4$, and studied the fixed points and their bifurcations in and close to the conformal window. We have found a fixed point merger (saddle-node bifurcation) in the Veneziano limit at $N_f/N_c = 4.049$ indicating the lower edge of the conformal window. This prediction is similar to the ones by \cite{Gukov,Kusafuka}, where the same model was analyzed. Furthermore, we have continued this bifurcation to the small $N_c$ regime, and found that this curve is more or less constant up to $(N_c=3.00,N_f=12.17)$. For $N_c\in\{1,2\}$ the window is choked off. On the other hand, there exists another fixed point merger (corresponding to the same merger in the Veneziano limit), which is more or less constant up to $(N_c=3.00,N_f=11.26)$. For $N_c=1$, these fixed points diverge instead of disappearing through a saddle-node bifurcation, while $N_c=2$ is an intermediate value where they merger at large coupling constants. Both mergers could indicate the lower edge of the conformal window. In both cases the transition at the lower edge of the conformal window is an infinite order phase transition, generated by the effective scalar four-fermi interaction for $N_c>2$. Furthermore, we have found a few extra sets of fixed points, that either disappear through saddle-node bifurcations or diverge.\\

The existence of multiple fixed points and multiple saddle-node bifurcations in this model, opens up many new questions about the lower edge of the conformal window. It might very well be the case that the infinite order perturbative beta function contains more fixed points apart from the well known Banks Zaks fixed point. This could result in a one or more transitions within the conformal window, where pairs of fixed points disappear through mergers. The lower edge is then reached when the last fixed point merger takes place. It would be interesting to try to relate such transitions to phase transitions that have been predicted in the quark gluon plasma. Furthermore, it would be interesting to extend the model with higher order effective interactions, and see which of the fixed points and saddle-node bifurcations remain to exist.\\

In this paper, we already extended the effective QCD model with an scalar field coupling through a Yukawa interaction to the fermions. This interaction turns out to destabilize the model. In this model we found that the 2 candidates for the conformal window both exist in this model as well. Although the lower edge of the conformal window is located at larger values of $N_f/N_c$ for such a model, as was already predicted in \cite{Terao}. The other fixed points on the other hand are pushed into a regime $\alpha_g<0$ and/or $\alpha_y<0$.\\

In addition, we have found that the model allows for complex scaling dimensions of the effective interactions. These complex scaling dimension contain both a scaling and a rotation of the associated operator. The rotation indicates that the (ir)relevant operators associated with the complex eigenvalue rotates in theory space, while the orbits move away from/towards the fixed point. Such behavour was early found for example in \cite{Weinrib,Hoogervorst}. For one pair of fixed points in the effective QCD model these complex scaling dimensions induced a clearly visible spiraling behavior in the RG-flow. This pair disappeared the the model extended with a scalar field and Yukawa coupling. It would be interesting to study these kind of exotic orbits in more detail, and maybe relate them to other exotic RG-flows that have been found, such as the bouncing solutions recently found in holographic studies \cite{ExoticRG}\\

Finally, we have shown that the use of bifurcation analysis in the study of RG-flows can be very useful tool in mapping out the fixed points and their transition in complicated model with multiple variables and parameters as was suggested in \cite{Gukov}. It would be interesting to use numerical tools such as the MATLAB package MatCont \cite{Matcont1,Matcont} in other studies of renormalization group flows. In particular for non-perturbative beta functions this can lead to new insights. An alternative to our numerical approach would be to use Conley index theory and describe the exit sets for RG-flows as was suggested in \cite{Gukov}.\\

\section*{Acknowledgements}

It is a pleasure to thank Sergei Gukov for helpful discussions and comments on the manuscript.  This work is partially supported by the Netherlands Organisation for Scientific Research (NWO)
under the VIDI grant 680-47-518 and the Delta-Institute for Theoretical Physics ($\Delta$-ITP), both funded by the Dutch Ministry of Education, Culture and Science (OCW).

\appendix

\section{Dynamical systems and bifurcation theory}\label{Ap:Bifurcations}
\subsection{Dynamical Systems}
This appendix presents a short overview of relevant terminology from the field of dynamical systems, partially following \cite{Brin}.  A \textit{dynamical system} is a tuple $(T,X, \Phi^t)$, where $X$ is a \textit{state space}, for \textit{time} $t \in T$, $\Phi^t:X\rightarrow X$ is a function defined on the state space, and typically $T=\mathbb{N}_{0}$ or $T=\mathbb{N}$ for \textit{discrete dynamical systems}, and $T=\mathbb{R}_{0}^{+}$ or $T=\mathbb{R}$ for \textit{continuous dynamical systems}.  The collection of all maps $\{\Phi^{t}\}_{t\in T}$ is called the \textit{flow} of the dynamical system and should satisfy
\begin{itemize}
	\item $\Phi^{0}(x) = x;$
	\item $\Phi^{s+t}(x) = \Phi^{s}(\Phi^{t}(x)), \, \forall x\in X,\, s,t\in T$.
\end{itemize}
Therefore, the flow forms a group, if $\Phi$ is invertible, and a semi-group otherwise. \\

For a point $x\in X$, one can define the \textit{orbit} $\mathcal{O}_{\Phi}(x) := \{\Phi^{t}(x)|t\in T)\}$. Furthermore, one can define the notion of an invariant set of a dynamical system.\\

\begin{mydef}
	Given a dynamical system $(T,X,\Phi^t)$,
	\begin{itemize}
	\item a \emph{forward invariant set} is a set $U \subset X$ such that $\Phi^{t}(U)\subseteq U$ $\forall t\geq0;$
	\item a \emph{backward invariant set} is a set $U \subset X$ such that $\Phi^{t}(U)\subseteq U$ $\forall t\leq0;$
	\item a set that is both forward and backward invariant is called an \emph{invariant set} of the dynamical system.
	\end{itemize}
\end{mydef}

If the map $\Phi^{t}$ is non-invertible and $T$ is only defined on positive numbers, every forward invariant set is called an invariant set of the dynamical system. It is assumed above that $\Phi^t(x)$ is defined for all $(t,x) \in T \times X$. If it is not the case, i.e., given $x \in X$, $\Phi^t(x)$ is only defined for a subset of $T$ that includes $t=0$, the dynamical system is called \textit{local}, and all definitions should be modified accordingly.\\

A natural way to generate a dynamical system is by a set of ordinary differential equations (continuous-time dynamical system) or a set of difference equations (discrete-time dynamical system). In this article, we consider dynamical systems generated by a set of differential equations, since those can be related to the beta functions of the renormalization group. Given the space $X=\mathbb{R}^{n}$, one can consider the system of autonomous ordinary differential equations
\begin{equation}\label{IntroBifAna:system}
\dot{x} = f(x),
\end{equation}
where $x\in\mathbb{R}^{n}$ and $f:\mathbb{R}^{n}\rightarrow\mathbb{R}^{n}$, which is assumed to be (sufficiently) smooth. If one defines the map $\Phi^{t}:\mathbb{R}^{n}\rightarrow\mathbb{R}^{n}$ by $\Phi^{t}(x_{0}) = x(t,x_{0})$, then the tuple $(\mathbb{R},\mathbb{R}^{n}, \Phi)$ is a (local) continuous-time dynamical system. The solutions for specified initial conditions, $x(0,x_{0})=x_{0}$, define the orbits and the visualization of the flow is called the \textit{phase portrait}. In the study of dynamical systems equilibria\footnote{In this article, we often use the term fixed point instead of equilibrium to adapt to the terminology used in renormalization group theory, while in the terminology of dynamical systems the term fixed point is reserved for the discrete analogue of an equilibrium.} and periodic, homoclinic, and heteroclinic orbits are of particular interest.\\

\begin{mydef}
	Given a continuous-time dynamical system $(\mathbb{R},\mathbb{R}^{n}, \Phi^t)$, then a point $x\in\mathbb{R}^{n}$ is called an \emph{equilibrium} if $\Phi^{t}(x)=x \; \forall t\in\mathbb{R}$.\\
\end{mydef}

\begin{mydef}
	Given a continuous-time dynamical system $(\mathbb{R},\mathbb{R}^{n}, \Phi^t)$, a point $x\in\mathbb{R}^{n}$, and an orbit $\mathcal{O}_{\Phi}(x)$, then the orbit is called
	\begin{itemize}
    	\item a \emph{periodic orbit} if $x$ is not an equilibrium and $\Phi^{T}(y)=y \;\forall y\in\mathcal{O}_{\Phi}(x)$ for some $T>0;$ minimal such $T$ is called the \emph{period}$;$
		\item a \emph{homoclinic orbit} if $\lim_{t\rightarrow-\infty} \Phi^{t}(x) = \lim_{t\rightarrow\infty} \Phi^{t}(x) = y$ and $\Phi^{t}(x) \neq y \; \forall  t\in\mathbb{R}$, where $y$ is an equilibrium$;$
		\item a \emph{heteroclinic orbit} if $\lim_{t\rightarrow-\infty} \Phi^{t}(x) = y \neq z = \lim_{t\rightarrow\infty} \Phi^{t}(x)$, where $y,z$ are equilibria.
	\end{itemize}
\end{mydef}

The topology of the phase portrait is characterized by its invariant sets and their stability. An invariant set is asymptotically stable, if all orbits starting in a neighborhood of the invariant set stay in this neighborhood and converge towards the invariant set if $t\rightarrow\infty$ and unstable otherwise. For generic equilibria the stability is determined by the eigenvalues of Jacobian matrix at the equilibrium. At the equilibrium, one can define the notions of a stable, unstable and critical eigenspace.\\

\begin{mydef}
	Given a continuous-time dynamical system defined by equation $(\ref{IntroBifAna:system})$, with an equilibrium point at $x_{0}\in\mathbb{R}^{n}$. Let $A$ be the Jacobian matrix of $f$ at $x_{0}$, $(\lambda_1,\lambda_{2},...,\lambda_{n})$ its eigenvalues and $\eta_1,\eta_{2},...,\eta_{n}$ the corresponding  $($generalized$)$ eigenvectors. Then 
	\begin{itemize}
	\item the \emph{stable eigenspace} is given by $T^{s} = \mathrm{span}\{\eta_{1}^{-},\eta_{2}^{-},...,\eta_{n_{-}}^{-}\};$
	\item the \emph{unstable eigenspace} by $T^{u} = \mathrm{span}\{\eta_{1}^{+},\eta_{2}^{+},...,\eta_{n_{+}}^{+}\}$, and 
	\item the \emph{critical eigenspace} by $T^{c} = \mathrm{span}\{\eta_{1}^{c},\eta_{2}^{c},...,\eta_{n_{c}}^{c}\}$,
	\end{itemize}
where $\eta_{i}^{+}$ label $($generalized$)$ eigenvectors that correspond to eigenvalues $\lambda_{i}^{+}$ with positive real part, $\eta_{i}^{-}$ $($generalized$)$ eigenvectors corresponding to eigenvalues $\lambda_{i}^{-}$ with negative real part and $\eta_{i}^{c}$ $($generalized$)$ eigenvectors corresponding to eigenvalues $\lambda_{i}^{c}$ with zero real part. Furthermore, $n_{+}+n_{-}+n_{c}=n$.\\
\end{mydef}

An equilibrium is called \textit{hyperbolic} if $\dim(T^{c})=0$, \textit{stable} if $\dim(T^{s})=n$ and \textit{unstable} if $\dim(T^{u})\geq1$. Furthermore, an equilibrium with $\dim(T^{s})=n$ or $\dim(T^{u})=n$ is a \textit{node} and an equilibrium with $\dim(T^{s})\geq1$ and $\dim(T^{u})\geq1$ is a \textit{saddle}. The classification of non-hyperbolic equilibria not satisfying these conditions is more involved, and won't be discussed further in this article. For a generic equilibrium in a continuous-time dynamical system, one can also define the notions of a stable and unstable manifold of the equilibrium.\\

\begin{mydef}
	Given a continuous-time dynamical system $(\mathbb{R},X,\Phi^t)$, and an equilibrium point $x\in X$ of $\Phi$, then 
	\begin{itemize}
	\item the \emph{stable manifold} of $x$ is defined as
	$W^{s}(\Phi,x) = \{y\in X : \lim_{t\rightarrow\infty}\Phi^{t}(y) = x\}$, and
	\item the \emph{unstable manifold} as
	$W^{u}(\Phi,x) = \{y\in X : \lim_{t\rightarrow-\infty}\Phi^{t}(y) = x\}$.
	\end{itemize}
\end{mydef}

For a dynamical system defined by (\ref{IntroBifAna:system}), the stable manifold is at the equilibrium tangent to the stable eigenspace, while the unstable manifold is at the equilibrium tangent to the unstable eigenspace. For nonhyperbolic equilibria, one can introduce the concept of a center manifold $W_{\textrm{loc}}^{c}(\Phi,x)$, which is $n_c$-dimensional and tangent to the critical eigenspace at the equilibrium. The existence of this locally defined manifold is implied by the Center Manifold Theorem.\\

\subsection{Bifurcation Analysis}
A continuous-time dynamical system defined by a set of ordinary differential equations can be analyzed by solving the ODE's numerically. However, if the differential equations also depend on a set of parameters, $f_{\alpha}(x)$, $\alpha\in\mathbb{R}^m$, one is often not interested in the exact solutions for specific values of $\alpha$, but in the topological properties of the phase portrait such as the number of equilibria, limit cycles and their stability or the existence of homo- and heteroclinic orbits and chaotic attractors. In general, changes of the phase portrait, when $\alpha$ is varied, occur smoothly except for some specific values of $\alpha$, where the topology of the phase portrait changes. Changes in the topology of the phase portrait are called \textit{bifurcations}. Topology changes often occur near or within invariant sets. Bifurcations can be divided in 2 subclasses:

\begin{itemize}
	\item local bifurcations: bifurcations that can be detected by considering small but non-shrinking neighborhoods of equilibria or limit cycles. 
	\item global bifurcations: bifurcations that cannot be detected in this way. Topology changes typically involve multiple invariant sets.
\end{itemize}

In addition, bifurcations may be called subcritical or supercritical, depending on the stability of the equilibria or limit cycles that appear or disappear at the bifurcation. Furthermore, bifurcations can be characterized by their codimension. The \textit{codimension} of a bifurcation in a generic system is given by the number of independent conditions that have to be satisfied for a bifurcation to happen, and for local bifurcations is often related to the number of critical eigenvalues of the Jacobian matrix at the equilibrium. Consequently, the codimension of a bifurcation is always smaller or equal to the dimension of the parameter space. In general, many types of bifurcations can be found. For an overview of possible bifurcations and their conditions, we refer to \cite{Kuznetsov}. In this appendix, we discuss bifurcations and related terminology that is relevant for this article. In doing so, we closely follow \cite{Kuznetsov}. Bifurcations in low dimensional systems and with low codimension can often be found using analytical techniques, but for higher dimensional systems and in particular for higher codimension of the bifurcation one often has to rely on numerical techniques. Several computer programs have been developed for bifurcation analysis. In this article we make use of the Matlab package Matcont \cite{Matcont1,Matcont}. Bifurcation software relies on the conditions that have to be satisfied in order for a bifurcation to happen.\\

In different dynamical systems one can expect different bifurcations, which are not immediately defined by the same conditions. However, one would like to have a way of categorizing bifurcation in various systems. For this, we need the notion of \textit{topological equivalence}. Consider two continuous-time dynamical systems:
\begin{align}
\dot{x} &= f(x,\alpha), \quad x\in\mathbb{R}^{n}, \; \alpha\in\mathbb{R}^{m}; \label{Intro:dynsys1} \\
\dot{y} &= g(y,\beta), \quad y\in\mathbb{R}^{n}, \; \beta\in\mathbb{R}^{m}. \label{Intro:dynsys2}
\end{align}

\begin{mydef}
	Dynamical systems $(\ref{Intro:dynsys1})$ and $(\ref{Intro:dynsys2})$ are \emph{topologically equivalent} if
	\begin{itemize}
	\item there exists a homeomorphism of the parameter space $p: \mathbb{R}^{m} \rightarrow \mathbb{R}^{m}$ s.t. $\beta=p(\alpha);$
	\item there exists a parameter-dependent homeomorphism of the phase space $h_{\alpha} : \mathbb{R}^{n} \rightarrow \mathbb{R}^{n}$ s.t. $y=h_{\alpha}(x)$, mapping orbits of the system $(\ref{Intro:dynsys1})$ at parameter values $\alpha$ onto orbits of the system $(\ref{Intro:dynsys2})$ at parameter values $\beta=p(\alpha)$, preserving the direction of time.
	\end{itemize}
\end{mydef}

Since many bifurcations are locally defined, one is often only interested in \textit{local topological equivalence} near specific points.\\

\begin{mydef}
	Dynamical systems $(\ref{Intro:dynsys1})$ and $(\ref{Intro:dynsys2})$ are \emph{locally topologically equivalent} near the origin, if there exists a map $(x,\alpha) \mapsto (h_{\alpha}(x),p(\alpha))$, defined in a small neighborhood of $(x,\alpha)=(0,0)\in\mathbb{R}^{n}\times\mathbb{R}^{m}$ such that
	\begin{itemize}
	\item $p: \mathbb{R}^{m} \rightarrow \mathbb{R}^{m}$ is a homeomorphism defined in a small neighbourhood of $\alpha=0$, $p(0)=0;$
	\item $h_{\alpha} : \mathbb{R}^{n} \rightarrow \mathbb{R}^{n}$ is a parameter-dependent homeomorphism defined in a small neighborhood $U$ of $x=0$, $h_{0}(0)=0$, and mapping orbits of the system $(\ref{Intro:dynsys1})$ in $U$ onto orbits of the system $(\ref{Intro:dynsys2})$ in $h_{\alpha}(U)$, preserving the direction of time.
	\end{itemize}
\end{mydef}

By coordinate translations one can easily generalize this definition to local topological equivalence near arbitrary points. Having these definitions, one can categorize bifurcations by topological equivalence to \textit{normal forms}.
Consider a polynomial continuous-time dynamical system
\begin{equation}\label{Intro:dynsys3}
\dot{z} = g(z,\beta;\sigma), \quad z\in \mathbb{R}^{n}, \, \beta\in \mathbb{R}^{k}, \, \sigma\in \mathbb{R}^{l},
\end{equation} 
which has an equilibrium $z=0$ at $\beta=0$, satisfying $k$ bifurcation conditions. Here, $\sigma$ is a vector of the coefficients of the polynomial $g(z,\beta;\sigma)$.\\

\begin{mydef}
	System $(\ref{Intro:dynsys3})$ is a \emph{topological normal form} for the bifurcation if any generic system $(\ref{Intro:dynsys1})$ with equilibrium $x_0$ satisfying the bifurcation conditions $\alpha_0$ is locally topologically equivalent near $(x_0,\alpha_0)$ to $(\ref{Intro:dynsys3})$ for some values of the coefficients $\sigma_i$.\\
\end{mydef}

We notice that the system (\ref{Intro:dynsys1}) can have higher dimension than is needed for the bifurcation to occur: $x\in\mathbb{R}^{n}$, while the normal form has $z^c\in\mathbb{R}^{n_c}$ with $n_c<n$. In such a case, there is a continuation of the critical center manifold. For nearby parameter values (\ref{Intro:dynsys1}) is locally topologically equivalent to
\begin{align}
\dot{z}^{c} &= g(z^c,\beta;\sigma),\label{CMR}\\
\dot{z}^{s} &= -z^{s},\\
\dot{z}^{u} &= +z^{u},
\end{align} 
where $g(z^c,\beta;\sigma)$ is a normal form on the center manifold and $z^s\in\mathbb{R}^{n_s}, z^u\in\mathbb{R}^{n_u}$ correspond to the stable and unstable eigenspaces.\\

We now give a short discussion of local bifurcations which are encountered throughout the article. We begin with bifurcations in scalar one-parameter ODEs
\begin{equation}
\label{1DIM}
\dot{x}=f(x,\alpha),\quad x \in \mathbb{R}, \alpha \in \mathbb{R}.
\end{equation}
In the $n$-dimensional case, each normal form below is the normal form (\ref{CMR})  on the one-dimensional center manifold.\\

\par
\noindent
\textbf{Saddle-Node Bifurcation}
\par
\begin{figure}[htbp]
\begin{center}
\includegraphics[width=2.0in]{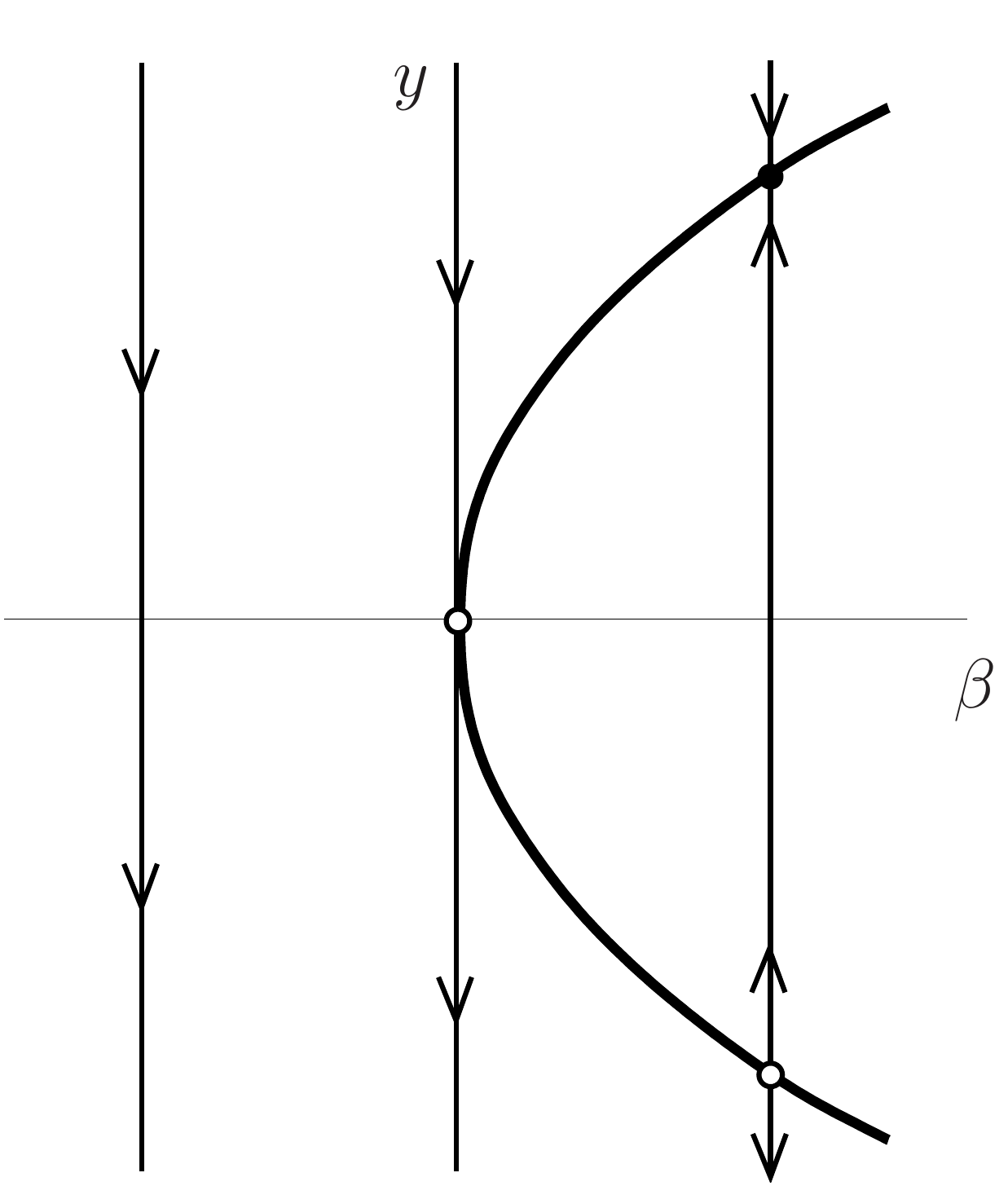}
\end{center}
\caption{Saddle-Node bifurcation in the system $\dot{y} = \beta - y^2$.}
\label{Fig:Bif_SaddleNode}
\end{figure}

This bifurcation is a codimension one bifurcation at which two equilibria collide and disappear, and is also known as \textit{fold} or \textit{limit point} bifurcation. The bifurcation occurs at $(\bar{x},\bar{\alpha})$, if the following conditions hold for the system (\ref{1DIM}):
\begin{itemize}
	\item $f(\bar{x},\bar{\alpha}) = 0,$
	\item $f_{x}(\bar{x},\bar{\alpha}) = 0$,
	\item $a = \frac{1}{2} f_{xx}(\bar{x},\bar{\alpha}) \neq 0$,
	\item $f_{\alpha}(\bar{x},\bar{\alpha}) \neq 0$,
\end{itemize}
where $a$ is the quadratic coefficient of the system (\ref{1DIM}). The system is then locally topologically equivalent near $(\bar{x},\bar{\alpha})$ to the normal form 
\begin{equation}\label{NF_SNBif}
\dot{y} = \beta \pm y^{2}, \quad y,\beta \in\mathbb{R}
\end{equation}
near $(y,\beta)=(0,0)$  (see Figure \ref{Fig:Bif_SaddleNode}).\\
\newpage
\par
\noindent
\textbf{Transcritical Bifurcation}
\par
\begin{figure}[htbp]
\begin{center}
\includegraphics[width=2.0in]{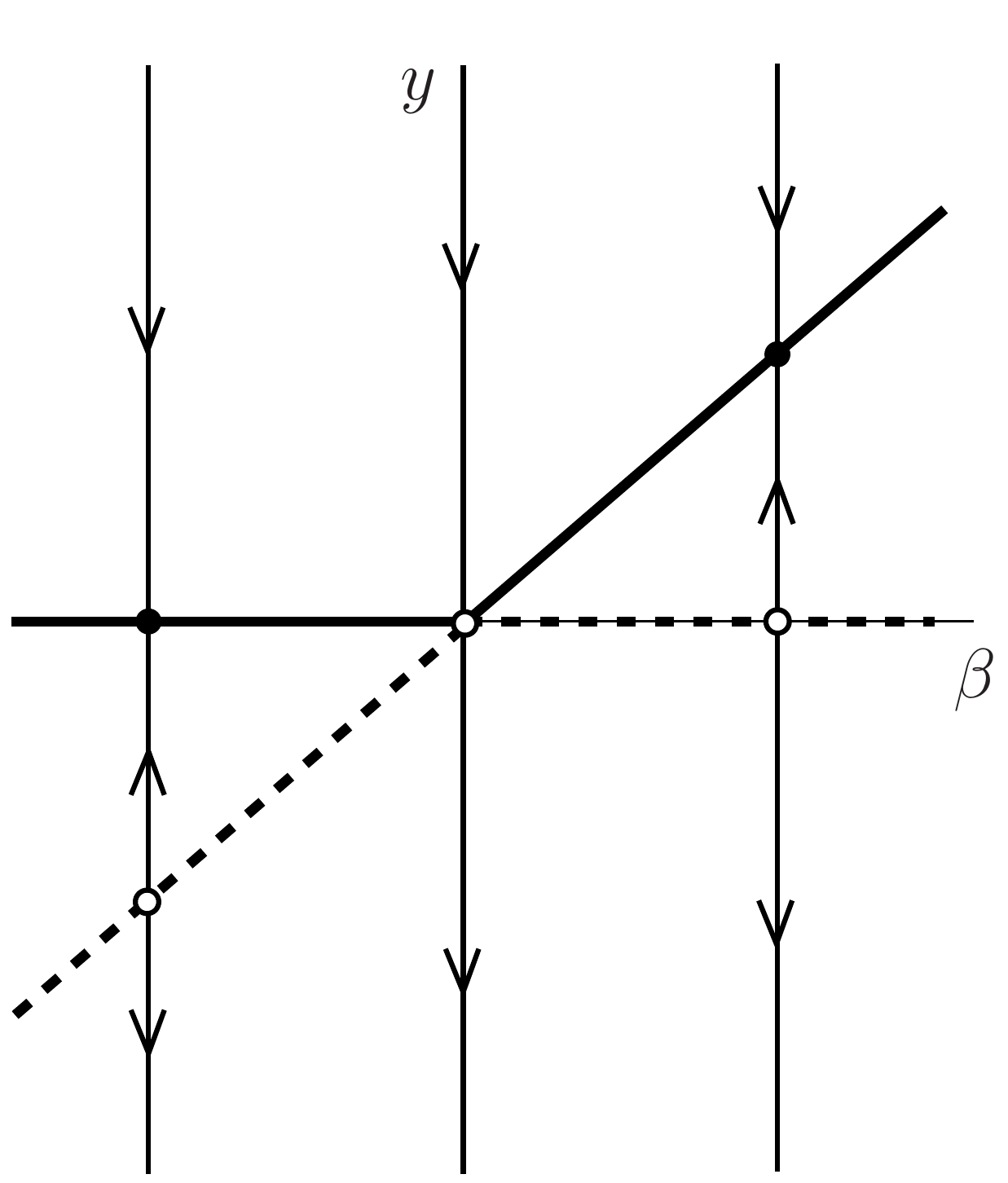}
\end{center}
\caption{Transcritical bifurcation in the system $\dot{y} = \beta y - y^{2}$.}
\label{Fig:Bif_Transcrit}
\end{figure}

This bifurcation  is a codimension one bifurcation in the class of systems that always have a trivial equilibrium, and is also a branching point: At the bifurcation, two equilibria collide and exchange their stability. The bifurcation occurs at $(\bar{x},\bar{\alpha})$, if the following conditions hold for the system (\ref{1DIM}):
\begin{itemize}
	\item $f(\bar{x},\bar{\alpha}) = 0$,
	\item $f_{x}(\bar{x},\bar{\alpha}) = f_{\alpha}(\bar{x},\bar{\alpha}) = 0$,
	\item $f_{xx}(\bar{x},\bar{\alpha}) \neq 0$, 
	\item $f_{x \alpha}(\bar{x},\bar{\alpha}) \neq 0$.
\end{itemize}
The system is then locally topologically equivalent near $(\bar{x},\bar{\alpha})$ to the normal form
\begin{equation}\label{NF_TCBif}
\dot{y} = \beta y \pm y^{2}, \quad y,\beta \in\mathbb{R}
\end{equation}
near $(y,\beta)=(0,0)$ (see Figure \ref{Fig:Bif_Transcrit}). 
Notice that we can transform the normal form of the tanscritical bifurcation to the normal form of the saddle-node bifurcation by the non-invertible coordinate transformation 
\begin{equation*}
(y,\beta) \mapsto  \left(y\pm \frac{\beta}{2}, \mp \frac{\beta^{2}}{4}\right).
\end{equation*}
This is called a nonversal unfolding of the saddle-node bifurcation.\\
\par
\noindent
\textbf{Pitchfork Bifurcation}
\par
\begin{figure}[htbp]
\begin{center}
\includegraphics[width=2.0in]{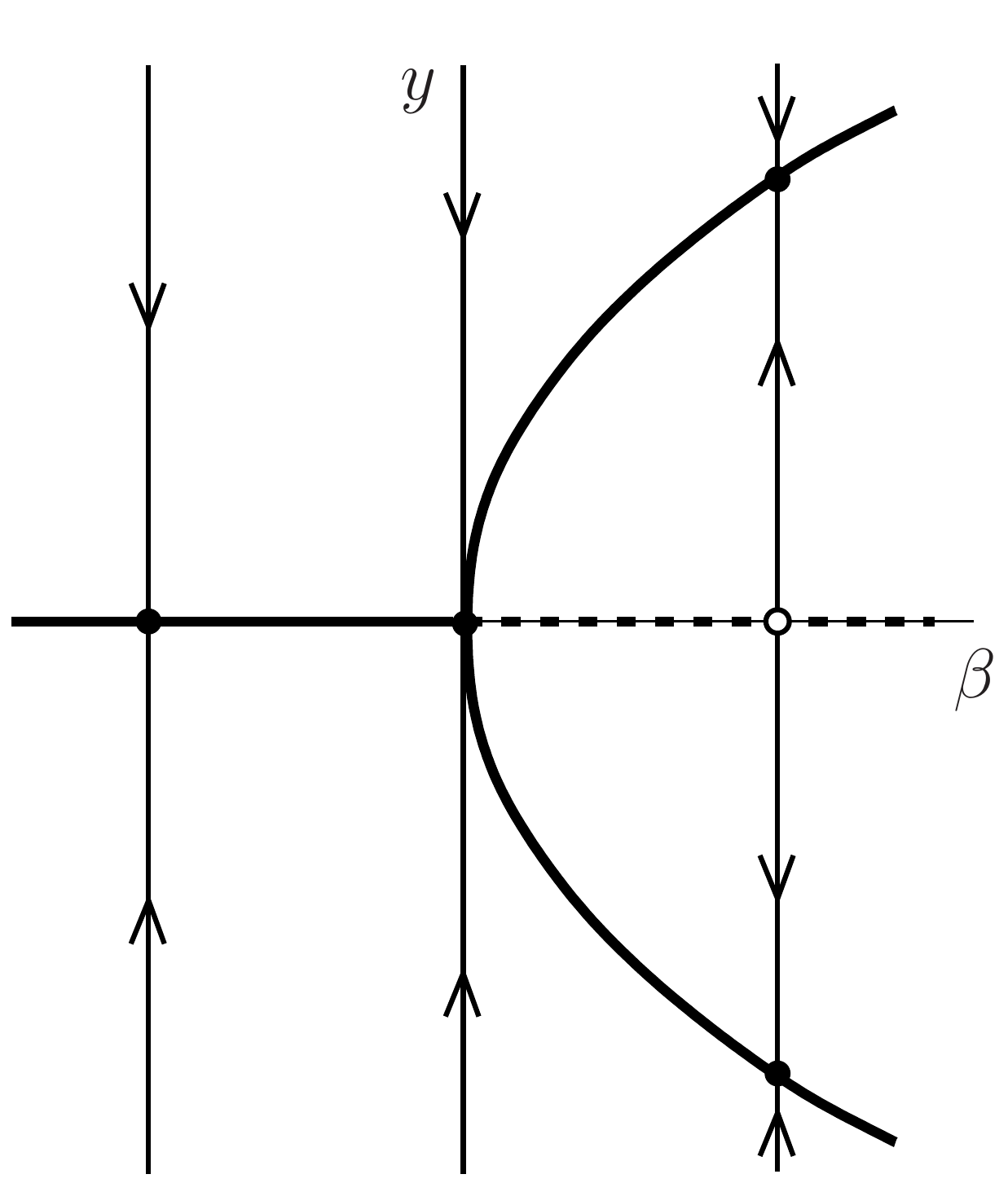}
\end{center}
\caption{Supercritical pitchfork bifurcation in the system $\dot{y} = \beta y - y^{3}$.}
\label{Fig:Bif_Pitchfork}
\end{figure}

This bifurcation  is a codimension one bifurcation for systems with reflectional symmetry, and is also a branching point: At the bifurcation one equilibrium splits into three equilibria. The bifurcation occurs at $(\bar{x},\bar{\alpha})$, if the following conditions hold for the system (\ref{1DIM}):
\begin{itemize}
	\item $f(\bar{x},\bar{\alpha}) = 0$
	\item $f_{x}(\bar{x},\bar{\alpha}) = f_{xx}(\bar{x},\bar{\alpha}) = f_{\alpha}(\bar{x},\bar{\alpha}) = 0$,
	\item $f_{xxx}(\bar{x},\bar{\alpha})\neq0$, 
	\item $f_{x\alpha}(\bar{x},\bar{\alpha}) \neq0$.
\end{itemize}
The system is then locally topologically equivalent near $(\bar{x},\bar{\alpha})$ to the normal form 
\begin{equation}
\dot{y} = \beta y \pm {y}^{3}, \quad y,\beta \in\mathbb{R}
\end{equation}
near $(y,\beta)=(0,0)$ (see Figure \ref{Fig:Bif_Pitchfork}). The $-$ sign gives the supercritical case where a stable equilibrium splits into two stable equilibria and an unstable equilibrium, and the $+$ sign is the subcritical case, where an unstable equilibrium splits into two unstable equilibria and a stable equilibrium.\\

We continue with two bifurcations in scalar two-parameter ODEs
\begin{equation}
\label{1DIM2}
\dot{x}=f(x,\alpha),\quad x \in \mathbb{R}, \ \alpha =(\alpha_1,\alpha_2) \in \mathbb{R}^2.
\end{equation}
In the $n$-dimensional case, each normal form below is the normal form (\ref{CMR})  on the corresponding one-dimensional center manifold.\\

\par
\noindent
\textbf{Cusp Bifurcation}
\par
\begin{figure}[htbp]
\begin{center}
\includegraphics[width=3.5in]{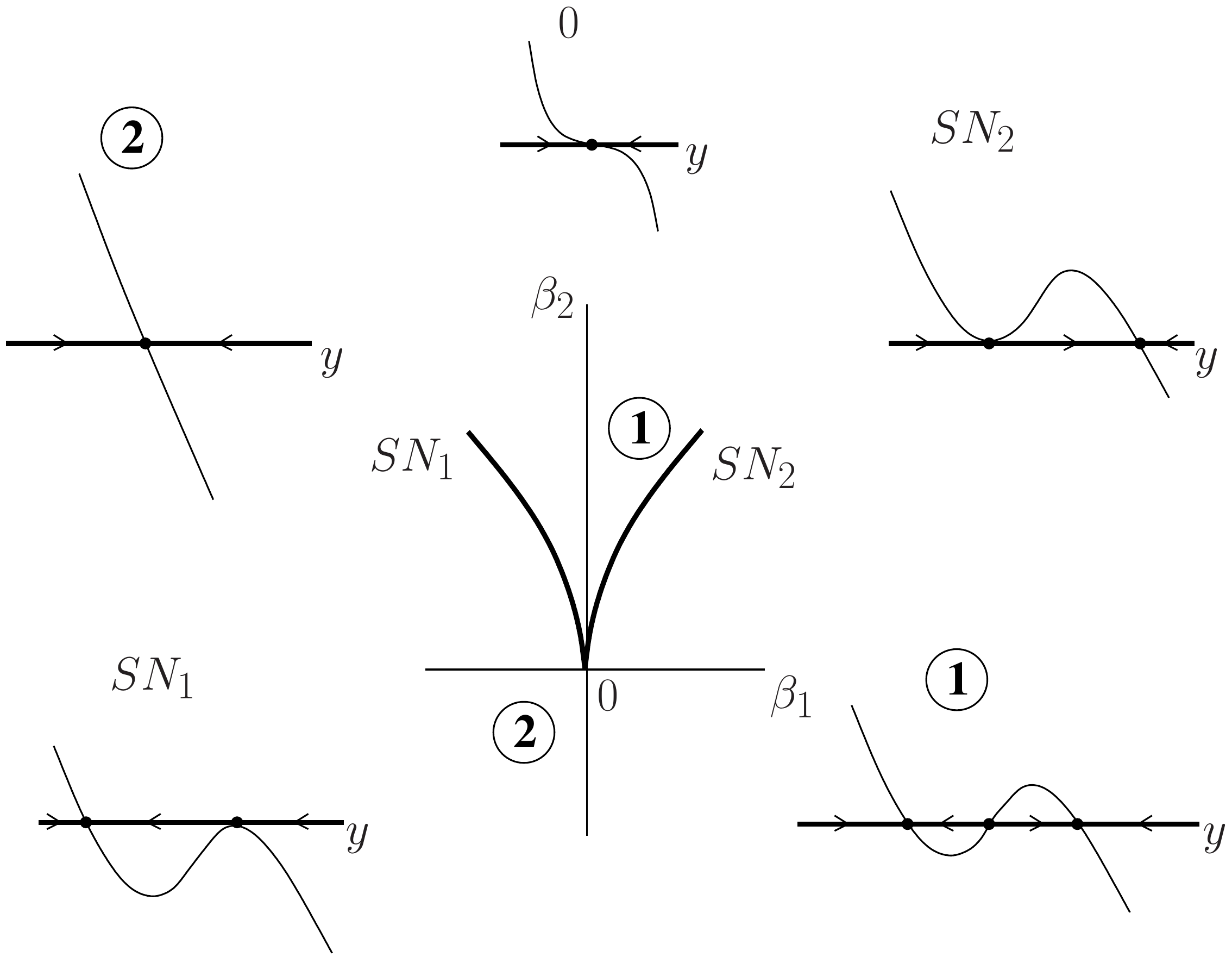}
\end{center}
\caption{Cusp bifurcation in the system $\dot{y} = \beta_1 + \beta_2 y - y^3$.}
\label{Fig:Bif_Cusp}
\end{figure}

This bifurcation is a codimension two bifurcation in generic systems, where two branches of saddle-node bifurcation meet tangentially. Nearby, the system can have three equilibria which collide and disappear at the saddle-node bifurcations. The bifurcation occurs at $(\bar{x},\bar{\alpha})$, if the following conditions hold for the system (\ref{1DIM2}):
\begin{itemize}
	\item $f(\bar{x},\bar{\alpha}) = 0$,
	\item $f_{x}(\bar{x},\bar{\alpha}) = f_{xx}(\bar{x},\bar{\alpha}) = 0$,
	\item $c = \frac{1}{6} f_{xxx}(\bar{x},\bar{\alpha}) \neq 0$,
	\item the map $(x,\alpha)\mapsto(f(x,\alpha),f_x(x,\alpha),f_{xx}(x,\alpha))$ is regular at $(x,\alpha)=(\bar{x},\bar{\alpha})$,
\end{itemize}
where $c$ is the cubic coefficient of the system (\ref{1DIM2}). The system is then locally topologically equivalent near $(\bar{x},\bar{\alpha})$ to the normal form 
\begin{equation}
\dot{y} = \beta_1 + \beta_2 y \pm y^{3}, \quad y, \beta_, \beta_2 \in \mathbb{R}
\end{equation}
near $(y,\beta)=(0,0)$  (see Figure \ref{Fig:Bif_Cusp}).\\

\par
\noindent
\textbf{Saddle-Node-Transcritical Bifurcation}
\par
This bifurcation is a codimension two bifurcation in the class of systems that always have a trivial equilibrium, where two branches of a transcritical bifurcation curve meet on a saddle-node bifurcation curve. There exist two types of this bifurcation. One can be seen as a nonversal unfolding of a Cusp bifurcation, while the other can be obtained as an unfolding of a degenerate Bogdanov-Takens bifurcation. Here, we discuss only the first case. \\
\begin{figure}[htbp]
\begin{center}
\includegraphics[width=3.5in]{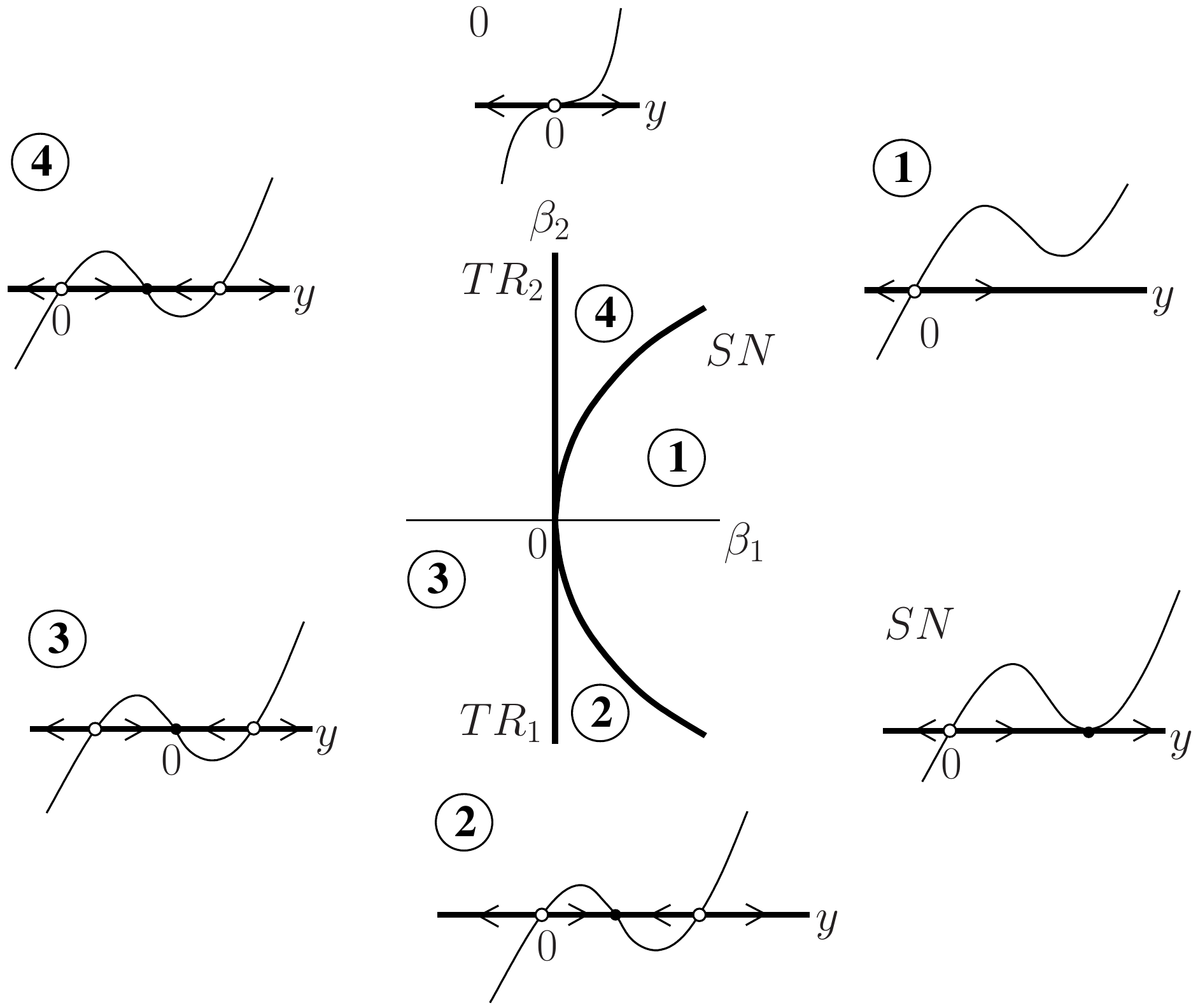}
\end{center}
\caption{Saddle-node-transcritical bifurcation in the system $\dot{y} = \beta_1 y + \beta_2 y^2 + y^3$.}
\label{Fig:Bif_SNT}
\end{figure}

Nearby, the system can have three equilibria: One pair has a transcritical bifurcation at one of the branches and another pair has a transcritical bifurcation at another branch. The last possible pair of equilibria collides and disappears at the saddle-node bifurcation curve. If the bifurcation occurs at $(\bar{x},\bar{\alpha})$, then the system (\ref{1DIM2}) is locally topologically equivalent near $(\bar{x},\bar{\alpha})$ to the normal form given by
\begin{equation}
\dot{y} = \beta_1 y + \beta_2 y^2 \pm y^{3}, \quad y,\beta_1,\beta_2\in\mathbb{R}
\end{equation}
near $(y,\beta)=(0,0)$  (see Figure \ref{Fig:Bif_SNT}). 
We then find a transcritical bifurcation along the line $\beta_1=0$, and a saddle-node bifurcation along $\beta_{1}=\pm\frac{\beta_{2}^{2}}{4}$. The normal form can be transformed to the normal form of the cusp bifurcation through the transformation
\begin{equation*}
(y,\beta_1,\beta_2) \mapsto \left(y\pm\frac{\beta_{2}}{3}, \frac{2\beta_{2}^3}{27} \mp \frac{\beta_{1} \beta_{2}}{3}, \beta_{1} \mp \frac{\beta_{2}^2}{3} \right).
\end{equation*}

Finally, we discuss for completeness the last codimension one local bifurcation in generic ODEs (\ref{Intro:dynsys1}), i.e. Andronov-Hopf bifurcation. Note that it does not occur in our RG-flows but might appear in other models. For this bifurcation, the center manifold is two-dimensional, so it is sufficient to consider (\ref{Intro:dynsys1}) with $n=2$ and $m=1$.\\
\newpage
\par
\noindent
\textbf{Andronov-Hopf Bifurcation}
\par
\begin{figure}[htbp]
\begin{center}
\includegraphics[width=4.5in]{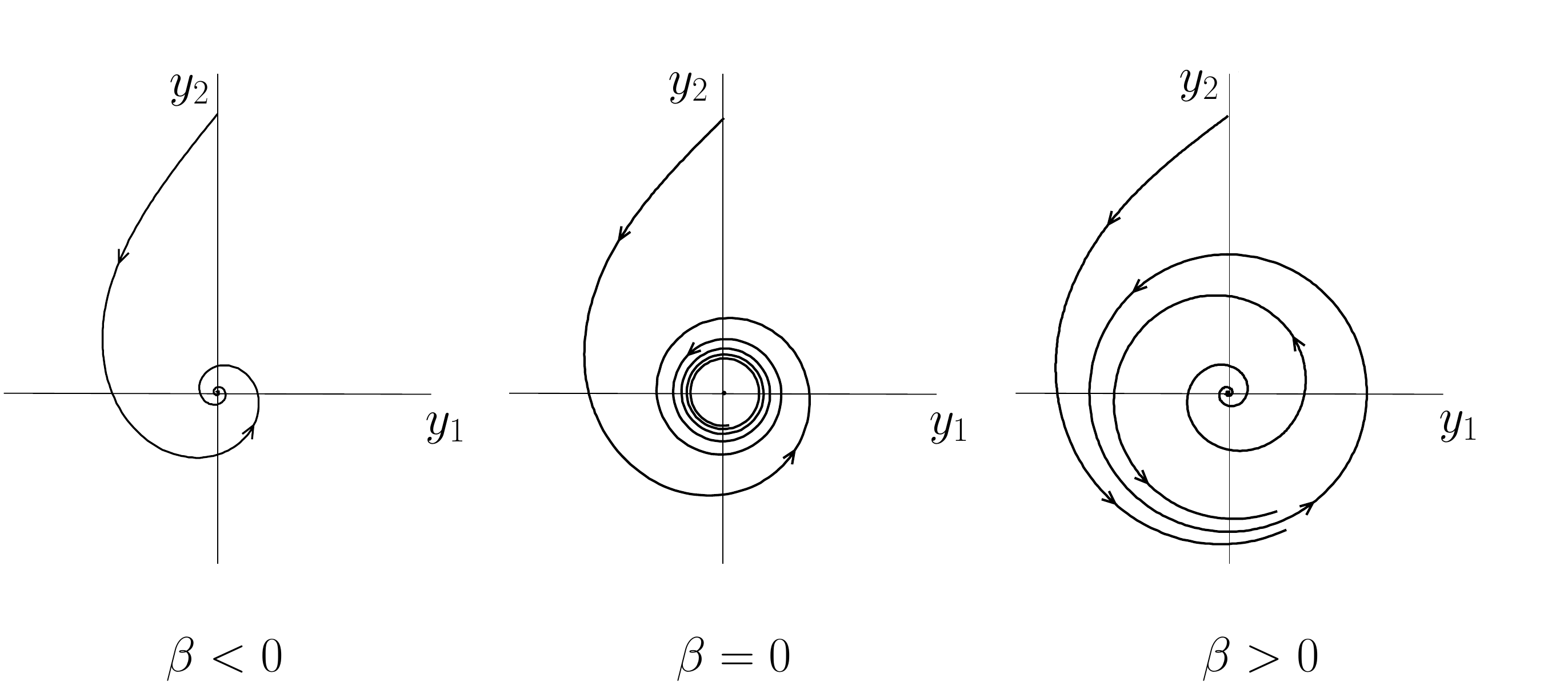}
\end{center}
\caption{Supercritical Andronov-Hopf bifurcation in the system 
$
\dot{y}_1 = \beta y_1 - y_2 - y_1 \left( y_{1}^{2} + y_{2}^{2} \right), \ 
\dot{y}_2 = y_1 + \beta y_2 - y_2 \left( y_{1}^{2} + y_{2}^{2} \right).
$
}
\label{Fig:Bif_Hopf}
\end{figure}

This bifurcation is a codimension one bifurcation, and implies the birth of a periodic orbit (\textit{limit cycle}). The bifurcation occurs at an equilibrium $(\bar{x},\bar{\alpha})$, if a pair of two conjugate complex eigenvalues of the Jacobian matrix crosses the imaginary axis. Generically, at the bifurcation the system is locally topologically equivalent near $(\bar{x},\bar{\alpha})$ to the normal form 
\begin{eqnarray*}
\dot{y}_1 &=& \beta y_1 - y_2 \pm y_1 \left( y_{1}^{2} + y_{2}^{2} \right), \\
\dot{y}_2 &=& y_1 + \beta y_2 \pm y_2 \left( y_{1}^{2} + y_{2}^{2} \right),
\end{eqnarray*}
near $(y_1,y_2,\beta)=(0,0,0)$. The plus sign corresponds to the subcritical case, where an unstable limit cycles is created and the minus sign to the supercritical case, where a stable limit cycle is created  (see Figure \ref{Fig:Bif_Hopf}).\\

\section{Beta functions in QCD$_4$}\label{Ap:QCD5}

\subsection{Feynman Rules for QCD$_4$ with a Four-Fermi Interaction}\label{Ap:FRules}
The theory as discussed in section 2 has three propagators:\footnote{All Feynman diagrams are generated with the Tikz-Feynman package \cite{diagrams}.}
\begin{figure}[H]
	\begin{center}
		\includegraphics[width=\linewidth]{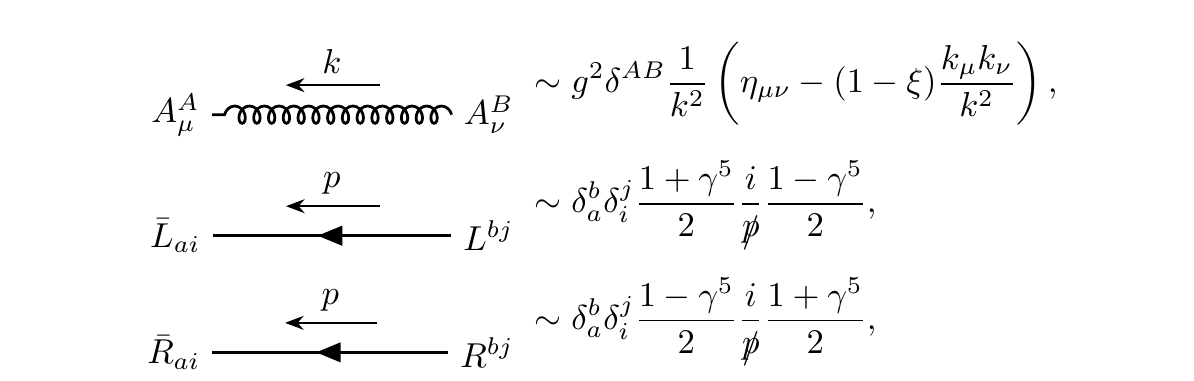}
	\end{center}
\end{figure}
\noindent where $\mu,\nu$ label the space-time indices, $A,B$ label the colors in the adjoint representation of $SU(N_c)$, $a,b$ label the color in the fundamental representation of $SU(N_c)$ and $\{i,j\}$ label the flavor in the fundamental representation of $SU(N_f)$. Furthermore, we take the Landau gauge in which the gauge parameter $\xi =0$.\\

The fermion-gluon interaction (gauge vertex) is given by:
\begin{figure}[H]
	\begin{center}
		\includegraphics[width=\linewidth]{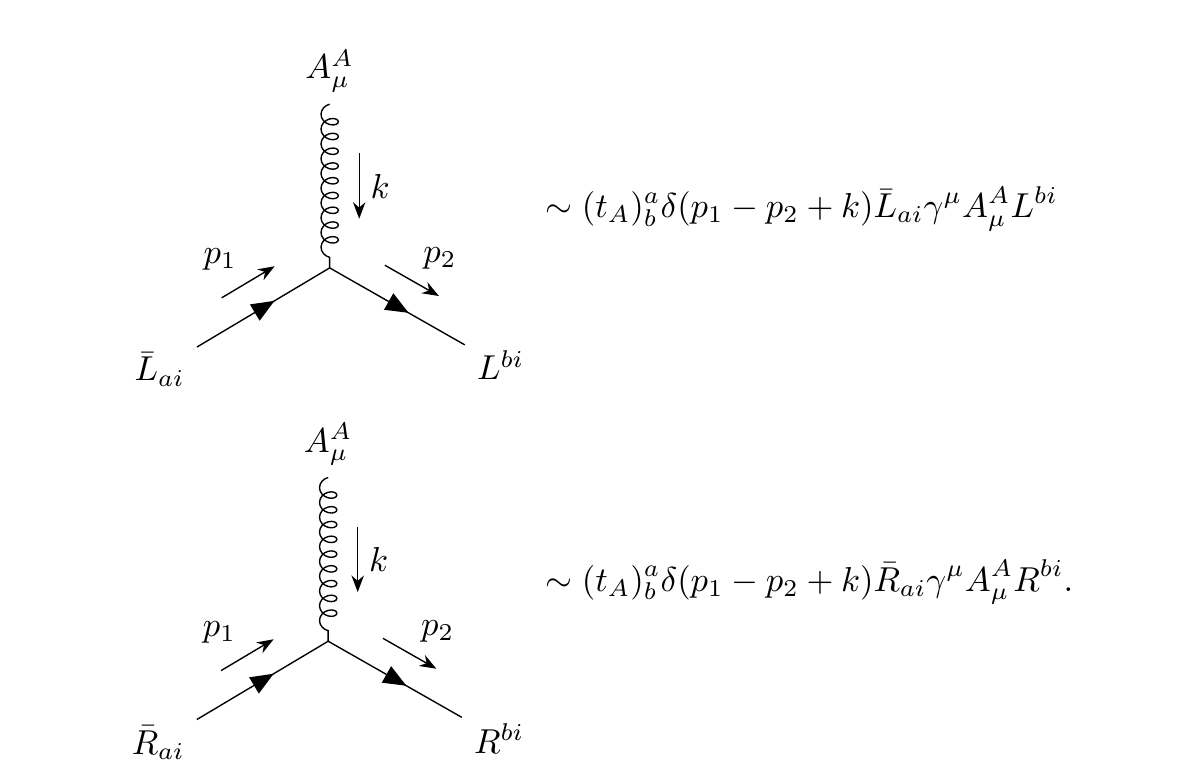}
	\end{center}
\end{figure}
\clearpage
Furthermore, there are four four-fermi interactions:
\begin{figure}[H]
	\begin{center}
		\includegraphics[width=\linewidth]{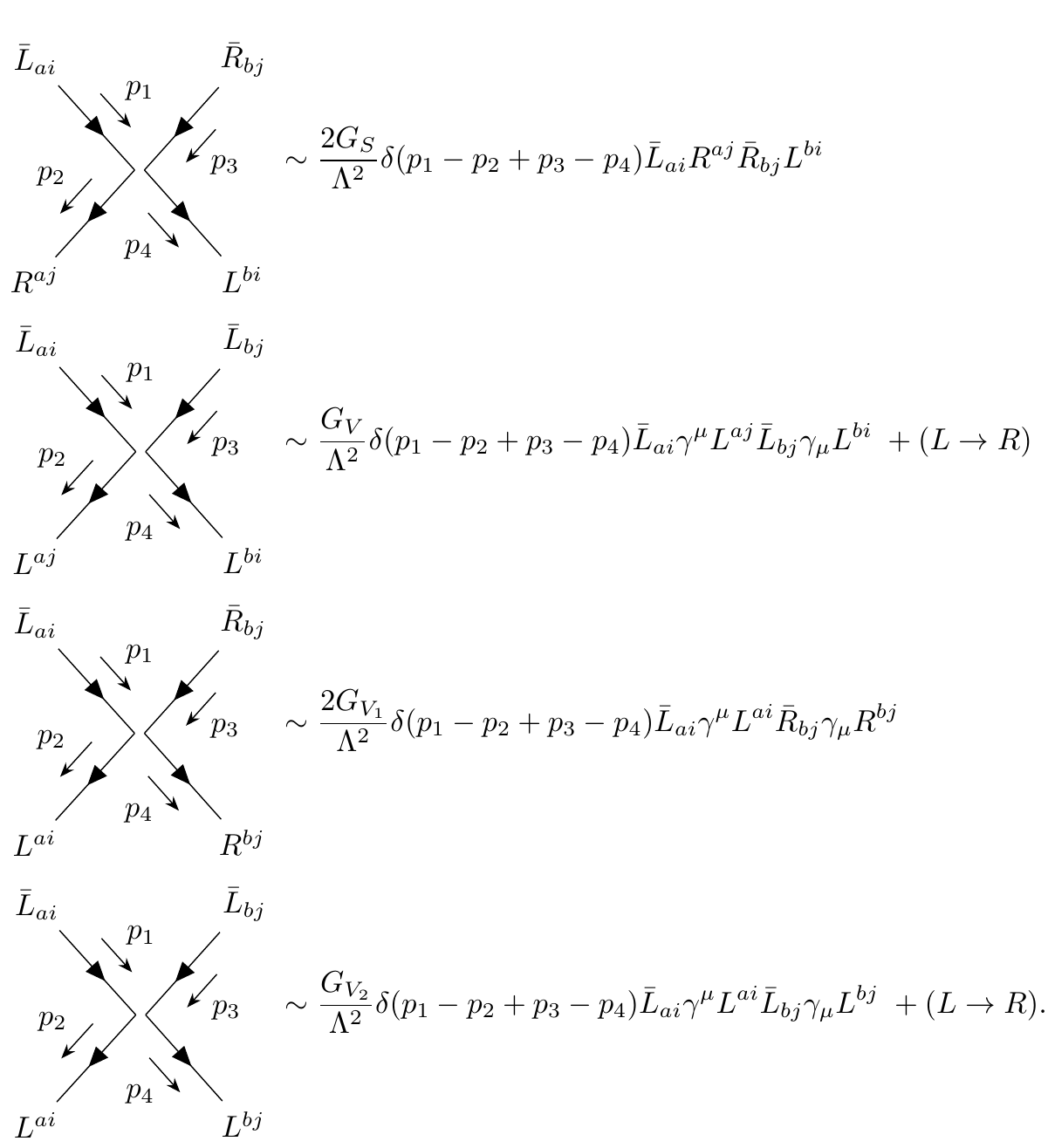}
	\end{center}
\end{figure}
Notice that the vertices have been split up in two parts. This is done to make clear along which lines the color charge is conserved, and will make the calculations in appendix \ref{Ap:FourFermi} more tractable, but has no further physical meaning.

\subsection{Beta Function for the Gauge Vertex}\label{Ap:GaugeVertex}
In this appendix, we reproduce the method used in \cite{Kusafuka} to calculate the beta function for the gauge vertex. We're only interested in gauge invariant contributions to the beta function, but the Wetterich equation does not necessarily respect gauge invariance. Therefore, we do not use the exact renormalization approach. Instead we start from the well known 2-loop perturbative beta function for QCD$_4$ with $N_c$ colors and $N_f$ massless flavors \cite{Caswell}:
\begin{equation}
\beta_{g}^{[2]} \equiv \Lambda \frac{d \alpha_{g}}{d \Lambda} = -2 b_0 \alpha_{g}^{2} - 2 b_1 \alpha_{g}^{3}
\end{equation}
with
\begin{align*}
b_0 &= \frac{11}{3} N_c - \frac{2}{3} N_f\\
b_1 &= \frac{34}{3} N_{c}^{2} - N_f \left( \frac{N_{c}^{2} -1}{N_c} + \frac{10}{3} N_c \right).
\end{align*}

Next, we would like to include effective interactions induced by the four-fermi couplings. These can be added perturbatively to the beta functions. The beta function is then given by
\begin{equation}
\Lambda \frac{d \alpha_g}{d\Lambda} = \beta_{g}^{[2]} + \Delta \alpha_{g},
\end{equation}
where the last term is represents the perturbative corrections induced by the effective couplings. These perturbations should be chosen such that they're gauge invariant and such that they represent perturbations, that are included in $\lim_{n\rightarrow\infty}\beta_{g}^{[n]}$, but not in $\beta_{g}^{[2]}$. There is one one-loop correction induced by the effective four-fermi interaction, which is represented by
\begin{figure}[H]
	\begin{center}
		\includegraphics[width=\linewidth]{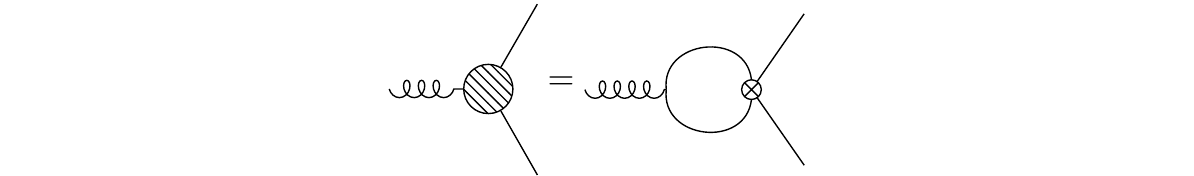}
	\end{center}
\end{figure}

However, this correction turns out to be gauge dependent \cite{GiesII}, and therefore we discard it. A possible two-loop correction is given by
\begin{figure}[H]
	\begin{center}
		\includegraphics[width=\linewidth]{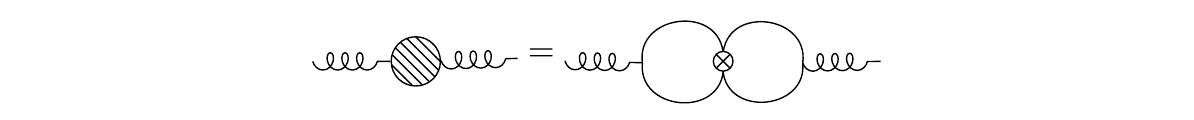}
	\end{center}
\end{figure}

For the effective four-fermi interaction one could take $\mathcal{O}_i$, $i\in\{S,V,V_1,V_2\}$. However, since the interactions should represent perturbations, which are included in $\lim_{n\rightarrow\infty}\beta_{g}^{[n]}$ the chirality should be the same in the whole diagram. This excludes the interactions $\mathcal{O}_S$ and $\mathcal{O}_{V_1}$. Furthermore, due to the tracelessness of the generators of $SU(N)$ the contribution due to $\mathcal{O}_{V_2}$ vanishes. Therefore, only $\mathcal{O}_{V}$ contributes, and its contribution is in the sharp cutoff limit given by
\begin{equation}
- \frac{N_c N_f g^4 G_V}{32 \pi^4} \delta\ln(\Lambda),
\end{equation}
yielding a contribution to the beta function of the form
\begin{equation}
\delta \alpha_g = 2 N_c N_f \alpha_{g}^{2} g_V \delta\ln(\Lambda).
\end{equation}

\subsection{Beta Functions for the Four-Fermi Interactions}\label{Ap:FourFermi}
We can find the beta functions of the model using the equation \cite{ERGEreview}
\begin{equation*}
	\frac{d}{dt}\Gamma_{\Lambda}[\phi] = - \Lambda \frac{d}{d\Lambda}\Gamma_{\Lambda}[\phi] = \sum_{n=1}^{\infty} \beta_{g_{n}}(t) \mathcal{O}_{n}(\phi)
\end{equation*}
by noticing that
\begin{equation*}
	\frac{d}{d\Lambda}\Gamma_{\Lambda}[\phi] = 
	\lim_{\delta\Lambda\rightarrow 0} \frac{\Gamma_{\Lambda+\delta\Lambda}[\phi] -\Gamma_{\Lambda}[\phi]}{\delta \Lambda} =:
	\lim_{\delta\Lambda\rightarrow 0} \frac{\delta\Gamma_{\Lambda}[\phi]}{\delta \Lambda}.
\end{equation*}
Furthermore,
\begin{equation*}
	\delta\Gamma_{\Lambda} = \int d^4 p \delta \mathcal{L}_{\textrm{QCD}} + \delta \mathcal{L}_{\textrm{4f}}
\end{equation*}
with
\begin{equation*}
	\delta \mathcal{L}_{\textrm{4f}} = \sum_{i\in\{S,V,V_1,V_2\}} \left( \frac{\delta G_{i}(\Lambda)}{\Lambda^{2(1+\eta)}} - \frac{2G_i(\Lambda)\delta\Lambda}{\Lambda^{3+2\eta}}\right)\mathcal{O}_{i},
\end{equation*}
whence
\begin{equation*}
	\Lambda \frac{d G_i}{d\Lambda} = \lim_{\delta\Lambda\rightarrow 0} 2(1+\eta) G_i + \left. \Lambda^{3 + 2\eta} \frac{\delta \Gamma _{\Lambda}}{\delta{\Lambda}}\right|_{\textrm{terms }\propto \mathcal{O}_{i}}.
\end{equation*}

For this Lagrangian the anomalous dimension is $\eta=0$ \cite{Kusafuka}. The last term can be evaluated using the Wetterich equation \cite{Wetterich}
\begin{equation}
 	\frac{d\Gamma_{\Lambda}[\phi]}{dt} = \frac{1}{2} \textrm{Tr}\left[ \left(\frac{\delta^{2} \Gamma_{\Lambda}}{\delta \phi \delta \phi} + \mathcal{R}_{\Lambda}\right)^{-1} \frac{d \mathcal R_{\Lambda}}{dt}\right],
\end{equation}
yielding
\begin{equation*}
	\lim_{\delta\Lambda \rightarrow 0} \Lambda \frac{\delta \Gamma _{\Lambda}}{\delta{\Lambda}} = \frac{1}{2} \textrm{Tr}\left[ \left(\frac{\delta^{2} \Gamma_{\Lambda}}{\delta \phi \delta \phi} + \mathcal{R}_{\Lambda}\right)^{-1} \frac{\delta \mathcal{R}_{\Lambda}}{\delta \Lambda}\right].
\end{equation*}

The right hand side is equal to the expansion in terms of all 1-particle-irreducible diagrams, so we only have to find all the 1-particle-irreducible diagrams proportional to the four point interactions. We'll evaluate the diagrams in the sharp cutoff limit, where\\ $\delta \mathcal{R}_{\Lambda} = \Lambda \delta(|p|-\Lambda) \delta\Lambda$. Furthermore, we use spherical coordinates in which
\begin{equation*}
	\int'dp := \int_{-\infty}^{\infty} \frac{d^4p}{(2\pi)^4} \Lambda \delta(|p|-\Lambda) = \frac{1}{(2\pi)^4} \int_{0}^{\infty} dp\, p^3 \int d\Omega \Lambda \delta(|p|-\Lambda),
\end{equation*}
where the $\Omega$ represents the spherical part. If there is no angular dependence this part is the surface area of a 3-dimensional sphere, which is $2\pi^2$. In the calculations of the diagrams we'll make use of the following integrals, which follow from symmetry arguments
\begin{align*}
	\int'dp \frac{p_{\mu}p_{\nu}}{p^{2n}} &= \frac{\Lambda^{6-2n}}{8\pi^2}\frac{1}{4} \eta_{\mu\nu}\\
	\int'dp \frac{p_{\mu}p_{\nu}p_{\rho}p_{\sigma}}{p^{2n}} &= \frac{\Lambda^{8-2n}}{8\pi^2} \frac{1}{24} (\eta_{\mu\nu} \eta_{\rho\sigma} + \eta_{\mu\rho} \eta_{\nu\sigma} + \eta_{\mu\sigma} \eta_{\nu\rho})\\
	\int'dp \frac{p_{\alpha}p_{\beta}p_{\mu}p_{\nu}p_{\rho}p_{\sigma}}{p^{2n}} &= \frac{\Lambda^{10-2n}}{8\pi^2} \frac{1}{192} (\eta_{\alpha\beta} \eta_{\mu\nu} \eta_{\rho\sigma} + \eta_{\alpha\beta} \eta_{\mu\rho} \eta_{\nu\sigma} + \eta_{\alpha\beta} \eta_{\mu\sigma} \eta_{\nu\rho} + \eta_{\alpha\mu} \eta_{\beta\nu} \eta_{\rho\sigma}\\
	& + \eta_{\alpha\mu} \eta_{\beta\rho} \eta_{\nu\sigma} + \eta_{\alpha\mu} \eta_{\beta\sigma} \eta_{\nu\rho} + \eta_{\alpha\nu} \eta_{\beta\mu} \eta_{\rho\sigma} + \eta_{\alpha\nu} \eta_{\beta\rho} \eta_{\mu\sigma} + \eta_{\alpha\nu} \eta_{\beta\sigma} \eta_{\mu\rho}\\
	& + \eta_{\alpha\rho} \eta_{\beta\mu} \eta_{\nu\sigma} + \eta_{\alpha\rho} \eta_{\beta\nu} \eta_{\mu\sigma} + \eta_{\alpha\rho} \eta_{\beta\sigma} \eta_{\mu\nu} + \eta_{\alpha\sigma} \eta_{\beta\mu} \eta_{\nu\rho} + \eta_{\alpha\sigma} \eta_{\beta\nu} \eta_{\mu\rho}\\
	& + \eta_{\alpha\sigma} \eta_{\beta\rho} \eta_{\mu\nu}).\\
\end{align*}
In addition, we make extensive use of the following identities
\begin{align*}
	\{\gamma^{\mu},\gamma^{\nu}\} &= 2\eta^{\mu\nu} I_{4}\\
	\gamma^{\mu}\gamma_{\mu} &= 4 I_{4}\\
	\gamma^{\mu}\gamma^{\nu}\gamma_{\mu} &= -2 \gamma^{\nu}\\
	\gamma^{\mu}\gamma^{\nu}\gamma^{\rho}\gamma_{\mu} &= 4 \eta^{\nu\rho} I_4\\
	\gamma^{\mu}\gamma^{\nu}\gamma^{\rho} &= \eta^{\mu\nu} \gamma^{\rho} + \eta^{\nu\rho} \gamma^{\mu} - \eta^{\mu\rho} \gamma^{\nu} - i \varepsilon^{\sigma\mu\nu\rho} \gamma_{\sigma} \gamma^{5}\\	
	\textrm{Tr}(\gamma^{\mu}\gamma^{\nu}) &= 4 \eta^{\mu\nu}\\
	\textrm{Tr}(\gamma^{5}) = \textrm{Tr}(\gamma^{\mu}\gamma^{\nu}\gamma^{5}) &= 0,\\
	\varepsilon_{\alpha\mu\nu\rho}\varepsilon^{\beta\mu\nu\rho} &= - 6 \delta_{\alpha}^{\beta}.
\end{align*}
Finally, we use the identities \cite{Kusafuka}
\begin{equation*}
	2 \sum_{A=1}^{N_c} \left(t^A\right)_{d}^{a} \left(t^A\right)_{b}^{c} = \delta_{b}^{a} \delta_{d}^{c} - \frac{1}{N_c} \delta_{d}^{a} \delta_{b}^{c}
\end{equation*}
and \cite{Kusafuka}
\begin{align*}
	\bar{L}_{a i} \gamma^{\mu} L^{a i} \bar{R}_{b j} \gamma_{\mu} R^{b j} &= - 2 \bar{L}_{a i} R^{b j} \bar{R}_{b j} L^{a i}\\
	\bar{L}_{a i} \gamma^{\mu} L^{b i} \bar{R}_{b j} \gamma_{\mu} R^{a j} &= - 2 \bar{L}_{a i} R^{a j} \bar{R}_{b j} L^{b i}\\
	\bar{L}_{a i} \gamma^{\mu} L^{a i} \bar{L}_{b j} \gamma_{\mu} L^{b j} &= \bar{L}_{a i} \gamma^{\mu} L^{b j} \bar{L}_{b j} \gamma_{\mu} L^{a i}, \quad (L\rightarrow R)\\
	\bar{L}_{a i} \gamma^{\mu} L^{b i} \bar{L}_{b j} \gamma_{\mu} L^{a j} &= \bar{L}_{a i} \gamma^{\mu} L^{a j} \bar{L}_{b j} \gamma_{\mu} L^{b i}, \quad (L\rightarrow R).
\end{align*}

Using the Feynman rules from appendix \ref{Ap:FRules}, we find the following contributing 1PI diagrams:
\begin{figure}[H]
	\begin{center}
		\includegraphics[width=\linewidth]{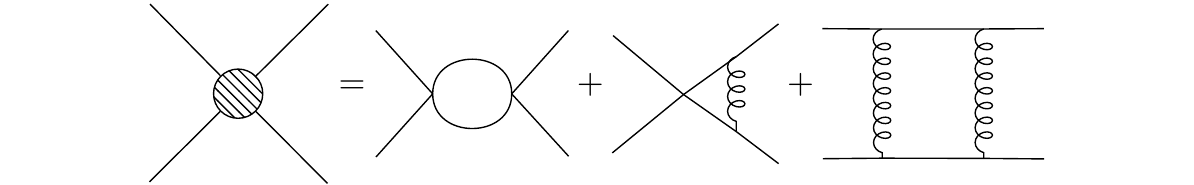}
	\end{center}
\end{figure}
There are 29 different diagrams of the first type, 12 of the second type and 4 of the third.
Five diagrams of the first type contribute to the beta function of $G_{S}$. One of those is proportional to $g_{S}^2$:
\begin{figure}[H]
	\begin{center}
		\includegraphics[width=\linewidth]{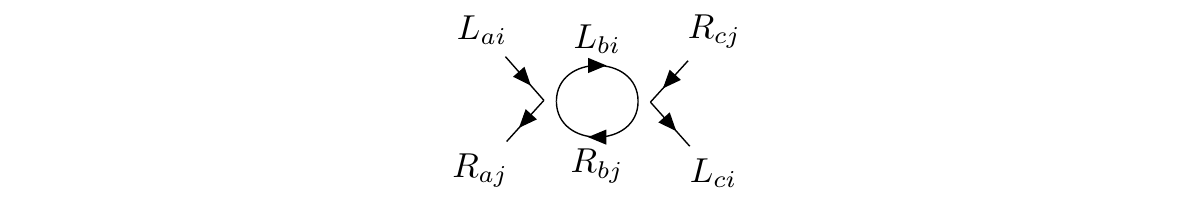}
	\end{center}
\end{figure}
\noindent which is evaluated as
\begin{align*}
	&\left(\frac{2G_s}{\Lambda^2}\right)^2 N_c \int' dp \bar{L}_{ai}R^{aj} \textrm{Tr} \left[ \frac{1-\gamma^5}{2} \frac{i}{\slashed{p}} \frac{1+\gamma^5}{2}\frac{1+\gamma^5}{2} \frac{i}{\slashed{p}} \frac{1-\gamma^5}{2} \right] \bar{R}_{cj}L^{ci} \\
	&= -\frac{1}{2} \frac{4G_{S}^{2}}{\Lambda^4} N_c \int'dp \frac{p_{\mu}p_{\nu}}{p^{4}} \textrm{Tr} \left[ \gamma^{\mu} (1+\gamma^5) \gamma^{\nu} \right] \frac{\mathcal{O}_{S}}{2}\\
	&= - \frac{G_{S}^{2}}{\Lambda^4} N_c \frac{\Lambda^2}{8\pi^{2}}\frac{\eta_{\mu\nu}}{4} \textrm{Tr} \left[ \gamma^{\mu} (1+\gamma^5) \gamma^{\nu} \right] \mathcal{O}_{S}\\
	&= - \frac{N_c G_{S}^{2}}{2\pi^2 \Lambda^2} \mathcal{O}_{S}.
\end{align*}
One diagram is proportional to $g_{S}g_{V}$:
\begin{figure}[H]
	\begin{center}
		\includegraphics[width=\linewidth]{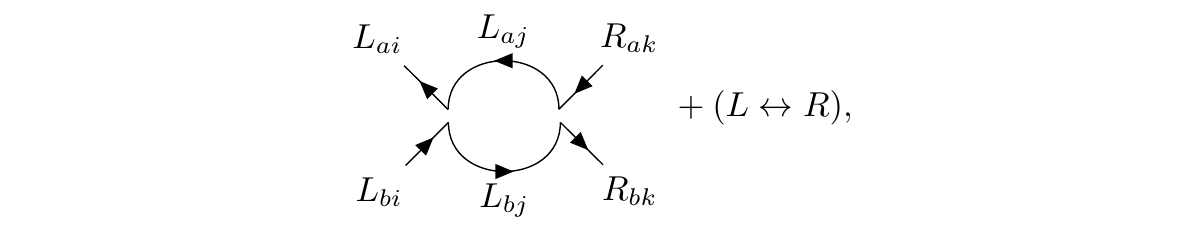}
	\end{center}
\end{figure}
\noindent which is evaluated as\footnote{Due to lengthiness of the calculations, we have suppressed further explicit calculations of the diagrams in this article.}
\begin{equation*}
	\frac{N_f G_{S}G_{V}}{2\pi^2 \Lambda^2} \mathcal{O}_{S}.
\end{equation*}
Two diagrams are proportional to $g_{S}g_{V_1}$:
\begin{figure}[H]
	\begin{center}
		\includegraphics[width=\linewidth]{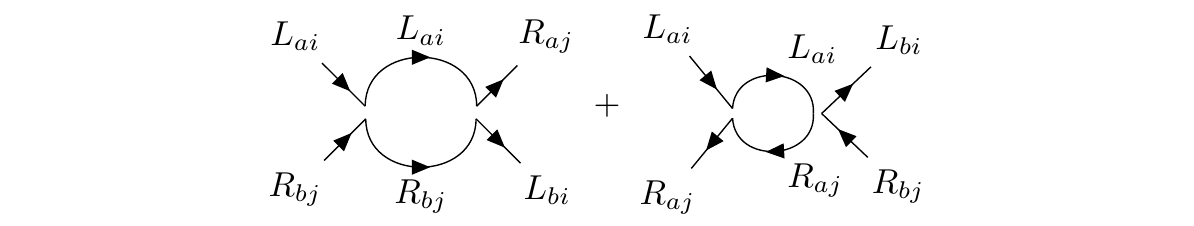}
	\end{center}
\end{figure}
\noindent which are evaluated as
\begin{equation*}
	-\frac{G_{S}G_{V_1}}{2\pi^2 \Lambda^2} \mathcal{O}_{S} \quad \textrm{and} \quad \frac{2 G_{S}G_{V_1}}{\pi^2 \Lambda^2} \mathcal{O}_{S}.
\end{equation*}
One diagram is proportional to $g_{S}g_{V_2}$:
\begin{figure}[H]
	\begin{center}
		\includegraphics[width=\linewidth]{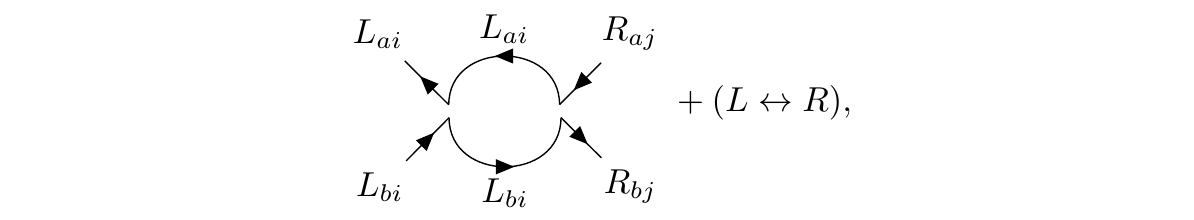}
	\end{center}
\end{figure}
\noindent which is evaluated as
\begin{equation*}
	\frac{G_{S}G_{V_2}}{2\pi^2 \Lambda^2} \mathcal{O}_{S}.
\end{equation*}

Five diagrams of the first type contribute to the beta function of $G_{V}$. One of those is proportional to $g_{S}^2$:
\begin{figure}[H]
	\begin{center}
		\includegraphics[width=\linewidth]{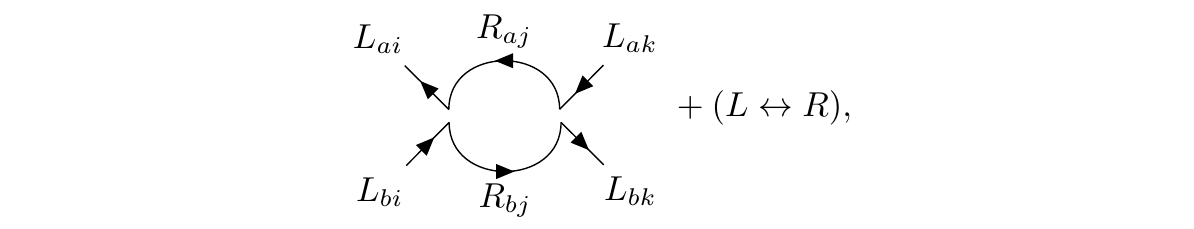}
	\end{center}
\end{figure}
\noindent which is evaluated as
\begin{equation*}
	\frac{N_f G_{S}^2}{16\pi^2 \Lambda^2} \mathcal{O}_{V}.
\end{equation*}
Two diagrams are proportional to $g_{V}^2$:
\begin{figure}[H]
	\begin{center}
		\includegraphics[width=\linewidth]{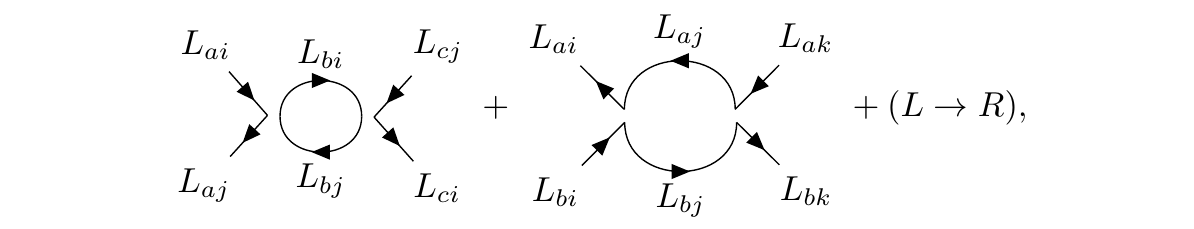}
	\end{center}
\end{figure}
\noindent which are evaluated as
\begin{equation*}
	\frac{N_c G_{V}^2}{4\pi^2 \Lambda^2} \mathcal{O}_{V} \quad \textrm{and} \quad \frac{N_f G_{V}^2}{4\pi^2 \Lambda^2} \mathcal{O}_{V}.
\end{equation*}
Two diagrams are proportional to $g_{V}g_{V_2}$:
\begin{figure}[H]
	\begin{center}
		\includegraphics[width=\linewidth]{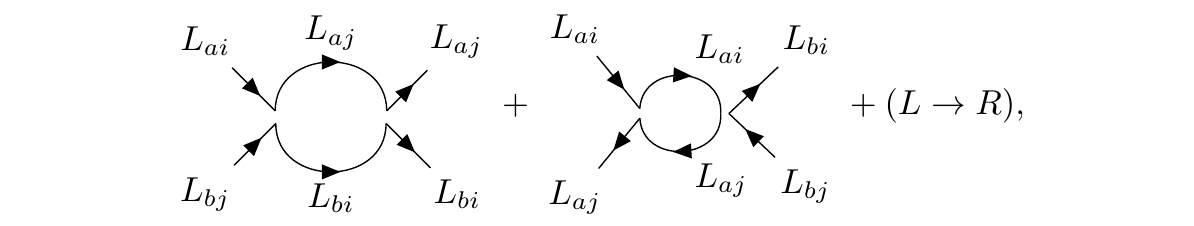}
	\end{center}
\end{figure}
\noindent which are evaluated as
\begin{equation*}
	-\frac{2 G_{V}G_{V_2}}{\pi^2 \Lambda^2} \mathcal{O}_{V} \quad \textrm{and} \quad \frac{G_{V}G_{V_2}}{2\pi^2 \Lambda^2} \mathcal{O}_{V}.
\end{equation*}

Nine diagrams of the first type contribute to the beta function of $G_{V_1}$. One of those is proportional to $g_{S}^{2}$:
\begin{figure}[H]
	\begin{center}
		\includegraphics[width=\linewidth]{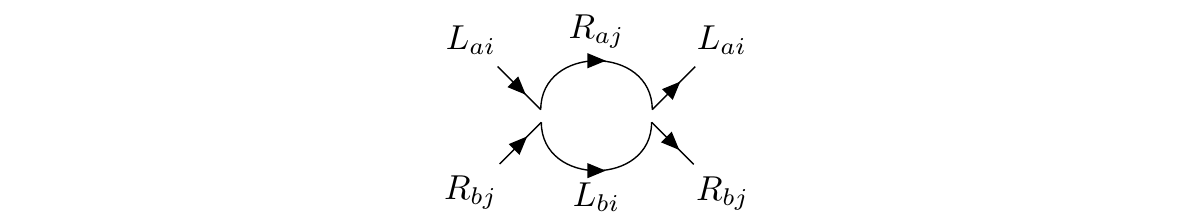}
	\end{center}
\end{figure}
\noindent which is evaluated as
\begin{equation*}
	- \frac{G_{S}^2}{16\pi^2 \Lambda^2} \mathcal{O}_{V_1}.
\end{equation*}
One diagram is proportional to $g_{S}g_{V}$:
\begin{figure}[H]
	\begin{center}
		\includegraphics[width=\linewidth]{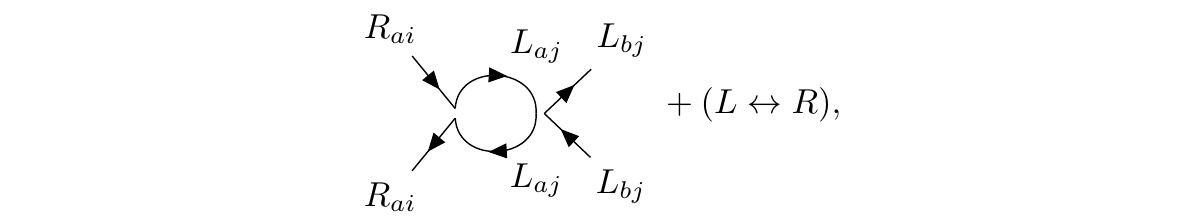}
	\end{center}
\end{figure}
\noindent which is evaluated as
\begin{equation*}
	- \frac{G_{S}G_{V}}{4\pi^2 \Lambda^2} \mathcal{O}_{V_1}.
\end{equation*}
Two diagrams are proportional to $g_{V_1}^2$:
\begin{figure}[H]
	\begin{center}
		\includegraphics[width=\linewidth]{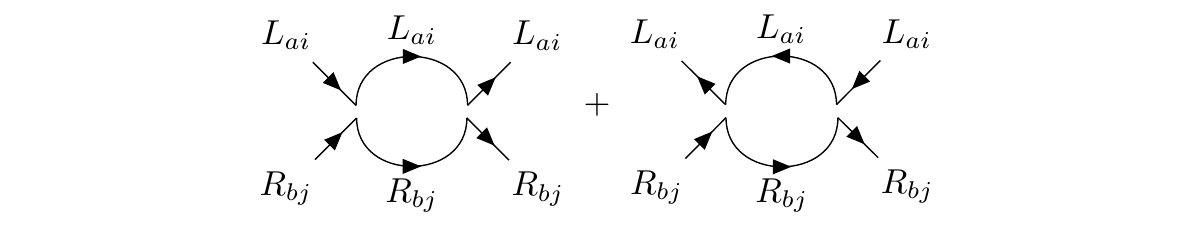}
	\end{center}
\end{figure}
\noindent which are evaluated as
\begin{equation*}
	\frac{G_{V_1}^2}{4\pi^2 \Lambda^2} \mathcal{O}_{V_1} \quad \textrm{and} \quad -\frac{G_{V_1}^2}{\pi^2 \Lambda^2} 	\mathcal{O}_{V_1}.
\end{equation*}
One diagram is proportional to $g_{S}g_{V_2}$:
\begin{figure}[H]
	\begin{center}
		\includegraphics[width=\linewidth]{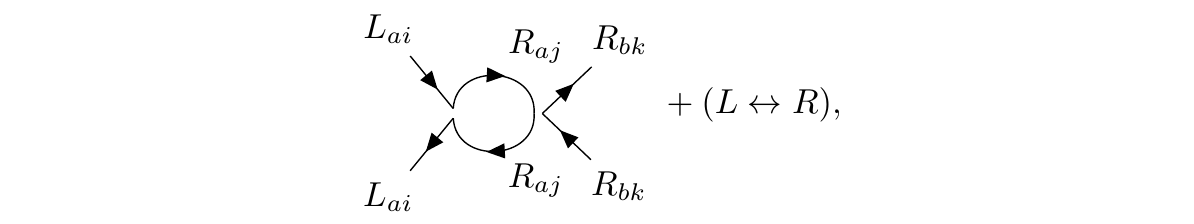}
	\end{center}
\end{figure}
\noindent which is evaluated as
\begin{equation*}
	- \frac{N_f G_{S} G_{V_2}}{4\pi^2 \Lambda^2} \mathcal{O}_{V_1}.
\end{equation*}
Two diagrams are proportional to $g_{V}g_{V_1}$:
\begin{figure}[H]
	\begin{center}
		\includegraphics[width=\linewidth]{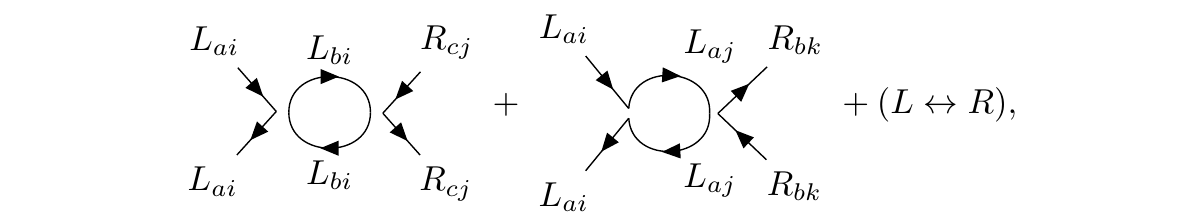}
	\end{center}
\end{figure}
\noindent which are evaluated as
\begin{equation*}
	\frac{N_c G_{V} G_{V_1}}{2\pi^2 \Lambda^2} \mathcal{O}_{V_1} \quad \textrm{and} \quad 	\frac{N_f G_{V} G_{V_1}}{2\pi^2 \Lambda^2} \mathcal{O}_{V_1}.
\end{equation*}
Two diagrams are proportional $g_{V_1}g_{V_2}$:
\begin{figure}[H]
	\begin{center}
		\includegraphics[width=\linewidth]{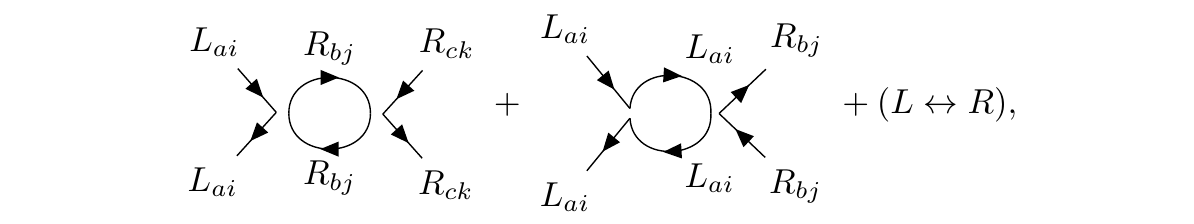}
	\end{center}
\end{figure}
\noindent which are evaluated as
\begin{equation*}
	\frac{N_c N_f G_{V_1} G_{V_2}}{2\pi^2 \Lambda^2} \mathcal{O}_{V_1} \quad \textrm{and} \quad \frac{G_{V_1} G_{V_2}}{2\pi^2 \Lambda^2} \mathcal{O}_{V_1}.
\end{equation*}

Ten diagrams of the first type contribute to the beta function of $G_{V_2}$. Two of those are proportional to $g_{V}^2$:
\begin{figure}[H]
	\begin{center}
		\includegraphics[width=\linewidth]{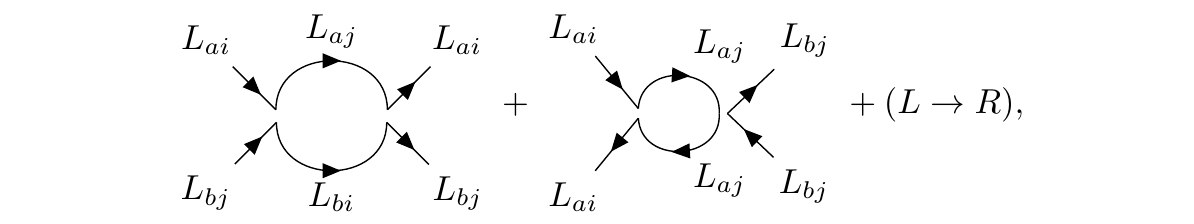}
	\end{center}
\end{figure}
\noindent which are evaluated as
\begin{equation*}
	-\frac{G_{V}^2}{\pi^2 \Lambda^2} \mathcal{O}_{V_2} \quad \textrm{and} \quad \frac{G_{V}^2}{4\pi^2 \Lambda^2} \mathcal{O}_{V_2}.
\end{equation*}
One diagram is proportional to $g_{V_1}^2$:
\begin{figure}[H]
	\begin{center}
		\includegraphics[width=\linewidth]{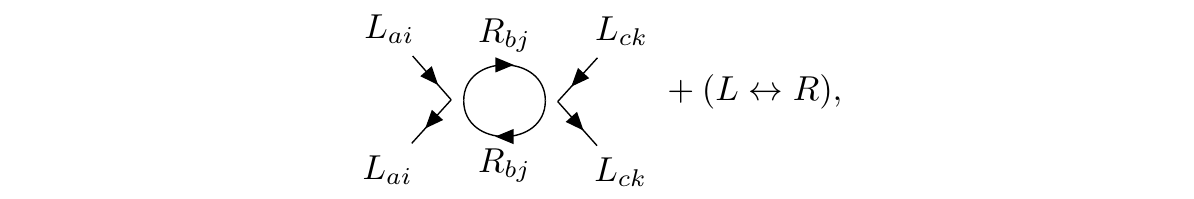}
	\end{center}
\end{figure}
\noindent which is evaluated as
\begin{equation*}
- \frac{N_c N_f G_{V_1}^2}{4\pi^2 \Lambda^2} \mathcal{O}_{V_2}.
\end{equation*}
Four diagrams are proportional to $g_{V_2}^2$:
\begin{figure}[H]
	\begin{center}
		\includegraphics[width=\linewidth]{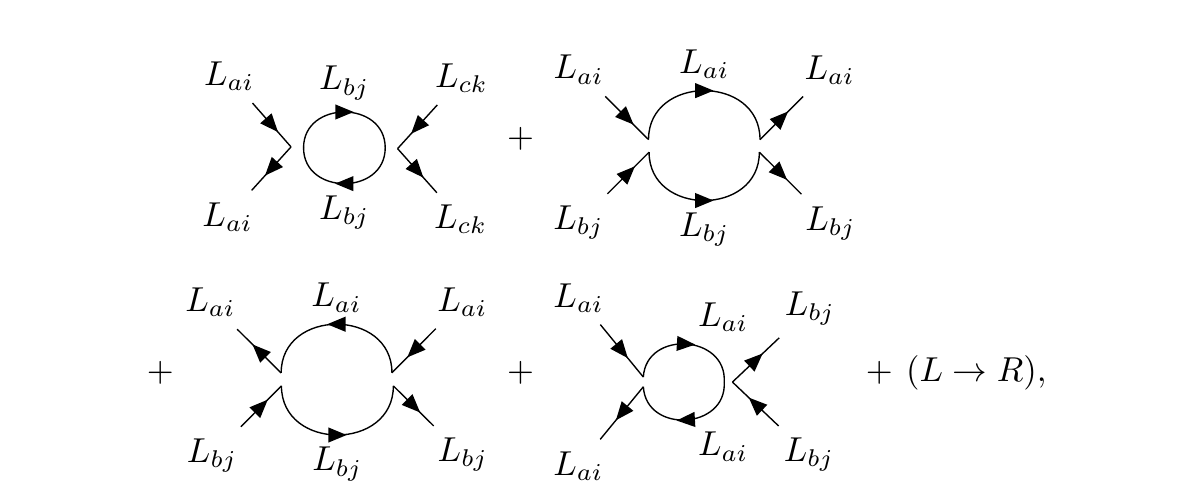}
	\end{center}
\end{figure}
\noindent which are evaluated as
\begin{equation*}
	\frac{N_c N_f G_{V_2}^2}{4\pi^2 \Lambda^2} \mathcal{O}_{V_2}, \quad -\frac{G_{V_2}^2}{\pi^2 \Lambda^2} \mathcal{O}_{V_2}, \quad \frac{G_{V_2}^2}{4\pi^2 \Lambda^2} \mathcal{O}_{V_2} \quad \textrm{and} \quad \frac{G_{V_2}^2}{4\pi^2 \Lambda^2} \mathcal{O}_{V_2}.
\end{equation*}
One diagram is proportional $g_{S}g_{V_1}$:
\begin{figure}[H]
	\begin{center}
		\includegraphics[width=\linewidth]{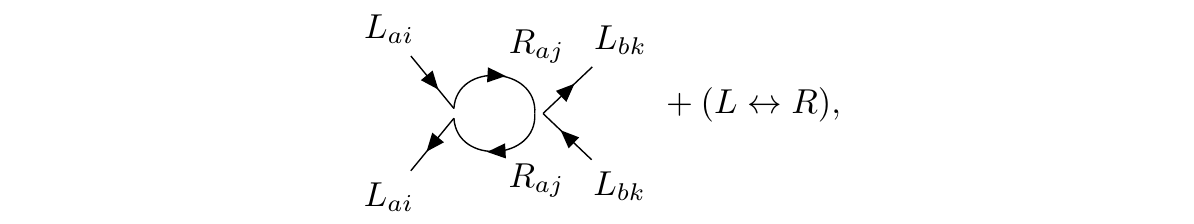}
	\end{center}
\end{figure}
\noindent which is evaluated as
\begin{equation*}
	-\frac{N_f G_S G_{V_1}}{4\pi^2 \Lambda^2} \mathcal{O}_{V_2}.
\end{equation*}
Two diagrams are proportional $g_{V}g_{V_2}$:
\begin{figure}[H]
	\begin{center}
		\includegraphics[width=\linewidth]{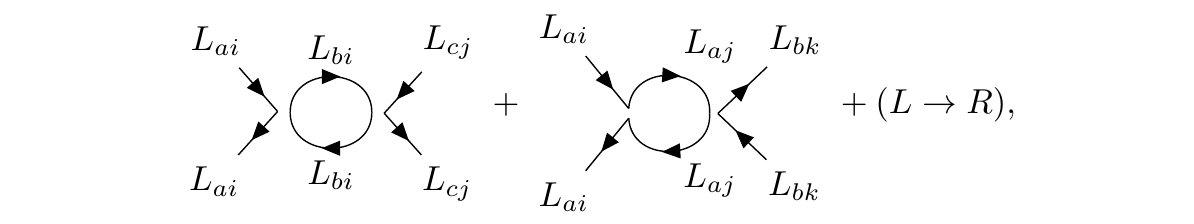}
	\end{center}
\end{figure}
\noindent which are evaluated as
\begin{equation*}
	\frac{N_c G_V G_{V_2}}{2\pi^2 \Lambda^2} \mathcal{O}_{V_2} \quad \textrm{and} \quad \frac{N_f G_V G_{V_2}}{2\pi^2 \Lambda^2} \mathcal{O}_{V_2}.
\end{equation*}

Three diagrams of the second type are proportional to $g_{S} \alpha_{g}$ and contribute to the beta functions of $G_{S}$ and $G_{V_1}$:
\begin{figure}[H]
	\begin{center}
		\includegraphics[width=\linewidth]{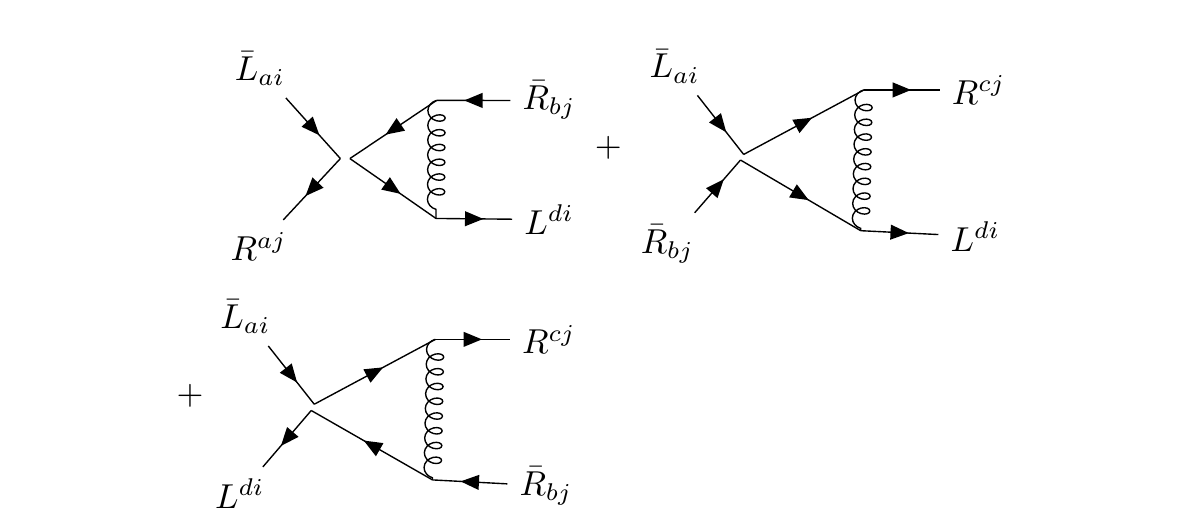}
	\end{center}
\end{figure}
\noindent The second and the third both evaluate to $0$ and the first is evaluated as
\begin{equation*}
	-3\left(N_c - \frac{1}{N_c}\right) \frac{G_S g^2}{8\pi^2 \Lambda^2} \mathcal{O}_S.
\end{equation*}
Three diagrams of the second type are proportional to $g_{V_1} \alpha_{g}$ and contribute to the beta functions of $G_{S}$ and $G_{V_1}$:
\begin{figure}[H]
	\begin{center}
		\includegraphics[width=\linewidth]{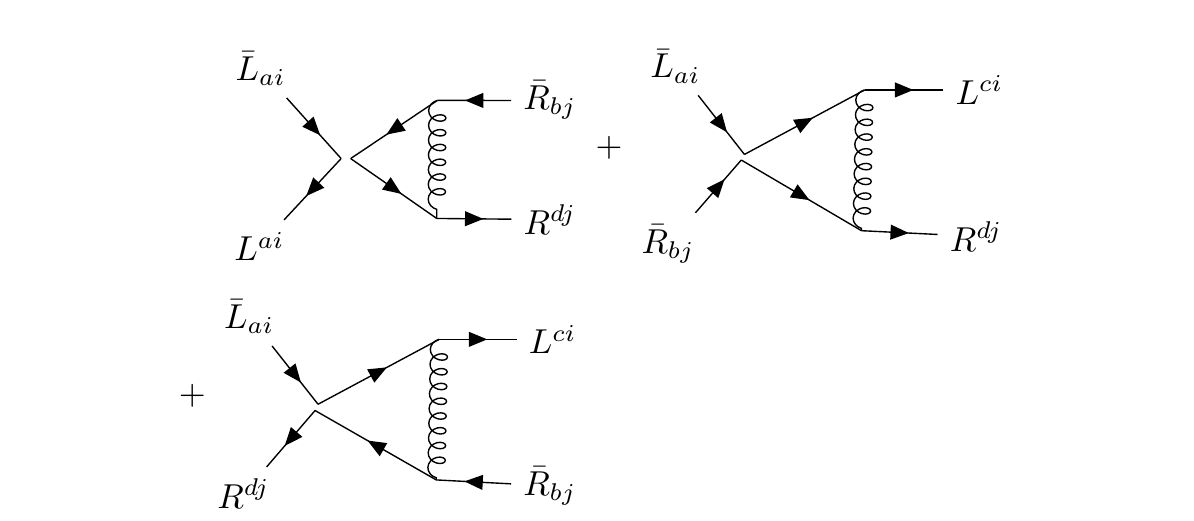}
	\end{center}
\end{figure}
\noindent The first and second both evaluate to $0$ and the third evaluates to
\begin{equation*}
	6\frac{G_{V_1} g^2}{8\pi^2 \Lambda^2} \mathcal{O}_S + \frac{3}{N_c}\frac{G_{V_1} g^2}{8\pi^2 \Lambda^2} \mathcal{O}_{V_1}.
\end{equation*}
Three diagrams of the second type are proportional to $g_{V} \alpha_{g}$ and contribute to the beta functions of $G_{V}$ and $G_{V_2}$:
\begin{figure}[H]
	\begin{center}
		\includegraphics[width=\linewidth]{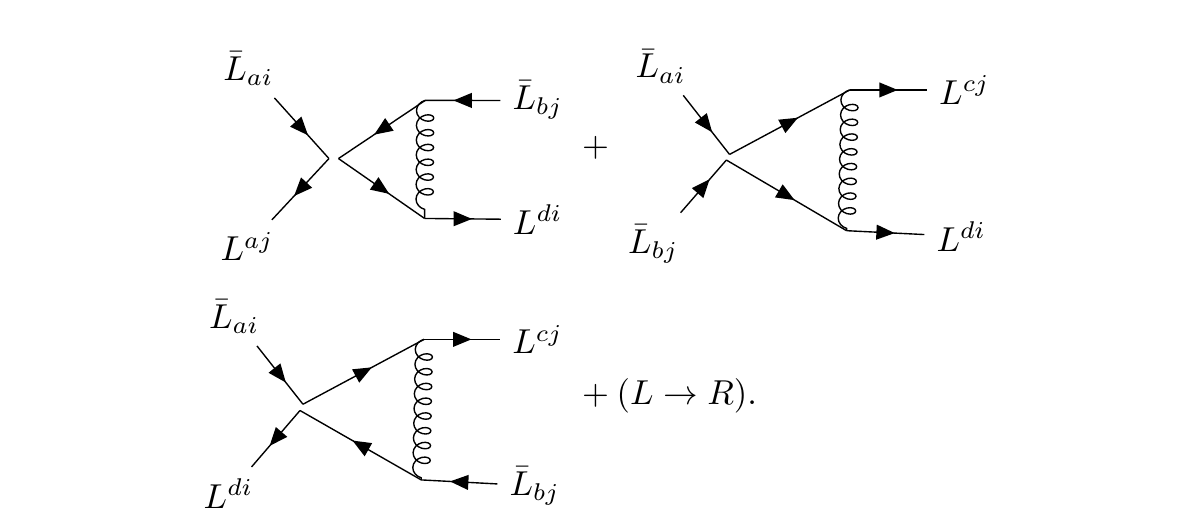}
	\end{center}
\end{figure}
\noindent The first and third evaluate to $0$ and the second is evaluated as
\begin{equation*}
	-\frac{3}{N_c}\frac{G_{V} g^2}{8\pi^2 \Lambda^2} \mathcal{O}_V + 3\frac{G_{V} g^2}{8\pi^2 \Lambda^2} \mathcal{O}_{V_2}.
\end{equation*}
Three diagrams of the second type are proportional to $g_{V_2} \alpha_{g}$ and contribute to the beta functions of $G_{V}$ and $G_{V_2}$:
\begin{figure}[H]
	\begin{center}
		\includegraphics[width=\linewidth]{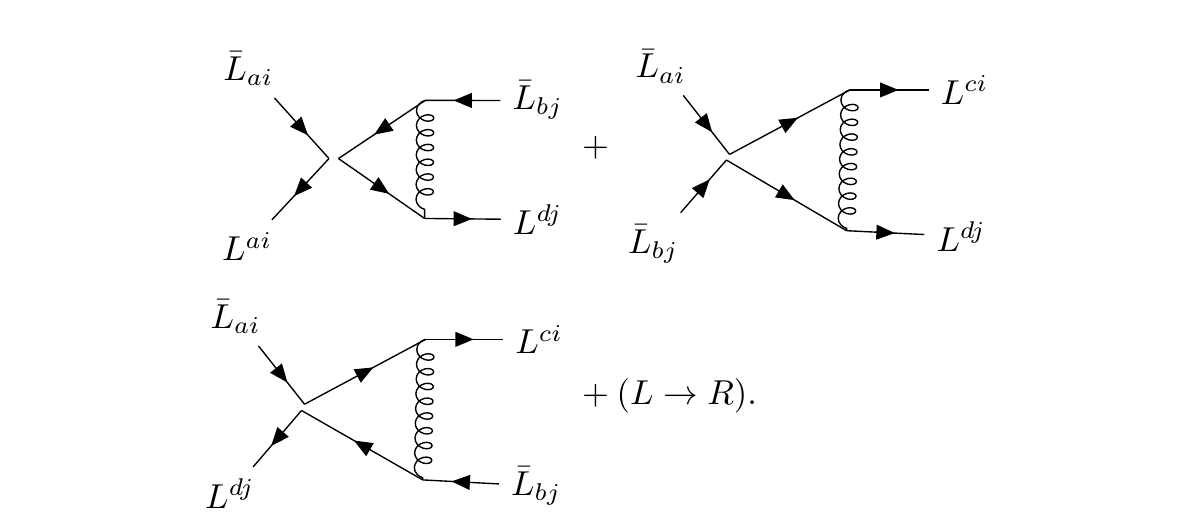}
	\end{center}
\end{figure}
\noindent The first and third evaluate to $0$, while the second is evaluated as
\begin{equation*}
	-\frac{3}{N_c}\frac{G_{V_2} g^2}{8\pi^2 \Lambda^2} \mathcal{O}_{V_2} + 3\frac{G_{V_2} g^2}{8\pi^2 \Lambda^2} \mathcal{O}_{V}.
\end{equation*}
\clearpage
Furthermore, the third type of correction is proportional to $\alpha_{g}^2$ and consists of 2 diagrams contributing to the beta functions of $G_{S}$ and $G_{V_1}$:
\begin{figure}[H]
	\begin{center}
		\includegraphics[width=\linewidth]{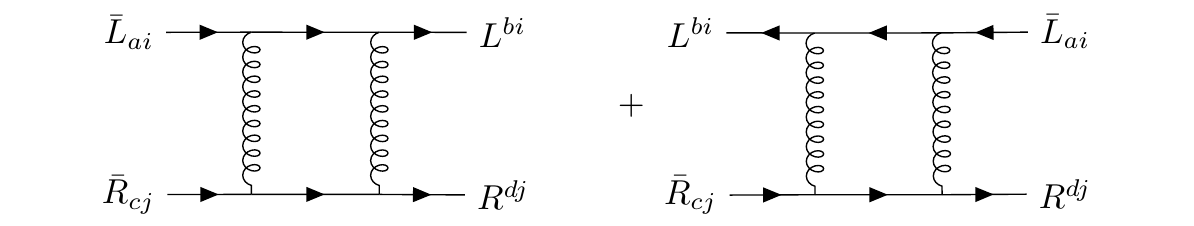}
	\end{center}
\end{figure}
\noindent which are evaluated as
\begin{align*}
	\frac{3}{16} \frac{2}{N_c} \frac{g^4}{8\pi^2\Lambda^2} \mathcal{O}_{S} + \frac{3}{32} \left( 1 + \frac{1}{N_c^2} \right) \frac{g^4}{8\pi^2\Lambda^2} \mathcal{O}_{V_1},\\
	-\frac{9}{16} \left(N_c - \frac{2}{N_{c}}\right) \frac{g^4}{8\pi^2\Lambda^2} \mathcal{O}_{S} + \frac{9}{32} \frac{1}{N_{c}^{2}} \frac{g^4}{8\pi^2\Lambda^2} \mathcal{O}_{V_1}.
\end{align*}
and two contributing to the beta functions of $G_{V}$ and $G_{V_2}$:
\begin{figure}[H]
	\begin{center}
		\includegraphics[width=\linewidth]{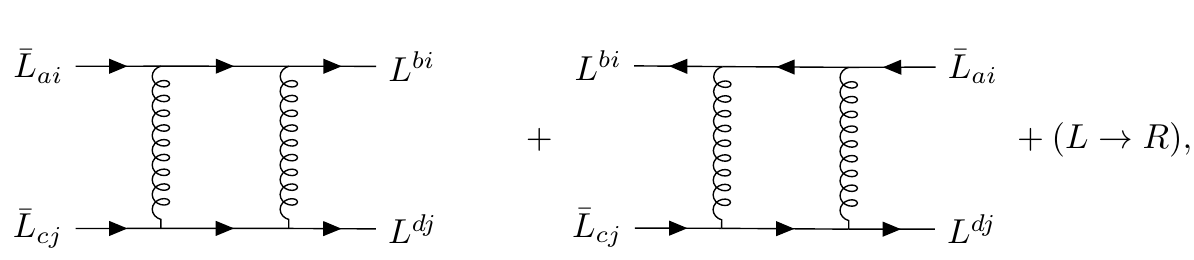}
	\end{center}
\end{figure}
\noindent which are evaluated as
\begin{align*}
	\frac{9}{32} \frac{2}{N_c} \frac{g^4}{8\pi^2\Lambda^2} \mathcal{O}_{V} - \frac{9}{32} \left(1 + \frac{1}{N_{c}^{2}} \right) \frac{g^4}{8\pi^2\Lambda^2} \mathcal{O}_{V_2},\\
	-\frac{3}{32} \left(N_c - \frac{2}{N_c}\right) \frac{g^4}{8\pi^2\Lambda^2} \mathcal{O}_{V} - \frac{3}{32} \frac{1}{N_{c}^{2}} \frac{g^4}{8\pi^2\Lambda^2} \mathcal{O}_{V_2}.
\end{align*}

\clearpage

\section{Beta Functions for QCD$_4$ with a scalar field}\label{Ap:QCD6}

When the scalar terms (\ref{eq:QCDYukA_Model1}) are added to the model (\ref{eq:QCD_Model1}), the set of beta functions is changed. As a consequence, new Feynman rules are added:
\begin{figure}[H]
	\begin{center}
		\includegraphics[width=\linewidth]{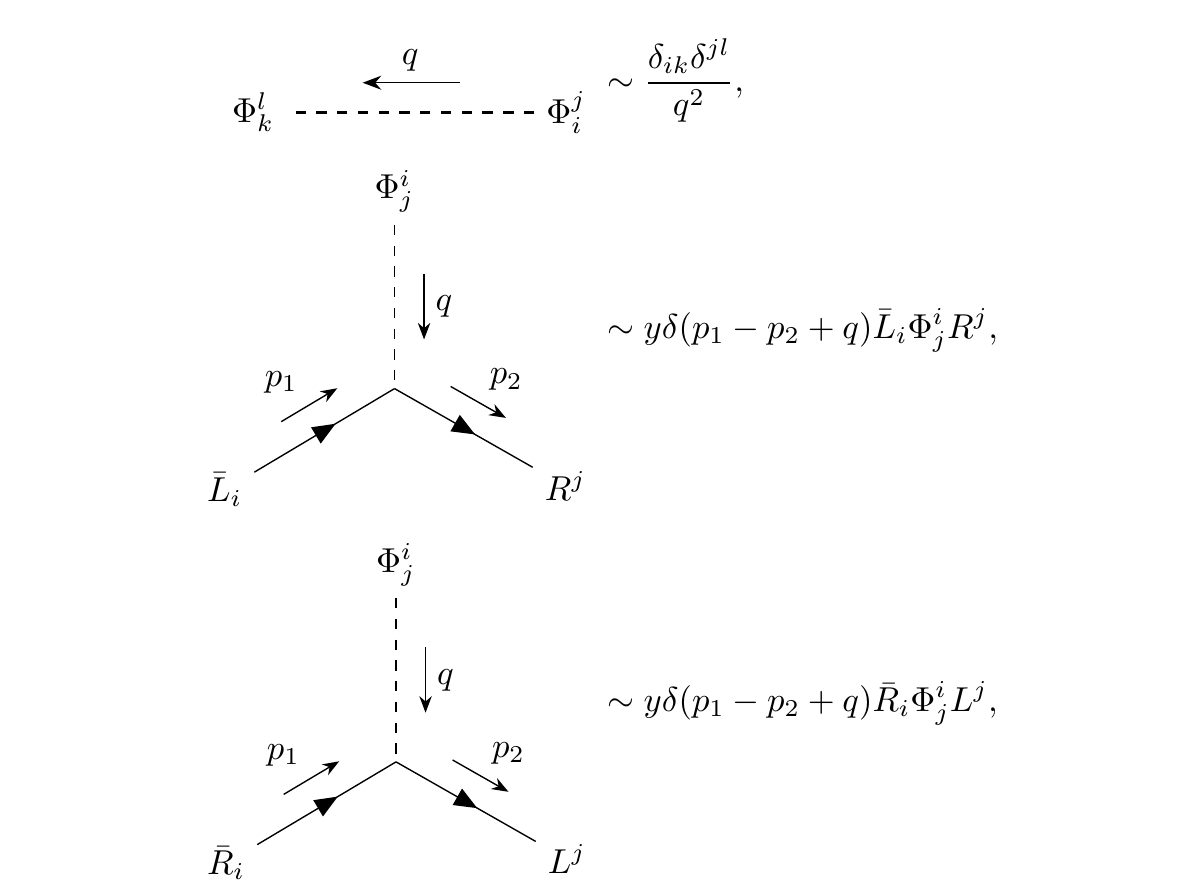}
	\end{center}
\end{figure}
First, we consider the function $\beta_{\alpha_{g}}$. Due to the addition of the scalar field terms, the two loop beta function for $\beta_{\alpha_{g}}$ is changed to \cite{VaughnI}
\begin{equation*}
	\beta_{g}^{[2]} \equiv \Lambda \frac{d \alpha_{g}}{d \Lambda} = -2 b_0 \alpha_{g}^{2} - 2 b_1 \alpha_{g}^{3} - 2 N_{f}^{2} \alpha_{y} \alpha_{g}^{2}.
\end{equation*}
The additional terms due to the four-fermion interactions can then be added as was done in appendix \ref{Ap:GaugeVertex}. The beta function $\beta_{\alpha_{y}}$ is given by \cite{VaughnII}
\begin{equation*}
	\Lambda \frac{d \alpha_y}{d \Lambda} = 2\alpha_y \left( \gamma_{\phi} + \gamma_{\bar{\psi}\psi} \right).
\end{equation*}
The anomalous dimension of the scalar field is given by $2 N_c \alpha_{y}$ \cite{VaughnI}. Using the four-fermi interactions, one can find a non-perturbative approximation to the anomalous dimension of the mass of the fermion field \cite{Terao} such that we get three contributing diagrams
\begin{figure}[H]
	\begin{center}
		\includegraphics[width=\linewidth]{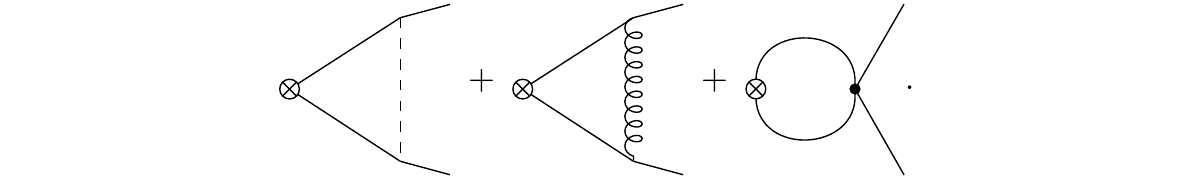}
	\end{center}
\end{figure}
\noindent The first two diagrams contribute $N_f \alpha_{y}$ and $- 6 \frac{N_{c}^{2} - 1}{2N_c}\alpha_{g}$, and two diagrams can be found that give a contribution to the second diagram:
\begin{figure}[H]
	\begin{center}
		\includegraphics[width=\linewidth]{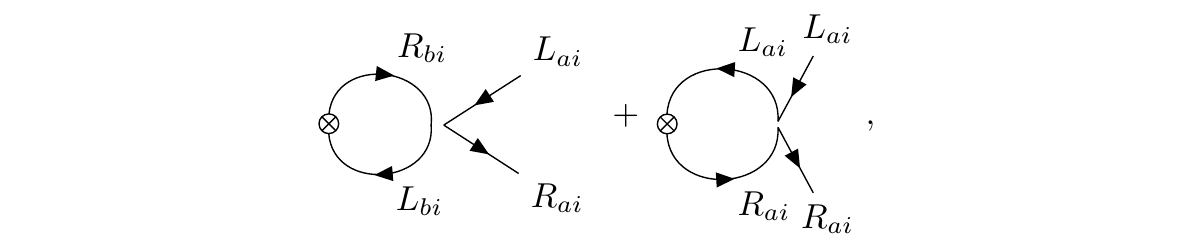}
	\end{center}
\end{figure}
\noindent which contribute $-2N_c g_S \quad \textrm{and} \quad 8 g_{V_1}$.\\

Next, we consider the beta functions four the four-fermi interactions, which are a slight variation of the ones found for the model without scalars in appendix \ref{Ap:FourFermi}. The anomalous dimension of the fermion fields is changed due to the presence of the scalar fields to $\eta=\alpha_y$ \cite{Terao}. Furthermore, we find additional contributions to
\begin{figure}[H]
	\begin{center}
		\includegraphics[width=\linewidth]{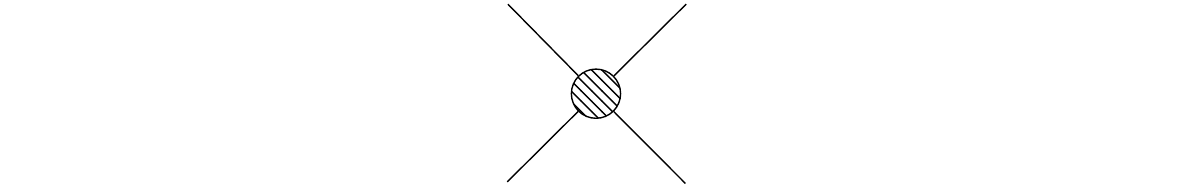}
	\end{center}
\end{figure}
\noindent represented by three types of diagrams
\begin{figure}[H]
	\begin{center}
		\includegraphics[width=\linewidth]{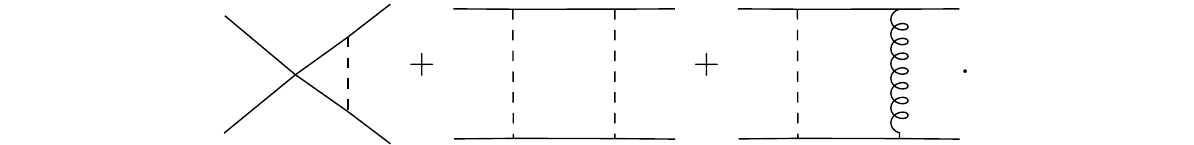}
	\end{center}
\end{figure}
\noindent Diagrams represented by the third type all evaluate to $0$. For the first type we find four diagrams that each contribute to one of the beta functions of the four-fermi interactions:
\begin{figure}[H]
	\begin{center}
		\includegraphics[width=\linewidth]{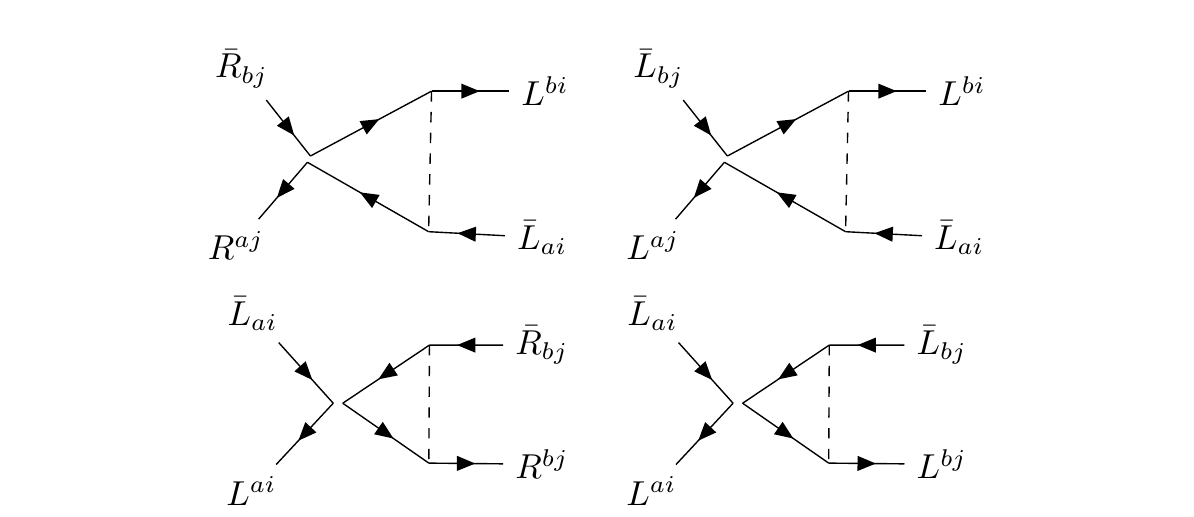}
	\end{center}
\end{figure}
\noindent contributing respectively $4 g_{V} \alpha_y$ to $\beta_{g_{S}}$, $g_{S} \alpha_y$ to $\beta_{g_{V}}$,  $-2 g_{V_2} \alpha_y$ to $\beta_{g_{V_1}}$ and $-2 g_{V_1} \alpha_y$ to $\beta_{g_{V_2}}$. Furthermore, for the second type there are 2 contributing diagrams:
\begin{figure}[H]
	\begin{center}
		\includegraphics[width=\linewidth]{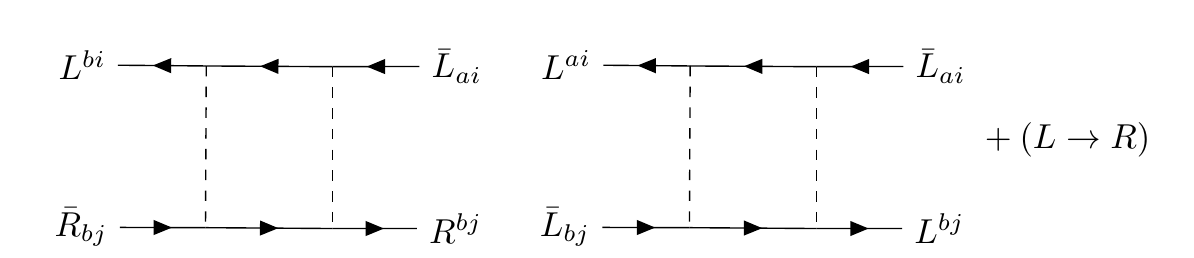}
	\end{center}
\end{figure}
\noindent contributing $2\alpha_{y}^{2}$ to $\beta_{g_{V_1}}$ and $2\alpha_{y}^{2}$ to $\beta_{g_{V_2}}$ respectively.

\end{document}